\documentclass[11pt]{article}

\usepackage{setspace,graphicx,epstopdf,amsmath,amsfonts,amssymb,amsthm,versionPO}
\usepackage{marginnote,datetime,enumitem,rotating,fancyvrb}
\usepackage{hyperref,float}
\usepackage{booktabs}
\usepackage[ruled, lined, linesnumbered, commentsnumbered, longend]{algorithm2e}
\usepackage[english]{babel}
\usepackage{natbib}
\usepackage{siunitx}
\usepackage{threeparttable}
\usepackage{subcaption}
\usepackage{array,xltabular,threeparttablex}
\usepackage{caption}
\usepackage{dcolumn}
\usepackage{tabularx}
\usepackage[T1]{fontenc}
\usepackage{multirow}
\usepackage{adjustbox}
\usepackage{pdflscape}
\usepackage{longtable}
\usepackage{tabu}
\usepackage{lscape}
\usepackage{afterpage}
\usepackage[autostyle, english = american]{csquotes}
\MakeOuterQuote{"}
\usepackage{makecell}

\newcolumntype{d}{D{.}{.}{4}}
\usdate

\excludeversion{notes}        % Include notes?
\includeversion{links}          % Turn hyperlinks on?

\iflinks{}{\hypersetup{draft=true}}

\ifnotes{%
\usepackage[margin=1in,paperwidth=10in,right=2.5in]{geometry}%
\usepackage[textwidth=1.4in,shadow,colorinlistoftodos]{todonotes}%
}{%
\usepackage[margin=1in]{geometry}%
\usepackage[disable]{todonotes}%
}

\makeatletter\let\chapter\@undefined\makeatother % Undefine \chapter for todonotes

\definecolor{darkgreen}{rgb}{0.0, 0.5, 0.0}
\definecolor{ashgrey}{rgb}{0.7, 0.75, 0.71}
\definecolor{babyblue}{rgb}{0.54, 0.81, 0.94}

\newcommand{\qlet}{\raisebox{-1pt}{\includegraphics[scale=0.042]{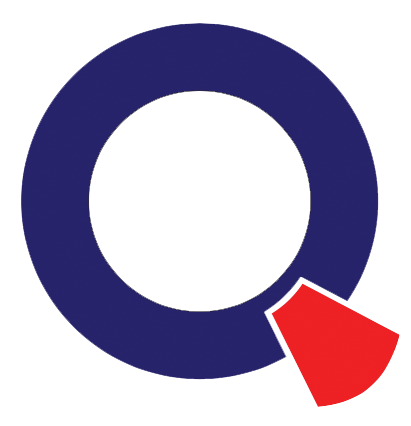}}}
\usepackage{indentfirst} % Indent first sentence of a new section.
\usepackage{jfe}          % JFE-specific formatting of sections, etc.

\newtheorem{assumption}{Assumption}[section]

\begin{document}

  %!TEX root = ../main.tex

\setlist{noitemsep}  % Reduce space between list items (itemize, enumerate, etc.)
%\onehalfspacing      % Use 1.5 spacing
% Use endnotes instead of footnotes - redefine \footnote command

\title{Networks of News and Cross-Sectional Returns$^*$
\footnotetext{$^*$ Financial supported by Deutsche Forschungsgemeinschaft via IRTG 1792 \enquote{High Dimensional Nonstationary Time Series}, Humboldt-Universit\"at zu Berlin, the European Union's Horizon 2020 research and innovation program \enquote{FIN-TECH: A Financial supervision and Technology compliance training programme} under the grant agreement No 825215 (Topic: ICT-35-2018, Type of action: CSA), the European Cooperation in Science \& Technology COST Action grant CA19130 - Fintech and Artificial Intelligence in Finance - Towards a transparent financial industry, the Yushan Scholar Program of Taiwan and the Czech Science Foundation's grant no. 19-28231X / CAS: XDA 23020303 are greatly acknowledged.
All supplementary and Python code can be found on Quantlet Repository: \href{https://github.com/QuantLet/NetworksofNews_CrossSectionalReturns}{\qlet NNCSR}
}
\footnotetext{$\dagger$ School of Business and Economics, Humboldt-Universit\"at zu Berlin. Spandauer Str. 1, 10178 Berlin, Germany. Email: junjie.hu@hu-berlin.de}
% \footnotetext{$\ddagger$ Department of Business and Economics, Berlin School of Economics and Law, Badensche Str. 52, 10825 Berlin, Germany. Email: packham@hwr-berlin.de}
\footnotetext{$\S$ Blockchain Research Center, Humboldt-Universit\"at zu Berlin, Germany; Wang Yanan Institute for Studies in Economics, Xiamen University, China; Sim Kee Boon Institute for Financial Economics, Singapore Management University, Singapore; Faculty of Mathematics and Physics, Charles University, Czech Republic; National Yang Ming Chiao Tung University, Taiwan. Email: haerdle@hu-berlin.de}
}

\author{Junjie Hu$^\dagger$, Wolfgang Karl Härdle$^\S$
% \\, Natalie Packham$^\ddagger$
%   Humboldt-Universit\"at zu Berlin\\
%   \\
  }

\date{First draft: April 2021\\This draft: October 2021}

\renewcommand{\thefootnote}{\fnsymbol{footnote}}

\singlespacing

\maketitle

\vspace{-.2in}
\begin{abstract}
\noindent 
We uncover networks from news articles to study cross-sectional stock returns. By analyzing a huge dataset of more than 1 million news articles collected from the internet, we construct time-varying directed networks of the S\&P500 stocks. The well-defined directed news networks are formed based on a modest assumption about firm-specific news structure, and we propose an algorithm to tackle type-I errors in identifying the stock tickers. We find strong evidence for the comovement effect between the news-linked stocks returns and reversal effect from the lead stock return on the 1-day ahead follower stock return, after controlling for many known effects. Furthermore, a series of portfolio tests reveal that the news network attention proxy, \emph{network degree}, provides a robust and significant cross-sectional predictability of the monthly stock returns. Among different types of news linkages, the linkages of within-sector stocks, large size lead firms, and lead firms with lower stock liquidity are crucial for cross-sectional predictability.

% The directed network uncovered from news articles is formed to study the cross-sectional stock returns. 

\end{abstract}

\medskip



\thispagestyle{empty}

\clearpage

\onehalfspacing
\setcounter{footnote}{0}
\renewcommand{\thefootnote}{\arabic{footnote}}
\setcounter{page}{1}

  %!TEX root = ../main.tex

\section{Introduction}

It is undeniable that the interdependency between firms is a major driver for stock returns; the practical problem is though how to capture this time-varying dependency structure. Among all the information sources, the fast-growing resource of news articles provides us via textual analysis a perspective to observe and define the connections between firms. Taking advantage of the advancement of computation and information technology, e.g., \enquote{Text Mining}, we are capable of capturing the linkages between firms from a huge dataset of more than 1 million news articles on NASDAQ.com. These linkages allow us to study cross-sectional stock returns based on news information.

Unlike the customer-supplier linkage in a supply chain, which has a natural representation of the linkage and direction, we first need to define the news linkage and its direction. With the directed linkages, we then study the impact of stock return via the news linkage and the structural information of the directed news networks.

We postulate a news linkage as a firm of main interest (\emph{lead}) pointing to a related firm (\emph{follower}) within a piece of news. By assuming that the firm-specific news has a structure of placing the firm of main interest on its headline and the related firms in its content, we uncover the directed news linkages based on the positions that those firms are identified, i.e., in the title or content. The assumption is modest because when writing firm-specific news reports, the authors will naturally emphasize the firm of interest on the headline and mention the other related firms in the content. Such definition on news linkage is different from the literature, where news linkage is formed if two firms are co-mentioned in one article (e.g., \cite{scherbina2015economic,chen2021media,guomedia,Ge2021}). Consequently, a directed news network is uncovered based on a collection of directed news linkages in a given period.

The task of firm tickers identification is the cornerstone of the later empirical analysis; however, the task is challenging as it could suffer from the type-I error if not handled carefully. We propose an algorithm that is dedicated to narrow down firm tickers identification of S\&P500 stocks only in the three most commonly seen and reliable scenarios with some specified exceptional cases. This algorithm is a trade-off between fuzzy and inefficient searching, dealing with the problem caused by the various ways of firm names written by a human. On the one hand, using keywords and tickers to identify firms is intuitive. However, this fuzzy searching falsely matches a huge amount of firms that are not in the news, i.e., also known as the type-I error.\footnote{We also did such an experiment, some firms were falsely identified for more than couple thousand times.} On the other hand, manually searching over the more than 1-million news articles in our sample is infeasible with limited time and resources. This algorithm will surely miss many firm occurrences, but for scientific prudence, a subset of the true set of news linkages gives conservative results.

Applying the proposed algorithm on a dataset of more than 1-million economic and finance related news articles published by 50+ online news publishers on the NASDAQ.com, we uncover the news network of S\&P500 stocks and conduct our empirical analysis focusing on two angles: \emph{Lead return}, the return of a linkage-weighted portfolio consisting of all the lead firms of a firm in a given news network, a similar definition can be found in \cite{COHEN_2008}, and \cite{scherbina2015economic}. The network attention proxy, \emph{network degree}, a well-known measure in complex network science and social network analysis (\cite{Albert_1999,barabasi1999emergence,jackson2010social}), intuitively counts the number of unique nodes that a firm directly connects with, therefore captures the \enquote{attention hype} of a firm.

Interestingly, we find that the detected news-linked firms, regardless of their business relationship, have a strong comovement effect on daily stock returns. We first calculate the daily stock residual return variables $r_{i,t}^{resid}$ from Fama-French 3-factor model (FF3, \cite{Fama_1993}) and Fama-French 5-factor model (FF5, \cite{FAMA20151}). As such variable $r_{i,t}^{resid}$ can be viewed as the part unexplainable by the known factors and firm alphas. By regressing $r_{i,t}^{resid}$ on the contemporaneous \emph{lead return} variables with control variables of firm characteristics, including market capitalization, book-market value, and firm liquidity, the econometric model shows that the daily stock residual return positively relates to the contemporaneous \emph{lead return} at 1\% level. The standard error is corrected by clustering on both individual and time. The regression result implies that the follower stock return moves with its lead stock return to the same direction on average over time. The results stay significant after considering whether the lead and follower firms belong to the same sector, though the comovement effect is more intensive between firms within the same sector. Moreover, the effect is significant at 1\% regardless of the sign of lead firm stock returns.

We further validate the comovement effect by forming portfolios based on \emph{lead return} of each firm. Of course, the portfolio test is infeasible as it requires knowing the stock returns of all lead firms and the news linkages on day-$t$ to form the portfolios on day-$t$, but the point here is to study the significance of stock returns comovement effects between the lead firms and follower firms. The trading rule is intuitive. First, all the S\&P500 constituents are sorted into quintiles in ascending order based on the contemporaneous \emph{lead return}. Then, the portfolios are equal-weighted and rebalanced each trading day. A significant difference in returns, averaging $165.15$ bps per day, between the quintile of highest \emph{lead return} and the lowest.

A reversal effect from \emph{lead return} on the 1-day ahead follower stock return emerges after controlling for the Fama-French factors and firm characteristics. We conduct the regressions of the 1-day ahead stock (residual) return variables $r_{i,t+1}$ $\left(r_{i,t+1}^{resid}\right)$ on the day-$t$ \emph{lead return} variables along with control variables including the firm characteristics and also contemporaneous stock return $r_{i,t}$. The regression results show that the \emph{lead return} has a significant negative impact on the 1-day ahead $r_{i,t+1}^{resid}$, despite the mixed effect on $r_{i,t+1}$, implying the existence of reversal effects from \emph{lead return} to stock return, nevertheless being dominated by many factors that offset the reversal effect. Furthermore, a daily rebalancing trading strategy based on sorting \emph{lead return} reveals that it gives only marginal cross-sectional predictability of 1-day ahead stock return. In the meanwhile, a strong reversal effect can be found with the accumulated residual stock returns.

We employ the \emph{network degree} as a proxy of measuring the \enquote{network attention hype} of a firm and find significant predictability of cross-sectional stock returns. \emph{Network degree} is motivated by our observation that a firm with more attention from news media has a higher likelihood of being found with other firms, i.e., a higher network degree. By conjecturing that a \enquote{hyped} firm, i.e., a firm with greater network attention, is more likely to outperform, we provide evidence that such information can be profitable with a simple long-short strategy. The S\&P500 stocks are sorted into quintiles each month based on the monthly aggregated network attention from the preceding month. The portfolio test shows that firms with higher network attention yield higher monthly returns in the next month. By investigating the attention variables of news networks formed by different types of news linkages, we find that the network attention, both being lead and follower, shows significant predictability; and the within-sector network attention variable carries much stronger predictability than the cross-sector network attention. Additionally, the lead firm size and stock liquidity are important considerations for predictability. Specifically, the network attention of the lead firm with a larger size and lower stock liquidity is more informative to predict the cross-sectional stock returns. To ensure that the predictability of \emph{network degree} is not only caused by some of the known effects, we provide robust evidence of controlling for firm size, value, liquidity, momentum, and also the selection of network formation window size.

% \subsection{Literature Review}

% The large literature in economics and finance discussing the network of firms can be roughly separated into three strands.
% Social networks have been studied in a range of disciplines spanning from computer science to macroeconomics and finance.

This paper contributes to a large literature that aims to distill interdependency information between firms from all the resources other than the stock market. For example, much of the literature focuses on the supply-chain networks and studies the risks implications at both macroeconomics level (e.g., \cite{gabaix2011granular,Acemoglu2012}) and the firm level (e.g., \cite{AHERN_2014,HERSKOVIC_2018,herskovic2020firm}), and also stock return predictability (e.g., \cite{COHEN_2008}). Moreover, some papers investigate the stock performance by utilizing common shareholders structure (e.g., \cite{ANTON2014}), air traveling connection (e.g., \cite{Da2021}).

Textual news information has been drawing attention in academia as powerful statistical tools, and commercial databases became readily available (e.g., \cite{Antweiler2004,da_fears,Zhang2016,boudoukh_information}). Recently, a new trend of studying the firm connection in text data has arisen. \cite{scherbina2015economic,chen2021media,guomedia} employ the Thomson Reuters News Analytics (TRNA) dataset and construct the linkages of firms co-mentioned in the same article, \cite{schwenkler2019network} employ the natural language processing tools to identify the firm tickers. \cite{scherbina2015economic} defines the lead-follower relationship between the news-linked firms by liquidity characteristic and identify the competitive relationships. As such, they show the predictability of returns of the lead firms. \cite{schwenkler2019network} investigate the market contagion by various interconnectivity measures. \cite{chen2021media} focus on using the strength of connection of firms to predict the stock returns correlations and how centrality matter for the predictability. \cite{Ge2021} calibrate a spatial factor model with the news-implied dependency and derive an asset pricing model.

Another strand of literature studying the interdependency of stocks focuses on modeling the stock price processes with statistical methods. For example, some papers study the connectedness of stocks by decomposing the covariance matrix (e.g., \cite{DIEBOLD2014119,Diebold_2008_Measuring}), some investigate the risk transmission between stocks (e.g., \cite{Hautsch_2014_Systemic,Haerdle_2016,Yang2017}), and some study the network effect with the vector autoregression framework (\cite{barigozzi2019nets,Zhu_2019}).

% The paper differentiates from the previous literature as follows. First, we propose a conservative algorithm to identify the firms and define the lead-follower relationship based on a modest assumption. Then we find the comovement effect between the news-linked firms followed by giving evidence on the marginal predictability of the returns of the lead firm. Finally, we show the persistent cross-sectional predictability of network degree.

The rest of the paper is organized as follows. 
Section \ref{sec:models} first defines the directed news networks and then introduces the two variables along with by the econometric and portfolio methods employed in the empirical analysis.
Section \ref{sec:empirical_network} details the proposed algorithm that identifies the tickers in S\&P500 constituents and summarizes the constructed empirical news network.
In Section \ref{sec:leadreturn_simul_pred} we give evidence on the comovement effect of the news-linked firms. Then we discuss the negative impact of \emph{lead return} on the 1-day ahead stock returns. 
In Section \ref{sec:networkdegree_predictability} we focus on the monthly predictability of the network attention proxy, \emph{network degree}, followed by the analysis of the importance of different types of linkages, and we conduct a series of robustness tests on the network attention predictability.

  % !TEX root = ../main.tex

\section{Methodologies and Models}\label{sec:models}

We first define a directed news network and then introduce the two variables derived from the news network, which are the two angles that relate the news networks with stock returns. Different from econometric literature, e.g., \cite{DIEBOLD2014119}, in which networks are uncovered under imposed models, the constructed adjacency matrix in this study are model-free and can be tested against econometric hypotheses. Finally, we introduce the econometric model and portfolio construction methods employed in the empirical analysis.

\subsection{A Directed Network on Firm-Specific News}\label{sec:newsnetwork_def}

% Why a directed network
News naturally embeds lead-follower relations since authors decide which firms to mention in headlines and will naturally mention the consequences for related firms. To define the lead-follower relationship, we first make a modest assumption on the structure of those firm-specific news articles.

\begin{assumption}
    The firm-specific news articles have the firm of main interest on the news headline and the related firms in the news content.
\end{assumption}

To understand the intuition behind this assumption, one can consider a situation where a news article is designated to cover firm $j$. The author most likely will mention firm $j$ on the article headline and mention the related firms in the content consequently for various purposes, such as potential competitors, cooperators, suppliers, customers, etc. We give a typical example of the firm-specific news in Figure \ref{fig:newsexample} of Appendix \ref{app:figures}, which demonstrate a piece of news about \emph{Apple}, where some other related firms, e.g., \emph{Intel}, \emph{Nike}, and \emph{Peleton}, are also mentioned in the content. Of course, not all the firm-specific news articles follow such structure; however, we argue that this assumption is naturally true for most firm-specific news and allows a clear definition of the lead-follower relationship.

Based on the assumption, we first define the firm that appears in the news title, firm $j$, as the \emph{"lead"} firm. Correspondingly in the same news article, each firm $i$ mentioned in the news content is defined as the \emph{"follower"} of firm $j$. Therefore, given a set of textual news articles aggregated in period $T$, $D_T=\{m1,\ldots,m_d\}$, we identify the set of news linkage pairs $l_{ij,T}$ of $(i,j)$ among $n$ firms, i.e. $i,j\in \{1,\ldots,n\}$ as

\begin{equation}
    l_{ij,T} \stackrel{\operatorname{def}}{=} \bigcup_{m_d\in D_T} \left\{\left(i,j\right)_{m_d} | j \text{ in } m_d \text{ title}, i \text{ in } m_d \text{ headline}, i\neq j\right\}.
\end{equation}

With the well-defined news linkage pairs, we then define the \emph{"News Network"}, $\mathcal{W}_T$ as follows. Given a weighted direct graph $\mathcal{G}=(\mathcal{V}, \mathcal{E})$ in a time period $T$ by the adjacency matrix $\mathcal{W}_T \stackrel{\operatorname{def}}{=} \left[\omega_{ij,T}\right]_{n\times n}=\left[n_{i,T}^{-1}a_{ij,T}\right]_{n\times n}$, where

\begin{equation}
    a_{ij,T} \stackrel{\operatorname{def}}{=} \#l_{ij,T},
\end{equation}

and $n_{i,T}=\sum^{n}_{j=1}a_{ij,T}$. The $a_{ij,T}$ counts the number of lead-follower pairs among all the articles in period $T$, and the adjacency matrix is row-normalized by $n_i$. Also, note that $\mathcal{W}_T$ is an asymmetric matrix with all zero diagonal values.

\subsubsection{Network Decomposition by Nodes}\label{sec:network_decomp}

The news networks can be decomposed by the nodes characteristics to further study the effects of various types of linkages. In this paper, we mainly study 3 types of decomposition. The first one is decomposition by sector. This decomposition tests whether a stock interacts with the stocks from different sectors differently compared to the ones from the same sector. Based on the sector defined by Global Industry Classification Standard (GICS), a news network $\mathcal{W}_T=\left[\omega_{ij,T}\right]$ can be fully decomposed into within-sector network $\mathcal{W}_T^{w}=\left[\omega_{ij,T}^{w}\right]$ and cross-sector network $\mathcal{W}_T^{c}=\left[\omega_{ij,T}^{c}\right]$, where

\begin{align}\label{def:within_cross_nn}
    \begin{split}
        \omega_{ij,T}^{w} &\stackrel{\operatorname{def}}{=} \mathbf{I}\{\text{i,j from same sector}\}\cdot\omega_{ij,T}\\
        \omega_{ij,T}^{c} &\stackrel{\operatorname{def}}{=} \mathbf{I}\{\text{i,j from diff. sectors}\}\cdot\omega_{ij,T},            
    \end{split}
\end{align}

and $\mathbf{I}$ is the indicator function, $T$ is the period of network formation.

The second type decomposition is by the firm size of the lead firm. It is intuitive to suspect that the impact from a large market capitalization firm is much stronger than that of a small size firm. We define big firms as the firms with market capital higher than the 70\% quantile of the market capitalization across all the S\&P500 stocks, $q^{mv}_{0.7})$. And small firms are the firms with market capital lower than the 30\% quantile of the market capitalization across all the S\&P500 stocks, $q^{mv}_{0.3})$. We take the average value of market capitalization over the network formation time window $T$ to avoid forward-looking bias. As such we have the news network consisting of only linkages with big size lead firm, $\mathcal{W}_T^{BigLead}=\left[\omega_{ij,T}^{BigLead}\right]$, and similarly, the news network of small lead firm, $\mathcal{W}_T^{SmallLead}=\left[\omega_{ij,T}^{SmallLead}\right]$, where

\begin{align}\label{def:big_small_lead_nn}
    \begin{split}
        \omega_{ij,T}^{BigLead} & \stackrel{\operatorname{def}}{=} \mathbf{I}\{mv_j >q^{mv}_{0.7}\}\cdot\omega_{ij,T}\\
        \omega_{ij,T}^{SmallLead} & \stackrel{\operatorname{def}}{=} \mathbf{I}\{mv_j <= q^{mv}_{0.3}\}\cdot\omega_{ij,T},
    \end{split}
\end{align}

The third decomposition is similar to the second one but with stock liquidity. A firm with higher stock liquidity is deemed to be more efficient in incorporating market information. Hence, decomposition by the lead firm's stock liquidity can test whether there is a difference in information transmission between a more liquid stock and a less liquid stock. Similar to firm size decomposition, the breakpoints of low liquidity and high liquidity are 30\% quantile $q^{turnover}_{0.3}$, and 70\% quantile $q^{turnover}_{0.7}$, of the average turnover rate over the network formation period $T$. Then, we define the network of high liquidity lead, $\mathcal{W}_T^{HighLead}=\left[\omega_{ij,T}^{HighLead}\right]$, and network of low liquid lead, $\mathcal{W}_T^{LowLead}=\left[\omega_{ij,T}^{LowLead}\right]$, where

\begin{align}\label{def:high_low_lead_nn}
    \begin{split}
        \omega_{ij,T}^{HighLead} & \stackrel{\operatorname{def}}{=} \mathbf{I}\{turnover_j > q^{turnover}_{0.7}\}\cdot\omega_{ij,T}\\
        \omega_{ij,T}^{LowLead} & \stackrel{\operatorname{def}}{=} \mathbf{I}\{turnover_j <= q^{turnover}_{0.3}\}\cdot\omega_{ij,T},
    \end{split}
\end{align}

\subsection{News Network Derived Variables}\label{sec:network_vars}
% [Connectedness within sector and across sectors]

Two variables, \emph{lead return} and \emph{network degree}, are derived from the news network along with stock return. The news network, $\mathcal{W}_T$, is uncovered with the definition mentioned above in period $T=[t-l:t]$, and stock return $r_{i,t}$ is the daily stock return of firm $i$ in day-$t$.

\subsubsection{News Network Lead Return}\label{sec:leadreturn}

The first network derived variable is aimed to investigate the impacts from lead stocks returns to follower stocks returns.
\emph{Lead return}, $\mathcal{LR}_{i,t}(\omega_{ij,t-l:t})$, of stock $i$ at day-$t$ is defined as the linear combination of all the returns of its lead companies $r_{j,t}$ given the news network $\mathcal{W}_{t-l:t}=\left[\omega_{ij,t-l:t}\right]$,

\begin{equation}
  \mathcal{LR}_{i,t}(\omega_{ij,t-l:t}) \stackrel{\operatorname{def}}{=} \sum_{j=1}^{n}\omega_{ij,t-l:t}\cdot r_{j,t},
\end{equation}

where $[t-l:t]$ is the time period of network formation, and $l$ is the window size. Note that $\omega_{ij,t-l:t}$ is a coefficient determined by the amount of linkages between firm $i$ and $j$, hence the network price proxy $\mathcal{LR}_{i,t}(\omega_{ij,t-l:t})$ also captures the attention effect. For instance, firm $j$ will be more influential to \emph{lead return} if there is a larger volume of news connecting $i$ and $j$. A similar definition can refer to the literature (e.g., \cite{COHEN_2008,scherbina2015economic}).

\emph{Lead return} is employed here for two reasons. First of all, it is the most intuitive way to synthesize the stock returns from the lead firms to each company $i$. Secondly, \emph{lead return} can be easily decomposed, e.g., by the characteristics of nodes and linkages in the news network, and its decomposed variants allow flexible econometric and portfolio tests. Finally, the econometric model with \emph{lead return} essentially tests the direct network effect between the lead stock return and follower stock return.

\begin{figure}[!htb]
    \centering
    \begin{minipage}{1\linewidth}
        \includegraphics[width=\textwidth]{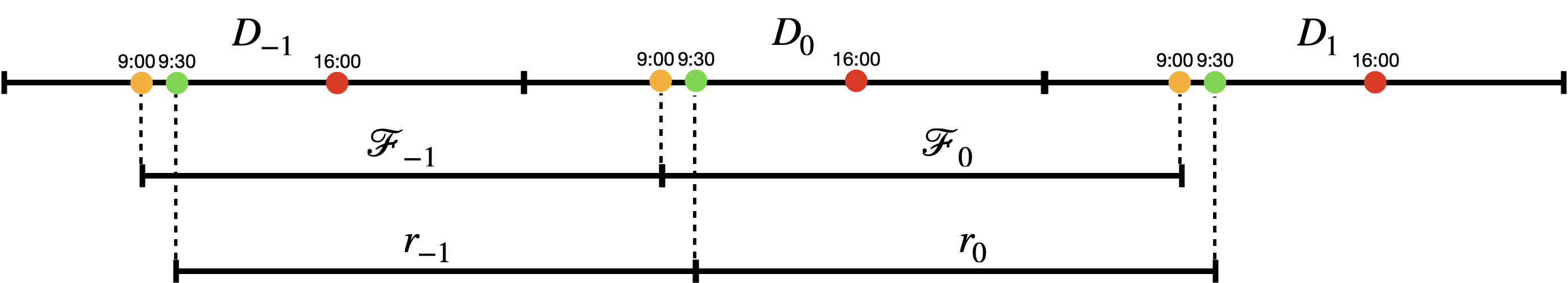}
        % \rule{\linewidth}{10em}
    \end{minipage}
    \caption{
        \textbf{Definition of Timeline of Information}
    \newline 
    \small    
This figure demonstrates the definition of the information timeline for computing stock returns and news networks formation. This upper line is the calendar day synchronized to the New York time, the middle line defines the information set on each day-$t$, $\mathcal{F}_t$, as 09:00 am on day-$t$ to 09:00 am next day, and the bottom line defines the stock return on day-$t$, $r_t$ with open prices at 09:30 am on day-$t$ and day-$t+1$.
}\label{fig:timeline_def}
\end{figure}

The timeline of the information used is carefully defined on a daily frequency to ensure no future information bias. As shown in Figure \ref{fig:timeline_def}, the information set on day-$t$ noted as $\mathcal{F}_t$ is defined as 09:00 am (day-$t$) until 09:00 am (day-$t+1$). Correspondingly, the stock logreturn on day-$t$ is defined by the logarithmic difference between open price at day-$t+1$ and day-$t$, i.e $r_t\stackrel{\operatorname{def}}{=}\log p_{t+1}^{open} - \log p_{t}^{open}$. The information used to construct factor on day-$t$ is up to half an hour before the market opens at day-$t+1$. Note that we synchronize the article arriving time with the local exchanges time (New York Time, i.e., Eastern Time Zone (UTC-05:00) with daylight saving time (UTC-04:00), in our case).

\emph{Lead return}, $\mathcal{LR}_{i,t}(\omega_{ij,t-l:t})$, is essentially a function of the network $\left[\omega_{ij,t-l:t}\right]$. With the network decomposition defined in Section \ref{sec:network_decomp}, \emph{Lead return} also have the decomposed variants, accordingly. 

We define the the \emph{within-sector lead return} as $\mathcal{LR}_{i,t}(\omega_{ij,t-l:t}^{w})$ and the \emph{cross-sector lead return} as $\mathcal{LR}_{i,t}(\omega_{ij,t-l:t}^{c})$, based on the decomposition in Equation \eqref{def:within_cross_nn}.
We further define the \emph{big size lead return} as $\mathcal{LR}_{i,t}(\omega_{ij,t-l:t}^{BigLead})$ and the \emph{small size lead return} $\mathcal{LR}_{i,t}(\omega_{ij,t-l:t}^{SmallLead})$ using the definition in Equation \eqref{def:big_small_lead_nn}. Similarly, according to Equation \eqref{def:high_low_lead_nn}, we define the \emph{high liquid lead return} as $\mathcal{LR}_{i,t}(\omega_{ij,t-l:t}^{HighLead})$ and \emph{low liquid lead return} as $\mathcal{LR}_{i,t}(\omega_{ij,t-l:t}^{LowLead})$.

Furthermore, given the sign of lead firm return, \emph{positive lead return} $\mathcal{LR}^{+}(\omega_{ij,t-l:t})$ and \emph{negative lead return} $\mathcal{LR}^{-}(\omega_{ij,t-l:t})$ are defined as,

\begin{equation}
  \mathcal{LR}_{i,t}^{+(-)}(\omega_{ij,t-l:t}) \stackrel{\operatorname{def}}{=} \sum_{j=1}^{n}\omega_{ij,t-l:t} \cdot \left|{r_{j,t}^{+(-)}}\right|.
\end{equation}

Notice that here we take the absolute value to compute $\mathcal{LR}^{-}(\omega_{ij,t-l:t})$. The decomposition by the sign of lead firm return helps us to understand if positive and negative returns of a lead firm impact differently on the follower firm.

\subsubsection{News Network Attention: Network Degrees}\label{sec:network_degree}

We employ the \emph{network degree} as a \enquote{network attention} proxy. The attention proxy, \emph{network degree}, $\mathcal{A}_{i,t}(\omega_{ij,t-l:t})$, for a firm $i$ in a network $\left[\omega_{ij,t-l:t}\right]$ is defined as

\begin{equation}
    \mathcal{A}_{i,t}(\omega_{ij,t-l:t}) = \#\left\{j: \omega_{ij,t-l:t} \neq 0\right\} + \#\left\{j: \omega_{ji,t-l:t} \neq 0\right\}, j=1,2,\cdots,n,
\end{equation}

where $\#\left\{j: \omega_{ij,t-l} \neq 0\right\}$ counts the number of non-zero entries in $i_{th}$-row in $\left[\omega_{ij,t-l:t}\right]$, likewise for $\#\left\{j: \omega_{ji,t-l} \neq 0\right\}$. This proxy is a well-known fundamental measure that counts the number of connections a node has, and the distribution of it describes how tight do nodes connect. In the following content, we use attention and \emph{network degree} interchangeably.

Compared with counting the number of firm-specific articles of a firm, \emph{network degree} counts the number of linkages of the firm, so that it reduces the noise caused by repetitive news on certain firms. The intuition of using \emph{network degree} is that a firm with higher \emph{network degree} implies more appearance with many other firms, either in headlines as lead firm or content as follower firm. For instance, in 2020, the automobile manufacturer Tesla had an unprecedented media exposure, when it is covered by media news in the headlines and mentioned in the news content of some other firms like General Motors and Ford. \footnote{Many automobile manufacturers like General Motors and Ford either released their electric vehicles or announced their plans for electric cars.}

The attention proxy, \emph{network degree}, $\mathcal{A}_{i,t}(\omega_{ij,t-l:t})$, is also a function of the given news network $\omega_{ij,t-l:t}$. Therefore, the attention variants can be defined based on the network decomposition in Section \ref{sec:network_decomp}. 
We define the \emph{network degree} of the network that consists linkages formed by firms from same (different) sector(s) as \emph{within(cross)-sector attention} noted as $\mathcal{A}_{i,t}(\omega_{ij,t-l:t}^{w(c)})$; \emph{network degree} of the network that consists linkages of big (small) size lead firm as \emph{big (small) lead attention} noted as $\mathcal{A}_{i,t}(\omega_{ij,t-l:t}^{BigLead(SmallLead)})$; and \emph{network degree} of the network that consists linkages of more (less) liquid lead firm as \emph{high (low) lead attention} noted as $\mathcal{A}_{i,t}(\omega_{ij,t-l:t}^{HighLead(LowLead)})$.

One would notice that \emph{network degree} consists of two parts that counting the row and column connections in the adjacency matrix $\left[\omega_{ij,t-l:t}\right]$. We define the \emph{lead attention} as $\mathcal{A}^{lead}(\omega_{ij,t-l:t}) \stackrel{\operatorname{def}}{=} \#\left\{j: \omega_{ji,t-l} \neq 0\right\}$, which essentially counts the number of different follower firms linked to lead firm $i$ in a time period $[t-l:t]$. Similarly, \emph{follower attention} is defined as $\mathcal{A}^{follower}(\omega_{ij,t-l:t}) \stackrel{\operatorname{def}}{=} \#\left\{j: \omega_{ij,t-l} \neq 0\right\}$ that counts the number of different lead firms that link to firm $i$.

\subsection{Econometric Model}\label{sec:eco_models}

To test the relationship between \emph{lead returns} and stock returns, the following panel model is employed.

\begin{equation}\label{eq:panel_reg_genereal}
  y_{i,t+h} = \alpha_i + \mathcal{LR}_{i,t}^{\top}(\omega_{ij,t-l:t})\beta + \sum_{k=1}^{m} \gamma_{k} \cdot control_{i,t}^{k} + \varepsilon_{i,t},
\end{equation}

where the dependent variable, $y_{i,t+h}$, is the simple $h$-day ahead stock return $r_{i,t+h}$ or the stock residual return $r_{i,t+h}^{resid}$ from an asset pricing model. The main regressors, $\mathcal{LR}_{i,t}^{\top}(\omega_{ij,t-l:t})$, is a vector of \emph{lead return} variables on the news network $[\omega_{ij,t-l:t}]$ constructed with $l$-day window size. Control variables can be employed to control for firm characteristics.

The parameters of fixed effect model in Equation \eqref{eq:panel_reg_genereal} are estimated by pooled OLS estimator with standard error corrected by clustering over individuals and time. The $\beta$ of our main interest estimates the averaged effect between \emph{lead return} variables and the dependent variables across all the individual stocks. Specifically, $\beta$ estimates the comovement effect between when $h=0$, and lead-lag effect when $h>=1$. $\alpha_i$ is employed to capture the possible fixed effects.

We employ the two stock residuals return variables from the Fama-Frech 3-factor model (FF3, \cite{Fama_1993}) and Fama-Frech 5-factor model (FF5, \cite{FAMA20151}). FF5 can be written as,

\begin{equation}\label{eq:resid_ff5}
    r_{i,t} - r_{t}^{f} = \theta_i + \theta_i^{m}(r_t^{m} - r_t^f) + \theta_i^{s} f_{t}^{smb} + \theta_i^{h} f_{t}^{hml} + \theta_i^{r} f_{t}^{rmw} + \theta_i^{c} f_{t}^{cma} + r_{i,t}^{resid},
\end{equation}

where $r_{i,t}$ is the simple stock return, $r_{t}^{f}$ is the risk-free return, $r_t^{m}$ is the market return. The 3-factor model includes risk premium factor $r_t^{m} - r_t^f$, size factor $f_{t}^{smb}$, and value factor $f_{t}^{hml}$. The 5-factor model adds profitability factor $f_{t}^{rmw}$ and investment factor $f_{t}^{cma}$ to the 3-factor model. With all coefficients estimated by OLS in full sample, one can easily have the residual return from FF5, $r_{i,t}^{resid}(FF5)$, and FF3, $r_{i,t}^{resid}(FF3)$.

The residual return from Equation \eqref{eq:resid_ff5} is the part of stock returns unexplained by those factors and idiosyncratic stock alpha. Therefore, $r_{i,t}^{resid}(FF5)$ exclude the effects that are caused by the the Fama-Frech 5-factor. Similarly, $r_{i,t}^{resid}(FF3)$ excludes the effects from Fama-French 3-factor.

\subsection{Portfolio Strategy by Sorting Characteristics}\label{sec:method_sorting}

The portfolio strategy of sorting a characteristic aims to test the cross-sectional effect of the characteristic without imposing some assumptions, e.g., linear restriction, in econometric models. 
The trading rule is simple. Consider a characteristic to test, $x_{i,t}$, for each individual stock $i$ at time $t$, all the stocks are sorted by $x_{i,t}$ into $K$ portfolios. Within each portfolio, stocks are equal-weighted or value-weighted. All the portfolios are held for one-period and then rebalanced for the next period.

The main point of portfolio strategy is to see whether a long-short trading rule based on sorting $x_{i,t}$ can deliver abnormal returns. In an efficient market, one shall not see an abnormal return with this simple trading strategy. While a positive abnormal return from the long-short portfolio suggests the existence of anomaly returns, which can be profited by investors.

\subsubsection{Double Sorting Method for Robustness}\label{sec:double_sorting_method}

Problem about sorting stocks based on their characteristic is that the characteristic only captures some of those known effects, e.g., size effect, therefore, it is imperative to test the robustness of the characteristic by controlling for those known effects.

The double-sorting method (\cite{ANG_2006}) is employed to relieve the results from the four possible effects including size, value, liquidity, and momentum. The double-sorting method is illustrated as follows. To control the effect from a characteristic, $c_{i,t}$, of firm $i$ at time $t$, the assets are first sorted into $K$ portfolios based on $c_{i,t}$. Then, within each portfolio, the assets are again sorted into $K$ portfolios by the characteristic $x_{i,t}$ in ascending order. Each portfolio in the second sort is formed into a new portfolio across the $K$ portfolios in the first sort. As such, the newly formed portfolios are relieved from the effect of $c_{i,t}$ since each new portfolio contains assets from all portfolios in the first sort.

  \section{Empirical News Networks of S\&P500 Stocks}\label{sec:empirical_network}

\subsection{Ticker Identification}\label{sec:linkages_identify}

Uncovering the networks from news articles is hindered by the lack of a reliable approach to recognize the firms. The challenges are twofold. First, most articles are written by financial analysts or reporters from different agencies, such as the media giant Reuters, the emerging research company Zacks. Each of them has its way of referring to a firm's name. Take the tech giant \emph{"Apple.Inc"} for example, Some articles have a strict form of reference, e.g. \emph{"(NASDAQ: AAPL)"}, others might allow any casual ways, e.g. \emph{"Apple"}. Second, recognizing a firm's name is a typical task that heavily relies on context awareness. For example, "Booking" can either refer to the company or a human act. Previous literature deals with the problem in two ways. One way is to directly use the Thomson-Reuters News Analytics (TRNA) that labels the companies mentioned in the news (\cite{scherbina2015economic,chen2021media,guomedia}). Recently, \cite{schwenkler2019network} employs the Natural Language Processing (NLP) tools and clustering algorithms to match the company names. However, neither of the two ways above is perfect in the sense of transparency and precision.

\subsubsection{An Algorithm for Identification}\label{sec:identification_algo}

We propose to tackle this problem with a conservative algorithm. The algorithm is designed to reduce the type-I error, i.e., to reduce the "false link" error and buy the "false unlink" error. We only identify the names/tickers of the given set of firms with the following three strategies, and for each strategy, we set multiple exceptional cases:
\begin{itemize}
    \item[] \emph{S1}: Tickers in the brackets.
    \begin{itemize}
        \item[] \emph{S1} is the main way among the three to identify firms as it is the safest way. We first manually configure a set of exceptional cases that can \emph{NOT} be identified. Note that this set is designed for the given set of firms to identify, e.g., in our scenario \{\enquote{PEG,} \enquote{COO,} \enquote{C,} \enquote{GPS}...\}, then we match the characters that are in the brackets. After differentiating the exceptional characters, we can have the clean tickers in the brackets.
    \end{itemize}
    \item[] \emph{S2}: Name segments of the firms.
    \begin{itemize}
        \item[] The reason of identifying firm name segments instead of their full names is that people rarely use the suffixes in the full name of firms, e.g., \emph{"JPMorgan Chase \& Co"} is usually referred to as \emph{"JPMorgan Chase"} or \emph{"JPMorgan"}. To identify the name segments, we first need to map the full names of all the firms to their n-gram name segments detailed in Algorithm\ref{algo:fullname_seg_map}. Then, we simply match the name segments in the text to identify corresponding firms. We carefully configure the redundant strings for the given set of firms to identify.
    \end{itemize}
    \item[] \emph{S3}: Tickers in plain text.
    \begin{itemize}
        \item[] \emph{S3} is employed to identify firms with long tickers like "GOOGL" only in the news headlines and only when neither \emph{S1} nor \emph{S2} could find any matches. In our experiment, we find the false identification error increases quite much if \emph{S3} is applied on the contents due to a lot of unexpected short initials, such as \emph{"HD"}, which is commonly short for \emph{"high-definition"} while it is also the ticker of the firm \emph{"Home Depot Inc"}.
    \end{itemize}
\end{itemize}

\begin{figure}[!htb]
\centering
\begin{minipage}{1\linewidth}
    \includegraphics[width=\textwidth]{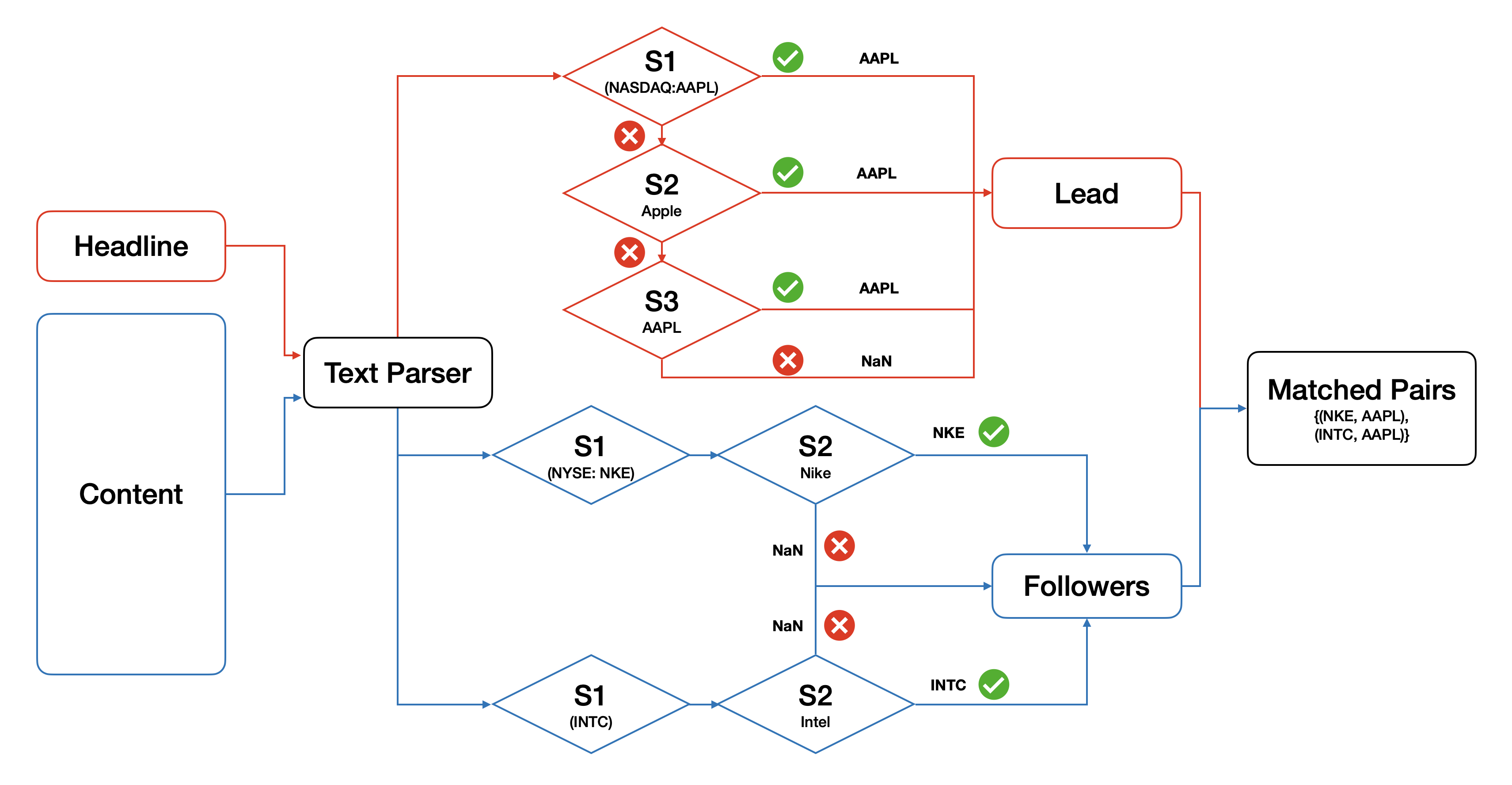}
    % \rule{\linewidth}{10em}
\end{minipage}
\caption{
    \textbf{Flowchart of Identification Algorithm}
\newline 
\small
A flowchart representation of how the algorithm treats the headline and content differently, and the priority of the three identification strategies.}\label{fig:algo_example}
\end{figure}

We treat headline text and content differently with the above three strategies. For headlines, we prioritize the strategies with the order of $S1 > S2 > S3$ by a weak assumption that the way all the firm names mentioned in the same headline is consistent, i.e., the lower priority strategy effects only when, the higher one finds no result. And for content text, only the first two strategies, $S1 and S2$, are employed as $S3$ could cause some apparent misidentifications. A flowchart of the identification process is illustrated in Figure \ref{fig:algo_example} with an example.
  
\subsubsection{Screening the Identified Pairs}

\begin{figure}[!htb]
\centering
\begin{minipage}{1\linewidth}
    \includegraphics[width=\textwidth]{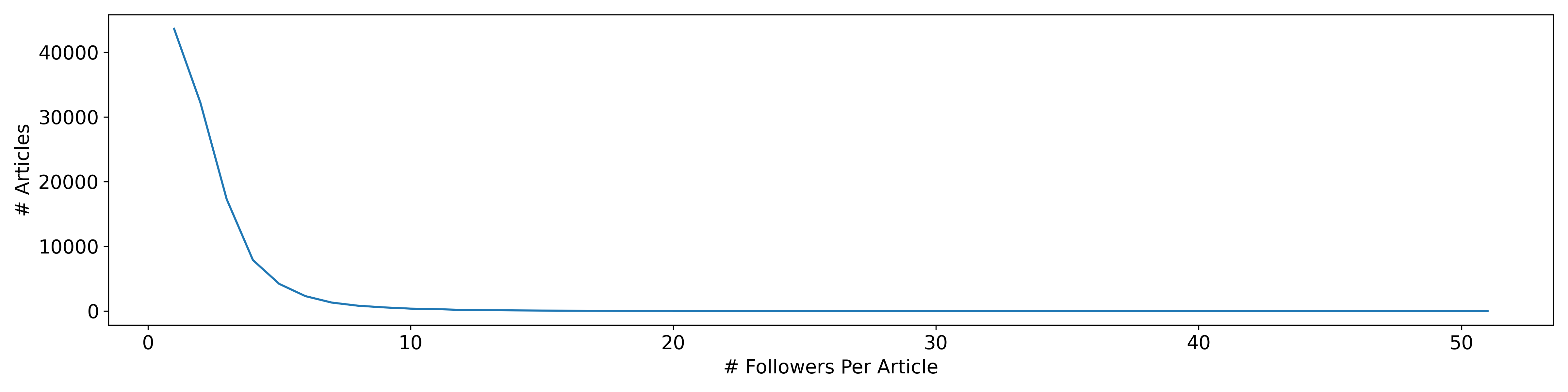}
    % \rule{\linewidth}{10em}
\end{minipage}
\caption{
    \textbf{Distribution of Followers in Each Article}
\newline 
\small
This figure shows the number of followers identified in each article before the screening. Most of the articles have the number of followers identified under 10.}\label{fig:followersnum}
\end{figure}

We further employ two steps to screen the articles after the identification so that our study is focused on single firm-specific news. Firstly, we screen out the articles that have more than one lead firm identified. This step mainly helps to eliminate those market summary reports that have few major companies on their headlines. Such reports do not fit in the scope of our definition of lead firm. Secondly, we remove the articles that have more than $10$ followers identified. This removal is aimed to overcome the errors caused by some of the market news mentioning an iconic firm in their headlines. A typical example is shown in the Appendix Figure \ref{fig:manyfollowerssample}, where 12 companies are identified in the content, apart from the single lead company. 
Apparently, articles like the sample are not firm-specific news. This truncation affects only a small number of linkages, as one can see that the frequency decreases dramatically as the number of followers increases shown in the follower distribution plot in Figure \ref{fig:followersnum}.

We provide pseudocode in Algorithm \ref{algo:tickers_identi} in Appendix. As the algorithm narrows down the scenarios that a firm could be mentioned, the design allows the errors that many linkages are not identified. However, for the purpose of testing the effects of the linkages, we will have more robust and conservative results. Surely, one can attempt to handle this task with more sophisticated tools, which, however, also introduce more statistical modeling errors.

\subsection{News Linkages Identified Among S\&P500 Constituents}

In this paper, we focus only on the firms that are the constituents of S\&P500. S\&P500 consists of the companies with large capitalization listed in the U.S stock market and is one of the most commonly studied indices. It is also the underlying asset of many financial instruments, including ETFs, futures, and options. The choice is a trade-off between generality and efficiency of identification. One could adapt the algorithm to consider a broader range of assets, e.g., small firms or even listed funds, which, however, would cause more false links to be mismatched even after a huge effort of specifying special cases.

To identify the linkages from the media news, we collect a huge dataset of public articles from more than 50 online publishers on nasdaq.com news platform. The early period is removed due to the sparsity of the news arrivals, and we utilize the 5-year sample period from Jan. 2016 until the end of 2020 and obtain 1,027,272 articles in total, averaging ~764 articles each trading day. Apart from the financial market and stock news as the main type of articles, the dataset also consists of analytical reports, comments, and even investor meeting notes and covers topics including global economics, commodities, etc. Note that all the articles are economic and financial news, and we keep all the articles of different topics as the input of the algorithm.

\begin{figure}[!htb]
    \centering
    \begin{minipage}{1\linewidth}
        \includegraphics[width=\textwidth]{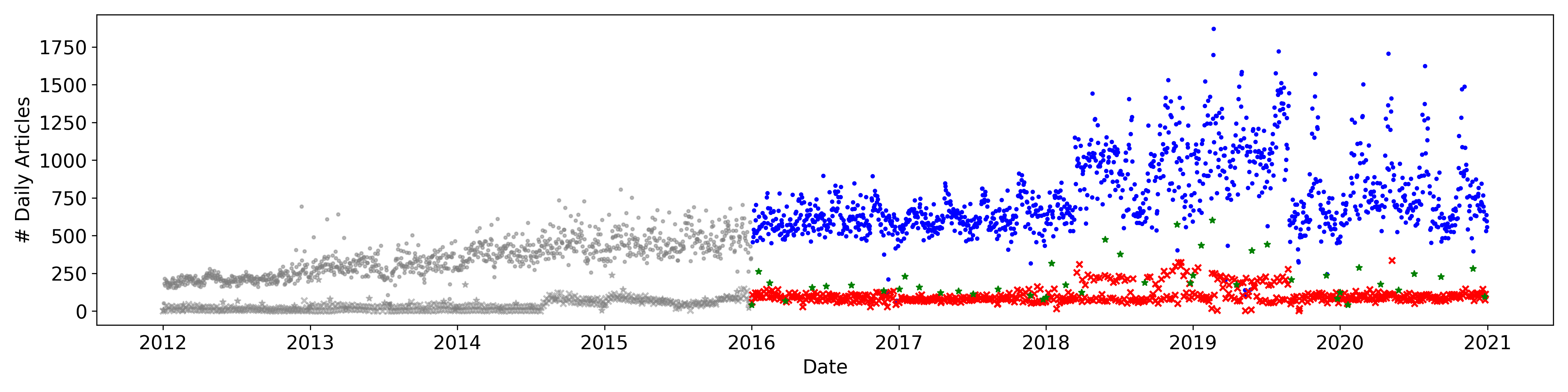}
        % \rule{\linewidth}{10em}
    \end{minipage}
    \caption{
        \textbf{Number of Articles in The Database}
    \newline 
    \small
    This plot shows the number of articles daily aggregated from Jan. 2012 until Dec. 2020.
    We only utilize sample starting from Jan. 2016, so the grey dots are the samples not used.
    The blue dots indicate the trading days, the red crosses represent the weekends, and the green stars are the market holidays.
}\label{fig:articlenum}
\end{figure}

As shown in Figure \ref{fig:articlenum}, along with the booming of digital news, the volume of daily articles gets higher. Apart from that, the volume of news articles evolves quite periodically.  More specifically, we can observe the 4-peak each year, caused by the massive quarterly and annual financial reports releases. We mark the holidays (in green star) and the weekends (in the red cross) in Figure \ref{fig:articlenum} to contrast the news volumes on the trading days. Thus, one can see that most articles are released on trading days.

\begin{figure}[!htb]
    \centering
    \begin{minipage}{1\linewidth}
        \includegraphics[width=\textwidth]{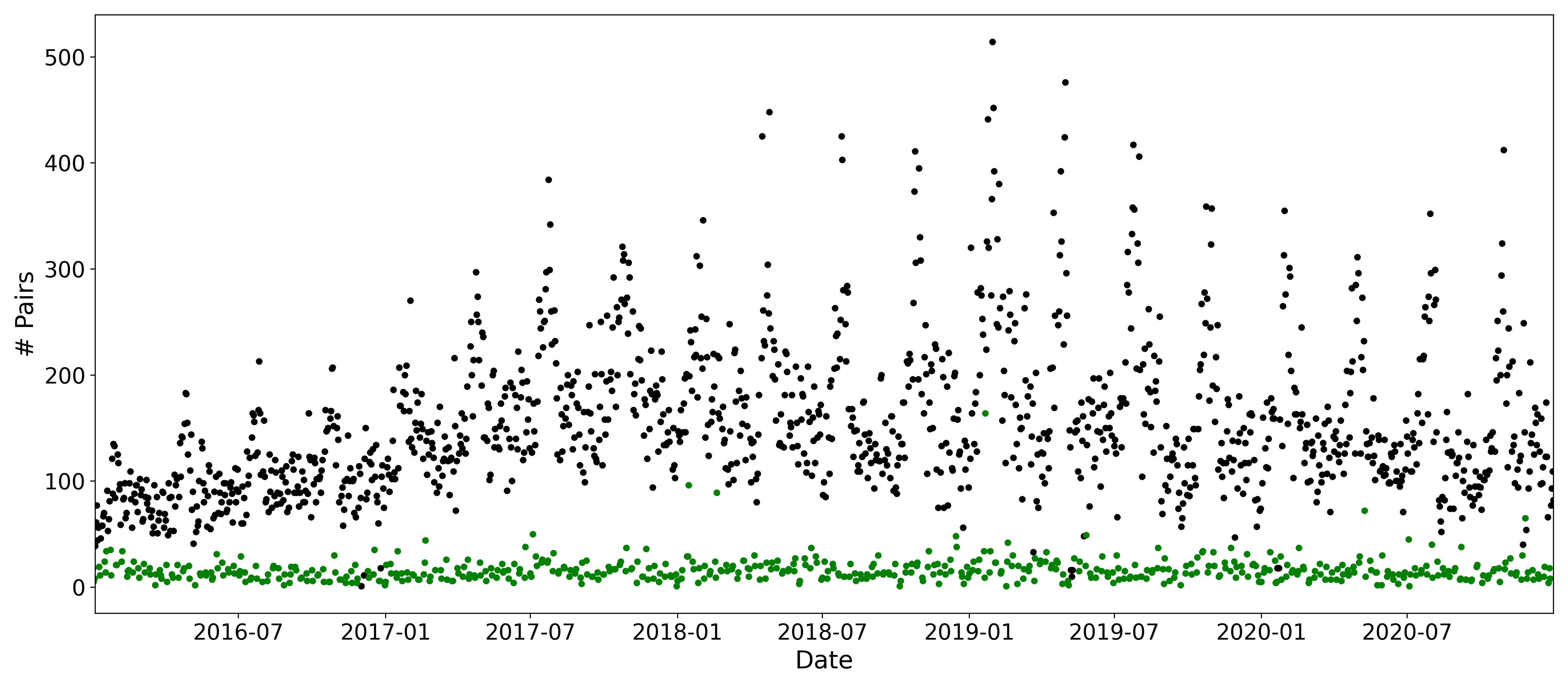}
        % \rule{\linewidth}{10em}
    \end{minipage}
    \caption{
        \textbf{Number of Linkages Identified Daily}
    \newline 
    \small
    This figure shows the number of linkages identified each day over the 5-year sample period, from Jan. 2016 to Dec. 2020. The dark color bar represents the trading days and the green bar indicates the weekends and market holidays.
}\label{fig:matchedparisnum}
\end{figure}

Applying the algorithm on the collected dataset, 100,960 articles out of the total 1,027,272 articles are identified with S\&P500 constituents, and we obtain 204,074 pairs of linkages in those articles during the whole 5-year sample period. Among that, 195,017 pairs of linkages are identified in trading days. As shown in Figure \ref{fig:matchedparisnum}, we can observe a strong periodicity of the daily linkages number. The number of pairs in trading days shows four local maximum each year, which is consistent with the daily volume of news arrivals.

A typical example of the articles that are identified with linkages of the news network is documented in Appendix Figure \ref{fig:newsexample}. The algorithm uncovers two pairs of linkages: \{(INTC, AAPL), (NKE, AAPL)\}. From the title, it is easy to recognize that this article is mainly covering the listed company, \emph{"Apple"}. And the article content also mentions \emph{"Intel"} and \emph{"Nike"}, both of which are the business partners of \emph{"Apple"}. One might have noticed that the listed company, \emph{"Peloton Interactive"}, is not identified as the company was not included in S\&P500 during that sample period.

One interesting fact is the fairly large amount of cross-sector linkage. The identified result shows that almost half of the linkages are cross-sector pairs, i.e., the lead and follower are from different sectors. Among the 204,074 pairs of linkages, 99,047 pairs are cross-sector pairs, and 105,027 pairs are intra-sector. The top-3 cross-sector linkage types are:

\begin{itemize}
    \item Communication Services / Information Technology: 15,071
    \item Financial / Information: 8,778
    \item Consumer Discretionary/ Information Technology: 7,993
\end{itemize}

\subsection{Summary of News Network Variables}

A news network, $\left[\omega_{ij,t-l:t}\right]_{n\times n}$, is formed with linkages aggregated in period from day-$t-l$ to day-$t$, and the $n$ nodes of the network are the monthly updated S\&P500 constituents obtained from Bloomberg. As such, the news networks is time-varying. The S\&P500 stocks market data is provided by Compustat on Wharton Research Data Services (WRDS).

We employ a 1-year window network, $\left[\omega_{ij,t-365:t}\right]$, to compute the \emph{lead return} variables, and a 1-month window network, $\left[\omega_{ij,t-30:t}\right]$, for the \emph{network degree} variables. Panel A in Table \ref{tab:summary_stats} documents the \emph{lead return} variables with 1-year window news networks along the daily return variables, and Panel B reports the \emph{network degree} variables with 1-month window news network, as well as the monthly stock return variables.

\begin{table}[!htb]
    % \small
    \setlength{\tabcolsep}{0pt}
    \begin{threeparttable}
      \caption{Summary Statistics}\label{tab:summary_stats}
      \begin{tabular*}{
        \linewidth}{@{\extracolsep{\fill}}>{\itshape}l*{7}{S[table-format=-1.3,table-number-alignment = center]}} 
          \toprule\toprule
          
          % & \multicolumn{8}{c}{\textit{Dependent variable:}} \\ 
          {}                          & \multicolumn{1}{c}{Mean} & \multicolumn{1}{c}{Std.} & \multicolumn{1}{c}{Median}
          & \multicolumn{1}{c}{10\%}  & \multicolumn{1}{c}{25\%} & \multicolumn{1}{c}{75\%} & \multicolumn{1}{c}{95\%} \\
          \cline{2-8}
          \\
          
          % \midrule
          \multicolumn{8}{l}{Panel A: Daily Return \& \emph{Lead Return} of 1-Year Window Network $T=[t-365:t]$}\\
          \cline{1-8}
          $r$                                    & 0.026  & 2.187 & 0.079 & -1.989 & -0.821 & 0.940 & 2.905 \\
          $r^{resid}(FF3)$                       & -0.016 & 2.077 & 0.019 & -1.872 & -0.818 & 0.831 & 2.677 \\
          $r^{resid}(FF5)$                       & -0.015 & 2.065 & 0.020 & -1.860 & -0.814 & 0.828 & 2.659 \\
          $\mathcal{LR}(\omega_{ij,T})$             & 0.033  & 1.497 & 0.045 & -1.346 & -0.513 & 0.656 & 2.011 \\
          $\mathcal{LR}(\omega_{ij,T}^{w})$         & 0.028  & 1.686 & 0.000 & -1.509 & -0.576 & 0.704 & 2.252 \\
          $\mathcal{LR}(\omega_{ij,T}^{c})$         & 0.036  & 1.475 & 0.000 & -1.303 & -0.455 & 0.619 & 2.003 \\
          $\mathcal{LR}^{+}(\omega_{ij,T})$         & 0.657  & 0.894 & 0.402 & 0.000  & 0.130  & 0.854 & 2.143 \\
          $\mathcal{LR}^{-}(\omega_{ij,T})$         & 0.625  & 1.002 & 0.319 & 0.000  & 0.093  & 0.761 & 2.232 \\
          $\mathcal{LR}(\omega_{ij,T}^{BigLead})$   & 0.042  & 1.418 & 0.006 & -1.292 & -0.467 & 0.637 & 1.955 \\
          $\mathcal{LR}(\omega_{ij,T}^{SmallLead})$ & -0.001 & 2.101 & 0.000 & -1.755 & -0.561 & 0.647 & 2.695 \\
          $\mathcal{LR}(\omega_{ij,T}^{HighLead})$  & 0.018  & 2.080 & 0.000 & -1.817 & -0.643 & 0.768 & 2.756 \\
          $\mathcal{LR}(\omega_{ij,T}^{LowLead})$   & 0.035  & 1.284 & 0.000 & -1.176 & -0.421 & 0.579 & 1.777 \\\\[0.15cm]

          \multicolumn{8}{l}{Panel B: Monthly Return \& \emph{Network Degree} of 1-Month Window Network $T=[t-30:t]$}\\
          \cline{1-8}
          
          $r$                                    & 0.623  & 9.848  & 1.275 & -9.432 & -3.527 & 5.640  & 13.976 \\
          $r^{resid}(FF3)$                       & -0.276 & 8.180  & 0.199 & -8.785 & -4.011 & 4.037  & 10.929 \\
          $r^{resid}(FF5)$                       & -0.253 & 8.126  & 0.222 & -8.753 & -3.934 & 4.062  & 10.886 \\
          $\mathcal{A}(\omega_{ij,T}))$            & 8.778  & 14.215 & 5.000 & 0.000  & 2.000  & 11.000 & 27.000 \\
          $\mathcal{A}^{lead}(\omega_{ij,T})$      & 4.390  & 5.743  & 3.000 & 0.000  & 1.000  & 6.000  & 14.000 \\
          $\mathcal{A}^{follower}(\omega_{ij,T})$  & 4.388  & 10.283 & 2.000 & 0.000  & 0.000  & 5.000  & 15.000 \\
          $\mathcal{A}(\omega_{ij,T}^{w})$         & 4.105  & 4.846  & 3.000 & 0.000  & 1.000  & 6.000  & 13.000 \\
          $\mathcal{A}(\omega_{ij,T}^{c})$         & 4.673  & 10.834 & 2.000 & 0.000  & 0.000  & 5.000  & 16.000 \\
          $\mathcal{A}(\omega_{ij,T}^{BigLead})$   & 4.838  & 10.016 & 1.000 & 0.000  & 0.000  & 5.000  & 22.000 \\
          $\mathcal{A}(\omega_{ij,T}^{SmallLead})$ & 1.156  & 2.617  & 0.000 & 0.000  & 0.000  & 1.000  & 6.000 \\
          $\mathcal{A}(\omega_{ij,T}^{HighLead})$  & 2.119  & 4.660  & 0.000 & 0.000  & 0.000  & 2.000  & 10.000 \\
          $\mathcal{A}(\omega_{ij,T}^{LowLead})$   & 3.287  & 6.678  & 1.000 & 0.000  & 0.000  & 4.000  & 14.000 \\
                    
          \bottomrule\bottomrule
      \end{tabular*}
    \begin{tablenotes}[flushleft]
      \setlength\labelsep{0pt} 
    \linespread{1}\small
    \item 
    Panel A reports the summary statistics of daily stock (residual) return variables and \emph{lead return} variables on a 1-year window size news network. Panel B reports the summary statistics of monthly (residual) return variables and \emph{attention} variables on a 1-month window news network. All statistics in Panel A are given in percentage, as well as the (residual) return variables in Panel B. Columns from left to right are mean, standard deviation, median, 10\% percentile, 25\% percentile, 75\% percentile, 95\% percentile of respective variables.
    \end{tablenotes}
    \end{threeparttable}
  \end{table}

In the literature of discussing the formation of networks, two of the models are the most prominent and well investigated. The first model named Erdős-Rényi network depicts the linkages being randomly formed with probability $p$, which is governed by the binomial distribution or Poisson distribution. Secondly, the scale-free network characterized by power-law distribution was discussed in the late 1990s and early 2000s. It is well documented that many networks in social science and economics carry with the scale-free property(\cite{barabasi1999emergence, Albert_1999, jackson2010social}). The power-law distribution can be written as, $p(d) = cd^{-\gamma}$, where $c$ is a constant scaler and $d$ is the degree of nodes. The degree exponent $\gamma$ is the coefficient in the linear function of $\log(d)$ when the distribution is represented in the log-log form, i.e $\log(p(d)) = \log(c) - \gamma \log(d)$.

\begin{figure}[!htb]
  \centering
  \begin{minipage}{1\linewidth}
      \includegraphics[width=\textwidth]{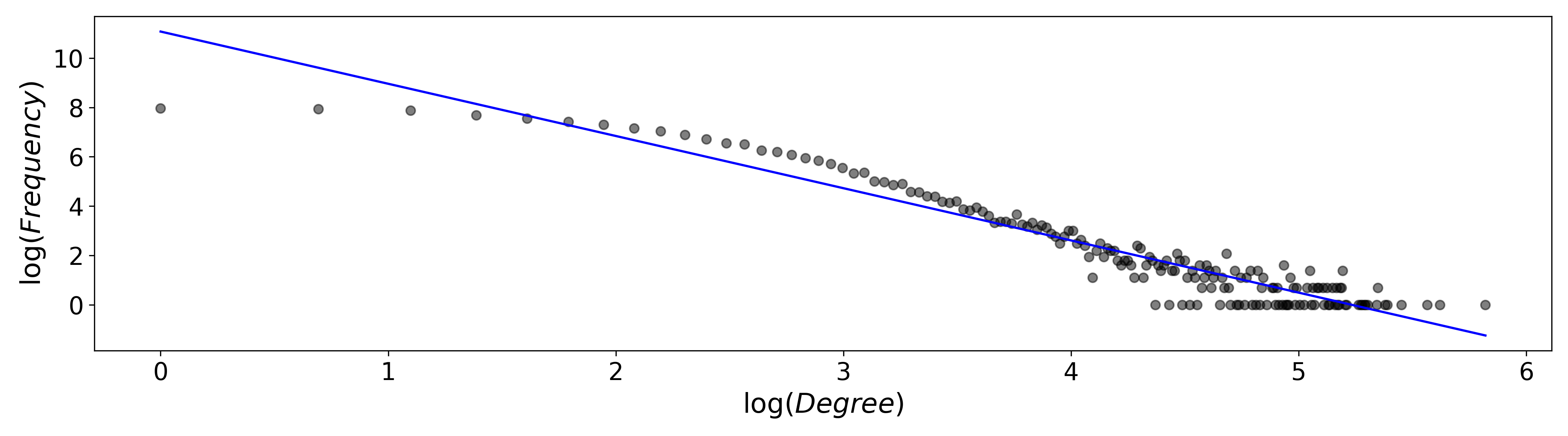}
    % \rule{\linewidth}{10em}
  \end{minipage}
  \caption{
        \textbf{Network Degree Distribution of Network Aggregated Monthly}
  \newline 
  \small
The log-log plot is the logarithmic frequency, $\log(frequency)$, against its logarithmic network degree, $\log(\mathcal{A})$. The solid line is the linear fitting estimated by OLS, and the slop coefficient $\gamma=2.12$ and $R^2=0.914$. Such a pattern is often deemed as power-law distribution.
        }\label{fig:degree_distribution}
\end{figure}

The degree distribution of the monthly aggregated news network falls quite well into the power-law distribution with $\gamma=2.12$ and $R^2=0.914$ as shown in Figure \ref{fig:degree_distribution} where the relationship between $\log(Frequency)$ and $\log(Degree)$ is approximated with a linear function. The power-law distributed news network implies the existence of highly connected nodes or hubs (\cite{Barabsi2016}), i.e., a portion of the companies are mentioned much more frequently than the other companies.

\begin{figure}[!htb]
  \centering
  \begin{minipage}{1\linewidth}
      \includegraphics[width=\textwidth]{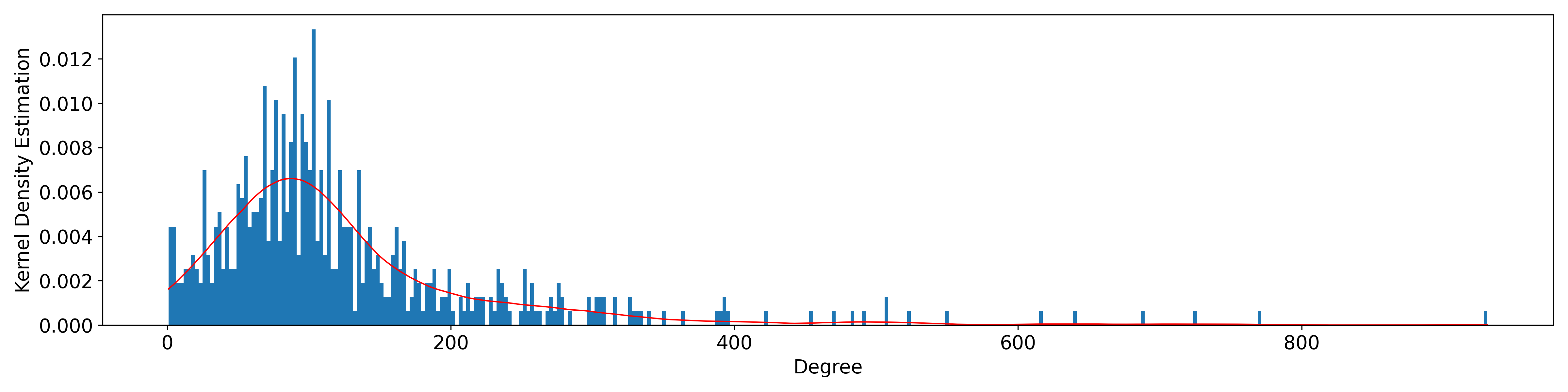}
    % \rule{\linewidth}{10em}
  \end{minipage}
  \caption{
        \textbf{Network Degree Distribution of Full-Sample Network}
  \newline 
  \small
This figure shows the network degree distribution of the network formed with full-sample data. The \textcolor{blue}{bars} represent the histogram and the \textcolor{red}{solid line} is the kernel density estimated with Epanechnikov kernel.
        }\label{fig:fullsample_degree_kde}
\end{figure}

Interestingly, when the news linkages are aggregated for the whole sample period, the degree distribution of the full sample news network can be well described as a randomized network as shown in Figure \ref{fig:fullsample_degree_kde}. Furthermore, the network degree of that network is closer to a Poisson distribution by Epanechnikov kernel. Note that the full-sample news network is constructed with all the historical S\&P500 constituents, i.e., network size of $596$.

  %!TEX root = ../main.tex

\section{\emph{Lead Returns} and Cross-Sectional Stock Returns} \label{sec:leadreturn_simul_pred}

The section shows how the returns of the lead firms, \emph{lead return}, interact with returns of the follower firms in two cases. We first investigate the contemporaneous relationship. We find a strong comovement effect between the news-linked stock daily returns, even after controlling for those known factors and firm characteristics. Then we focus on the case of 1-day ahead stock return, in which we give evidence of the reversal effect from \emph{lead return} on the 1-day ahead stock return; however, the negative impact is offset.

\subsection{Simultaneous Returns Test}\label{sec:linkage_valid}

The fundamental question on the constructed news network is how the linkages between firms identified relate to the performance of the stock returns. In this subsection, we provide evidence to justify the validity of news linkages by both econometric and portfolio tests.

Surprisingly, we found a strong stock price comovement effect after controlling for the known factors and firm characteristics, without recognizing an economic relationship between the news-linked firms, for example supplier-customer relationship. This fact implies two statements. First, the linkages in news articles are informative in stock returns. Second, the lead and follower firms are more likely to have a positive economic correlation than a rivalry relationship, as their daily stock returns move in the same direction.

\begin{figure}[!htb]
  \centering
  \begin{minipage}{1\linewidth}
    \includegraphics[width=\textwidth]{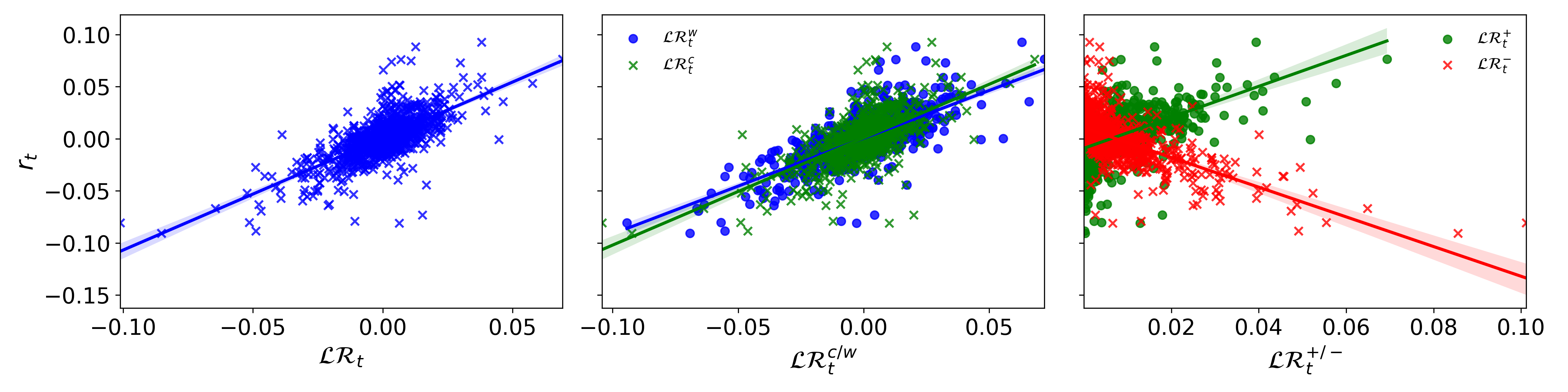}
    % \rule{\linewidth}{10em}
  \end{minipage}
  \caption{
    \textbf{Stock Return Against Simultaneous Lead Return, AAPL}
  \newline 
  \small
  Plots from left to right are AAPL daily return $r_{t}$ against its lead return $\mathcal{LR}(\omega_{ij,t-365:t})$, sector decomposed variants $\mathcal{LR}\left(\omega_{ij,t-365:t}^{w/c}\right)$, and the return sign decomposed variants $\mathcal{LR}^{+/-}(\omega_{ij,t-365:t})$, respectively. 
  The solid lines are estimated by OLS with bootstrap of 95\% confidence interval.}\label{fig:scatterplot_simul_pp_sector_sign}
\end{figure}

The left panel of Figure \ref{fig:scatterplot_simul_pp_sector_sign} shows that the daily return of Apple, $r_t$, positively correlates with its contemporaneous \emph{lead return} formed with a 1-year rolling window network, $\mathcal{LR}(\omega_{ij,t-365:t})$. The strong positive correlation stays after decomposing \emph{lead return} by sector, $\mathcal{LR}(\omega_{ij,t-365:t}^{w(c)})$, as shown in the middle of Figure \ref{fig:scatterplot_simul_pp_sector_sign}.  It is shown in the right plot that the effect remains true when \emph{lead return} is separated by the sign of lead firm returns, $\mathcal{LR}^{+(-)}(\omega_{ij,t-365:t})$.

Motivated by this scatter plot, we conjecture that the news linkages capture the comovement effect of stock returns. Therefore, we test the relationship between \emph{lead returns} and the contemporaneous stock returns by the panel model in Equation \ref{eq:panel_reg_genereal} with horizon $h=0$.

\begin{table}[htb]
  \footnotesize
  \setlength{\tabcolsep}{0pt}
  \begin{threeparttable}
    \caption{Estimation of Stock Returns Comovement Effects}\label{tab:simul_exret_regres_sign_sector}
    \begin{tabular*}{\linewidth}{@{\extracolsep{\fill}}>{\itshape}l *{5}{S[table-format=1.2,table-number-alignment = center]} l *{5}{S[table-format=1.2,table-number-alignment = center]}}
        \toprule\toprule
        
        % & \multicolumn{8}{c}{\textit{Dependent variable:}} \\ 
         {} & \multicolumn{5}{c}{$r_{i,t}^{resid}(FF3)$} & {} & \multicolumn{5}{c}{$r_{i,t}^{resid}(FF5)$} \\ 
          \cline{2-6} \cline{8-12}
{}                                                        & \multicolumn{1}{c}{(1)}      & \multicolumn{1}{c}{(2)}      
& \multicolumn{1}{c}{(3)}       & \multicolumn{1}{c}{(4)}      & \multicolumn{1}{c}{(5)}      & {} & \multicolumn{1}{c}{(6)}      
& \multicolumn{1}{c}{(7)}      & \multicolumn{1}{c}{(8)}       & \multicolumn{1}{c}{(9)}      & \multicolumn{1}{c}{(10)} \\

\multirow{2}{*}{$\mathcal{LR}_{i,t}(\omega_{ij,T})$}             & 0.752    &          &           &          &          & {} & 0.744    &          &           &          & \\
{}                                                        & (56.984) &          &           &          &          & {} & (58.020) &          &           &          & \\[0.15cm]
\multirow{2}{*}{$\mathcal{LR}_{i,t}(\omega_{ij,T}^{w})$}         &          & 0.622    &           &          &          & {} &          & 0.610    &           &          & \\
{}                                                        &          & (36.463) &           &          &          & {} &          & (36.702) &           &          & \\[0.15cm]
\multirow{2}{*}{$\mathcal{LR}_{i,t}(\omega_{ij,T}^{c})$}         &          & 0.121    &           &          &          & {} &          & 0.127    &           &          & \\
{}                                                        &          & (11.496) &           &          &          & {} &          & (12.503) &           &          & \\[0.15cm]
\multirow{2}{*}{$\mathcal{LR}_{i,t}^{+}(\omega_{ij,T})$}         &          &          & 0.741     &          &          & {} &          &          & 0.732     &          & \\
{}                                                        &          &          & (54.584)  &          &          & {} &          &          & (55.576)  &          & \\[0.15cm]
\multirow{2}{*}{$\mathcal{LR}_{i,t}^{-}(\omega_{ij,T})$}         &          &          & -0.761    &          &          & {} &          &          & -0.755    &          & \\
{}                                                        &          &          & (-51.768) &          &          & {} &          &          & (-52.526) &          & \\[0.15cm]
\multirow{2}{*}{$\mathcal{LR}_{i,t}(\omega_{ij,T}^{BigLead})$}   &          &          &           & 0.529    &          & {} &          &          &           & 0.526    & \\
{}                                                        &          &          &           & (41.797) &          & {} &          &          &           & (42.255) & \\[0.15cm]
\multirow{2}{*}{$\mathcal{LR}_{i,t}(\omega_{ij,T}^{SmallLead})$} &          &          &           & 0.214    &          & {} &          &          &           & 0.211    & \\
{}                                                        &          &          &           & (17.092) &          & {} &          &          &           & (17.480) & \\[0.15cm]
\multirow{2}{*}{$\mathcal{LR}_{i,t}(\omega_{ij,T}^{HighLead})$}  &          &          &           &          & 0.246    & {} &          &          &           &          & 0.244 \\
{}                                                        &          &          &           &          & (16.860) & {} &          &          &           &          & (17.262) \\[0.15cm]
\multirow{2}{*}{$\mathcal{LR}_{i,t}(\omega_{ij,T}^{LowLead})$}   &          &          &           &          & 0.536    & {} &          &          &           &          & 0.532 \\
{}                                                        &          &          &           &          & (40.903) & {} &          &          &           &          & (41.498) \\[0.15cm]
\bottomrule\bottomrule
 
    \end{tabular*}
  \begin{tablenotes}[flushleft]
    \setlength\labelsep{0pt}
  \linespread{1}\small
  \item 
  A 1-year window $T=[t-365:t]$ for news network formation. The left (right) panel reports the dependent variable of residual returns from Fama-French 3(5)-factor models. Estimated coefficients of \emph{lead returns} for each model are reported in each column, respectively. $t$-statistics are reported in the parentheses, with standard error corrected by clustering on individual and time. Control variables, including market value $\log(MV)$, book-market ratio $B/M$, and turnover rate $volume/share$, are employed in each model. All the financial statement related variables are 2 quarters lagged.
  \end{tablenotes}
  \end{threeparttable}
\end{table}

One would naturally believe that the comovement of stock returns can be explained by the market trend, the within-sector comovement, and etc. Here we employ the two stock residual return variable, $r_{i,t}^{resid}(FF3)$ and $r_{i,t}^{resid}(FF5)$, so that we estimate the comovement effect between \emph{lead return} and the part of stock returns unexplained by those factors. Four fundamental variables of stocks are used as control variables, including the daily market value in the logarithmic form $\log(MV)$, daily book-market value $B/M$, daily turnover rate $volume/share$. Following \cite{FAMA20151}, the financial statement data including, market share, and book equity $B$, are two quarters lagged values when they are used to compute the control variables.

We first test \emph{lead returns} formed with 1-year rolling window, $\mathcal{LR}(\omega_{ij,t-365:t})$. Due to the sparsity of the news arrivals, the 1-year window is a reasonable choice between testing the persistent linkages and capturing the innovative linkages. The results are consistent with a 6-month rolling window. Column (1) in Table \ref{tab:simul_exret_regres_sign_sector} reports that \emph{lead return} carries positive ($0.752$) comovement effect, with the significant ($t$-statistics$=56.984$) relationship between $\mathcal{LR}(\omega_{ij,t-365:t})$ and the residual return from FF3 $r_{i,t}^{resid}(FF3)$ at 1\% level. This implies that a one-standard-deviation increase in \emph{lead return} ($1.497\%$) accompanies with an average $112$ bps ($1.497\%*0.752$) of excessive return increase across all the stocks on the same trading day. Furthermore, a robust check with FF5 reported in Column (6) shows the consistent result. It is evidenced that the follower daily stock returns comove with their news-linked lead firms daily stock returns.

As shown in the Column (2) of Table \ref{tab:simul_exret_regres_sign_sector}, the regression replaces \emph{lead return} from Column (1) with its decomposed variants by sector. Both \emph{within-sector lead returns} and \emph{cross-sector lead returns} positively correlated with the residual returns from FF3 of individual stocks at 1\% level. The $\mathcal{LR}_{i,t}(\omega_{ij,t-365:t}^{w})$ shows a stronger positive relationship ($0.622$, $t$-statistics$=36.463$) compared to the $\mathcal{LR}_{i,t}(\omega_{ij,t-365:t}^{c})$ ($0.121$, $t$-statistics$=11.496$), implying that the comovement effect between news-linked firms is more prominent within the same sector. However, the cross-sector comovement effect is not ignorable.

The comovement effect still exists after accounting for the sign of lead firm returns. As reported in Column (3) of Table \ref{tab:simul_exret_regres_sign_sector}, the regression shows that \emph{positive lead return} $\mathcal{LR}_{i,t}^{+}(\omega_{ij,t-365:t})$ positively relates ($0.741$, $t$-statistics$=54.584$) with $r_{i,t}^{resid}(FF3)$ and \emph{negative lead return} $\mathcal{LR}_{i,t}^{-}(\omega_{ij,t-365:t})$ has negative relationship ($-0.761$, $t$-statistics$=-51.768$) with $r_{i,t}^{resid}(FF3)$.

Then, we see interesting results after testing the firm size and firm liquidity decomposed variants reported in Column (4)-(5) of Table \ref{tab:simul_exret_regres_sign_sector}, which shows the asymmetric comovement effects depending on the size and stock liquidity of lead firms. Although all the \emph{lead return} variables have significant relationship between the stock residual returns, one can see the \emph{big size lead return} $\mathcal{LR}_{i,t}(\omega_{ij,t-365:t}^{BigLead})$ has a much stronger comovement effect ($0.529$, $t$-statistics$=41.797$) with stock residual return compared with \emph{small size lead return} $\mathcal{LR}_{i,t}(\omega_{ij,t-365:t}^{SmallLead})$ ($0.214$, $t$-statistics$=17.092$), and surprisingly, the \emph{low liquidity lead return} $\mathcal{LR}_{i,t}(\omega_{ij,t-365:t}^{LowLead})$ shows stronger comovement effect ($0.536$, $t$-statistics$=40.903$) as opposed to a weaker comovement effect from \emph{high liquidity lead return} $\mathcal{LR}_{i,t}(\omega_{ij,t-365:t}^{HighLead})$ ($0.246$, $t$-statistics$=16.860$).

Consistently, with replacing the residual returns from FF3 with the ones from FF5, all the above conclusions stay true and significant at 1\% level as shown in Columns from (6)-(10) in Table \ref{tab:simul_exret_regres_sign_sector}.

\subsubsection{Infeasible Portfolio Test}

The econometric insight of the result above make it evident that if a firm has higher \emph{lead return}, it has a higher stock return at the same day. We apply an infeasible portfolio test in the real stock market to relieve the model restrictions and draw a consistent conclusion.

We conduct the portfolio test by sorting the S\&P500 constituents based on the contemporaneous \emph{lead returns} as the method described in Section \ref{sec:method_sorting}. The news networks are constructed with a 1-year rolling window. Also, the 6-month rolling window networks (unreported) deliver similar results. The stocks are sorted into quintiles, and the portfolios are rebalanced each day. Firms in quintile 1 have the lowest \emph{lead return} values, while firms in quintile 5 have lead firms performing the best. One can notice that none of the portfolios is feasible. However, the point of those infeasible portfolios is meant to prove the comovement effect between the news-linked firms with a simple trading rule.

\begin{table}[!htb]
   \small
  \setlength{\tabcolsep}{0pt}
  \begin{threeparttable}
    \caption{Infeasible Portfolios of Sorting Contemporaneous \emph{lead return} $\mathcal{LR}(\omega_{ij,t-365:t})$}\label{tab:portfolio_comparison_simul}
    \begin{tabular*}{
      \linewidth}{@{\extracolsep{\fill}}>{\itshape}
      l
      *{5}{S[table-format=1.2, table-number-alignment = center]}
      *{2}{S[table-format=2.2, table-number-alignment = center]}
      l
      *{2}{S[table-format=2.2, table-number-alignment = center]}
      }
    \toprule
    \toprule
    \multicolumn{1}{c}{}  
    &\multicolumn{5}{c}{}
    &\multicolumn{2}{c}{FF3}  
    &\multicolumn{1}{c}{}  
    &\multicolumn{2}{c}{FF5}
    \\
    \cmidrule{7-8}
    \cmidrule{10-11}
    \multicolumn{1}{c}{Rank}  
    &\multicolumn{1}{c}{Mean} 
    &\multicolumn{1}{c}{SR} 
    &\multicolumn{1}{c}{\%MV} 
    % &\multicolumn{1}{c}{Size} 
    &\multicolumn{1}{c}{B/M} 
    &\multicolumn{1}{c}{Liquidty}
  
    &\multicolumn{1}{c}{$\alpha$}  
    &\multicolumn{1}{c}{$R^2$}  
    &\multicolumn{1}{c}{}  
    &\multicolumn{1}{c}{$\alpha$}  
    &\multicolumn{1}{c}{$R^2$}
    \\
    \midrule
    % \cline{1-2}
      {}  & \multicolumn{1}{c}{(1)}  & \multicolumn{1}{c}{(2)}  & \multicolumn{1}{c}{(3)}   
      & \multicolumn{1}{c}{(4)}   & \multicolumn{1}{c}{(5)}  & \multicolumn{1}{c}{(6)}     
      & \multicolumn{1}{c}{(7)}   & {} & \multicolumn{1}{c}{(8)}     & \multicolumn{1}{c}{(9)} \\[0.15cm] % & {}    
    
    % \multicolumn{2}{l}{$\mathcal{LR}(\omega_{ij,t-365:t})$} \\ 
    1   & -82.41 & -9.14 & 16.69 & 44.13 & 1.02 & -85.03   & 15.32 & {} & -85.02   & 15.73 \\ %      & 23.77
    {}  & {}     & {}    & {}    & {}    & {}   & (-25.42) & {}    & {} & (-25.60) & {} \\[0.15cm] % & {}
    2   & -25.74 & -3.44 & 21.14 & 40.35 & 0.87 & -28.31   & 18.22 & {} & -28.40   & 18.61 \\ %      & 23.95
    {}  & {}     & {}    & {}    & {}    & {}   & (-10.78) & {}    & {} & (-10.86) & {} \\[0.15cm] % & {}
    3   & 3.97   & 0.48  & 22.66 & 39.62 & 0.85 & 1.47     & 18.28 & {} & 1.38     & 18.76 \\ %      & 24.00
    {}  & {}     & {}    & {}    & {}    & {}   & (0.59)   & {}    & {} & (0.56)   & {} \\[0.15cm] % & {}
    4   & 31.30  & 4.18  & 21.88 & 39.54 & 0.86 & 28.86    & 18.55 & {} & 28.78    & 19.10 \\ %      & 23.97
    {}  & {}     & {}    & {}    & {}    & {}   & (11.21)  & {}    & {} & (11.28)  & {} \\[0.15cm] % & {}
    5   & 82.73  & 9.59  & 17.64 & 43.05 & 0.99 & 80.46    & 16.62 & {} & 80.37    & 17.29 \\ %      & 23.80
    {}  & {}     & {}    & {}    & {}    & {}   & (24.73)  & {}    & {} & (24.99)  & {} \\[0.15cm] % & {}
    % 5-1 & 165.15 & 24.79 & {}    & {}    & {}   & 165.49   & 0.29  & {} & 165.39   & 0.84 \\ %       & {}
    % {}  & {}     & {}    & {}    & {}    & {}   & (45.18)  & {}    & {} & (45.44)  & {} \\ %         & {}
    
      \bottomrule
      \bottomrule
    
    \end{tabular*}
  \begin{tablenotes}[flushleft]
    \setlength\labelsep{0pt} 
  \linespread{1}\small
  \item 
  The table reports the performance of the equal-weighted portfolios sorted by \emph{lead return}, $\mathcal{LR}(\omega_{ij,t-365:t})$, computed with news linkages aggregated in a 1-year rolling window. All the portfolios are infeasible. Column (1)-(5) are average simple daily return in bps; average annualized Sharpe-ratio, average market value proportion in percentage, average book-to-market ratio in percentage, average volume-to-share ratio in percentage. The last 4 columns report the alpha, $R^2$, and robust Newey-West $t$-statistics in parentheses against Fama-French 3-factor and Fama-French 5-factor, respectively.  
\end{tablenotes}
  \end{threeparttable}
\end{table}

\begin{figure}[!htb]
  \centering
  \begin{minipage}{1\linewidth}
    \includegraphics[width=\textwidth]{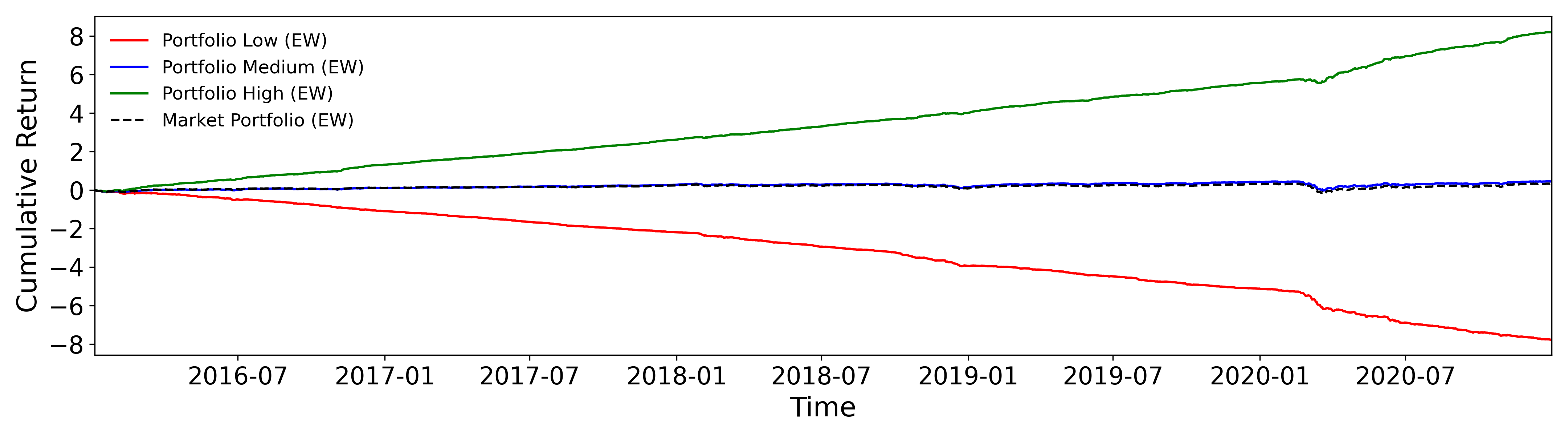}
      % \rule{\linewidth}{10em}
  \end{minipage}
  \caption{
      \textbf{Portfolios Cumulative Returns of Sorting \emph{Lead Return} $\mathcal{LR}(\omega_{ij,t-365:t})$}
  \newline 
  \small
  S\&P500 constituents are sorted into quintile equal-weighted portfolios. The three solid lines in \textcolor{red}{red}, \textcolor{blue}{blue}, \textcolor{green}{green} refer to cumulative returns of portofolios with lowest, medium, and highest \emph{lead return} value. The black dash line is the benchmark cumulative return of an equal-weighted S\&P500 market portfolio.
}\label{fig:accu_rts_rt0_ew_m1}
\end{figure}

As reported in the first two columns of Table \ref{tab:portfolio_comparison_simul}, the average simple returns of the equal-weighted portfolios monotonic increase from $-82.41$ bps to $82.73$ bps per day, and the annualized daily Sharpe-ratio increases from $-9.14$ to $9.95$. The last four columns of Table \ref{tab:portfolio_comparison_simul} under the heading FF3 and FF5 report the estimated alpha and $R^2$ for each portfolio in Fama-French 3 and 5-factor models, and the robust Newey–West $t$-statistics of the alpha is in the parentheses. Portfolio 5 has a significant positive alpha ($80.46$ bps) against the Fama-French 3-factor model compared with portfolio 1, which has a significant negative alpha ($-85.03$ bps). Columns (3)-(5) under the heading "\%MV", "B/M", "Liquidity" report the market value proportion, book-to-market ratio, volume-to-share ratio for each quintile portfolio. One can see that there is no monotonic pattern for all those characteristics. For example, both the quintile 1 and quintile 5 portfolios have relatively lower market value, higher book-to-market ratio, and higher stock liquidity. Figure \ref{fig:accu_rts_rt0_ew_m1} showcases the cumulative returns of three of the quintile portfolios (\textcolor{red}{lowest}, \textcolor{blue}{medium}, and \textcolor{darkgreen}{highest}) benchmarked by an equal-weighted S\&P500 market portfolio. The cumulative returns of quintile 1 and quintile 5 diverge quickly over time, implying that a portfolio with a higher simultaneous \emph{lead return} value consistently outperforms the portfolio with lower simultaneous \emph{lead return}.

We also report robust check on other \emph{lead return} variants including \emph{within(cross)-sector lead returns} $\mathcal{LR}(\omega_{ij,t-365:t}^{w(c)})$, \emph{positive(negative) lead returns} $\mathcal{LR}^{+(-)}(\omega_{ij,t-365:t})$ in Appendix, see Table \ref{tab:portfolio_comparison_simul_sector} and Table \ref{tab:portfolio_comparison_simul_sign}. All the robust check results draw a consistent conclusion. However, it is worth noting that the difference between quintile 5 and quintile 1 sorted by \emph{cross-sector lead return} (reported in the bottom panel of Table \ref{tab:portfolio_comparison_simul_sector}) is much smaller than that of those sorted by \emph{within-sector lead return}. Those evidences strongly suggest that the comovement effect does exist between firms from different sectors, though it is less significant.

\subsection{Lead Return Portfolio and Predictability}\label{sec:leadreturn_predictability}

\subsubsection{Momentum v.s Reversal}

With the evidence on stock return comovement between news-linked firms, it is appealing to investigate how the returns of a lead firm impact the returns of its follower stocks in the future. It is documented in the early studies, e.g., \cite{COHEN_2008}, that the customers' stock returns of a firm have a momentum effect on the firm's monthly stock returns. But the momentum effect is not true in our case for two plausible reasons. Firstly, information transmitting much faster causes higher efficiency in the current stock market than 10-15 years ago. Secondly, caused by the fast information transmission, the \emph{"overreaction"} phenomenon become significant, which explains the strong comovement effect between news-linked firms.

\begin{figure}[!htb]
  \centering
  \begin{minipage}{1\linewidth}
    \includegraphics[width=\textwidth]{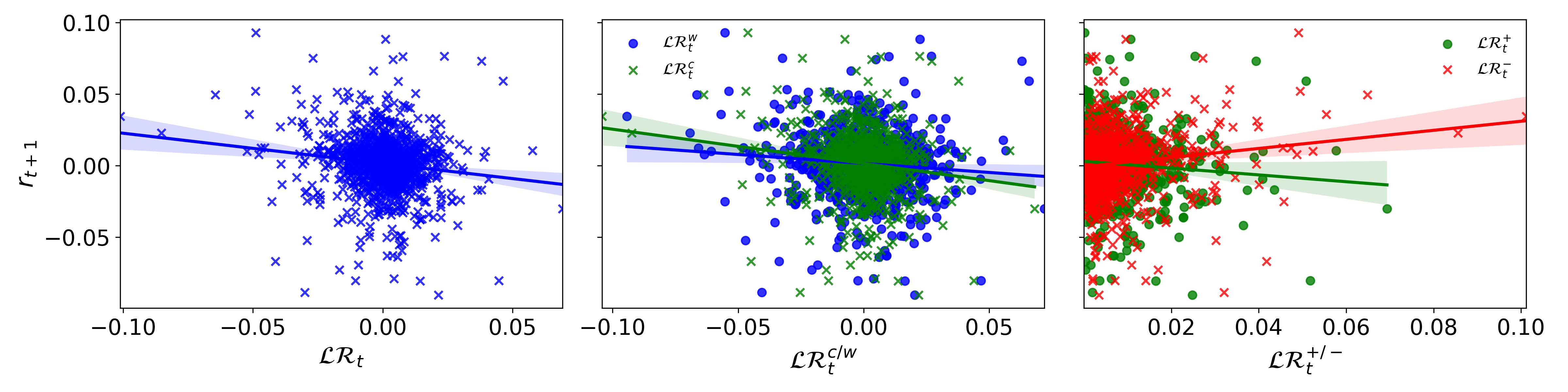}
      % \rule{\linewidth}{10em}
  \end{minipage}
  \caption{
      \textbf{Stock Return Against 1-Day Ahead \emph{lead return}, AAPL}
  \newline 
  \small
  Plots from left to right are AAPL 1-day ahead daily return $r_{t+1}$ against \emph{lead return} $\mathcal{LR}(\omega_{ij,t-365:t})$, sector decomposed variants $\mathcal{LR}\left(\omega_{ij,t-365:t}^{w/c}\right)$, and the return sign decomposed variants $\mathcal{LR}^{+/-}(\omega_{ij,t-365:t})$, respectively. The solid lines are estimated by OLS with bootstrap of 95\% confidence interval.}\label{fig:scatterplot_pred1_pp_sector_sign}
\end{figure}

As shown in Figure \ref{fig:scatterplot_pred1_pp_sector_sign}, three scatter plots of the 1-day ahead stock returns $r_{t+1}$ of Apple against multiple of its \emph{lead return} variables, implying that there is potential reversal effect of \emph{lead returns}. However, such a potential reversal effect could vanish for other stocks. To test the average effect across all S\&P500 constituents, we conduct the panel regression in Equation \eqref{eq:panel_reg_genereal} with $h=1$. In addition to the firm characteristics employed testing comovement effect in \ref{sec:linkage_valid}, we also use the 1-day lagged stock return $r_{t}$ as a control variable. 1-year rolling window size is used to construct the news networks.

Panels from left to right in Table \ref{tab:pred1_panel_res} are estimations of the impacts of \emph{lead returns} on 1-day ahead stock return $r_{t+1}$, stock residual return from FF3, $r_{t+1}^{resid}(FF3)$, and stock residual return from FF5, $r_{t+1}^{resid}(FF5)$, respectively. Column (1) of Table \ref{tab:pred1_panel_res} reports that \emph{lead return} shows insignificant effect ($0.003$) on the 1-day ahead stock return with ($t$-statistics=$0.601$), suggesting that one-standard deviation increase in $\mathcal{LR}(\omega_{ij,t-365:t})$ will cause only a $0.4$ bps ($1.497\%*0.003$) stock return increase in then next day $r_{t+1}$.

\begin{table}[!htb]
  \footnotesize
  \setlength{\tabcolsep}{0pt}
  \begin{threeparttable}
    \caption{Estimation of \emph{lead returns} Impact on 1-Day Ahead Stock Returns}\label{tab:pred1_panel_res}
    \begin{tabular*}{
      \linewidth}{@{\extracolsep{\fill}}>{\itshape}
      l 
      *{3}{S[table-format=1.3,table-number-alignment = center]} l 
      *{3}{S[table-format=1.3,table-number-alignment = center]} l 
      *{3}{S[table-format=1.3,table-number-alignment = center]}} 
      \toprule\toprule
      
      {} & \multicolumn{3}{c}{$r_{i,t+1}$} & {}  & \multicolumn{3}{c}{$r_{i,t+1}^{resid}(FF3)$} & {} & \multicolumn{3}{c}{$r_{i,t+1}^{resid}(FF5)$} \\
      \cline{2-4} \cline{6-8} \cline{10-12}
      {} & \multicolumn{1}{c}{(1)} & \multicolumn{1}{c}{(2)} & \multicolumn{1}{c}{(3)} &
      {} & \multicolumn{1}{c}{(4)} & \multicolumn{1}{c}{(5)} & \multicolumn{1}{c}{(6)} &
      {} & \multicolumn{1}{c}{(7)} & \multicolumn{1}{c}{(8)} & \multicolumn{1}{c}{(9)}\\
      
\multirow{2}{*}{$\mathcal{LR}_{i,t}(\omega_{ij,T})$}     & 0.003   &          & {}      & {} & -0.127    &           &          & {} & -0.125    &           & \\
{}                                                & (0.601) &          & {}      & {} & (-24.626) &           &          & {} & (-24.227) &           & \\[0.15cm]
\multirow{2}{*}{$\mathcal{LR}_{i,t}(\omega_{ij,T}^{w})$} &         & 0.010    & {}      & {} &           & -0.061    &          & {} &           & -0.062    & \\
{}                                                &         & (2.323)  & {}      & {} &           & (-12.452) &          & {} &           & (-13.036) & \\[0.15cm]
\multirow{2}{*}{$\mathcal{LR}_{i,t}(\omega_{ij,T}^{c})$} &         & -0.011   & {}      & {} &           & -0.087    &          & {} &           & -0.082    & \\
{}                                                &         & (-2.811) & {}      & {} &           & (-20.445) &          & {} &           & (-19.787) & \\[0.15cm]
\multirow{2}{*}{$\mathcal{LR}_{i,t}^{+}(\omega_{ij,T})$} &         &          & 0.032   & {} &           &           & -0.066   & {} &           &           & -0.064 \\
{}                                                &         &          & (5.124) & {} &           &           & (-9.697) & {} &           &           & (-18.650) \\[0.15cm]
\multirow{2}{*}{$\mathcal{LR}_{i,t}^{-}(\omega_{ij,T})$} &         &          & 0.021   & {} &           &           & 0.178    & {} &           &           & 0.176 \\
{}                                                &         &          & (3.503) & {} &           &           & (27.138) & {} &           &           & (56.187) \\[0.15cm]
% $Control$                                         & $\text{yes}$ & $\text{yes}$ & $\text{yes}$ & {} & $\text{yes}$ & $\text{yes}$ & $\text{yes}$ & {} & $\text{yes}$ & $\text{yes}$ & $\text{yes}$\\
      
      \bottomrule\bottomrule
      
    \end{tabular*}
  \begin{tablenotes}[flushleft]
   \setlength\labelsep{0pt} 
  \linespread{1}\small
  \item 
  A 1-year window size news network formation $T=[t-365:t]$. The first panel under the heading of $r_{t+1}$ reports the estimation of the impact from \emph{lead return} on stock returns. The middle (right) panel contains the estimation of the impact from \emph{lead returns} on stock residual returns from FF3 (FF5). Estimated coefficients of \emph{lead return} variables for each model are reported in corresponding column, and $t$-statistics is reported in the parentheses, with standard error corrected by clustering on individuals and time. Control variables, including market value $\log(MV)$, book-market ratio $B/M$, and turnover rate $volume/share$ and 1-day lagged return $r_t$, are employed in each model. All the financial statement related variables are 2 quarters lagged.
\end{tablenotes}
  \end{threeparttable}
\end{table}

We can see the mixed results when investigating the \emph{lead return} decomposed by sector and sign of lead returns. Column (2) of Table \ref{tab:pred1_panel_res} reports the impact from the sector decomposed \emph{lead return} variables. The \emph{cross-sector lead return} shows significant reversal impact on 1-day ahead return with an estimated coefficient of $-0.011$ ($t$-statistics=$-2.811$) at 1\% level, whereas the \emph{within-sector lead return} indicates significant momentum effect on 1-day ahead stock return with an estimated coefficient of $0.010$ ($t$-statistics=$2.323$) at 1\% level. Column (3) of Table \ref{tab:pred1_panel_res} reports the estimation when \emph{lead return} is decomposed by the sign of lead firm return. The \emph{positive lead return} component shows significant momentum effect with an estimated coefficient of $0.032$ ($t$-statistics=$5.124$), while the \emph{negative lead return} component indicates significant reversal effect with an estimated coefficient of $0.021$ ($t$-statistics=$3.503$), both at 1\% level.

A possible interpretation of the ambiguous reversal effect is the influences from other more dominant factors, such as the Fama-French factors. Columns (4)-(9) of Table \ref{tab:pred1_panel_res} report the estimation in which the dependent variable, 1-day ahead stock return, $r_{t+1}$, is replaced by the stock residual returns, $r_{t+1}^{resid}$, unexplained by the Fama-French factor models. One can see the consistent and significant reversal effect of \emph{lead return} variables on residual stock return. For instance, Column (4) of Table \ref{tab:pred1_panel_res} shows that $\mathcal{LR}(\omega_{ij,t-365:t})$ impacts negatively on $r_{t+1}^{resid}(FF3)$ with coefficient estimated $-0.127$ ($t$-statistics=$-24.626$), implying a one-standard-deviation increase in $\mathcal{LR}(\omega_{ij,t-365:t})$ will decrease the residual stock return from FF3 by $19$ bps ($1.497\%*0.127$). The estimated results strongly suggest that \emph{lead returns} contributes negatively to the stock returns in the next trading day. However, the reversal effect is offset by the momentum effect caused by other factors and firm characteristics.

\subsubsection{Portfolio Test on lead returns}

A portfolio test is performed to relieve the linearity restriction in the econometric model by sorting S\&P500 stocks into quintile based on the preceding day \emph{lead returns}. Here we mainly analyze the results for equal-weighted portfolios, and results from value-weighted portfolios are consistent.

\begin{equation}\label{eq:n_agg}
  \mathcal{LR}_{agg}(\omega_{ij,t-365:t}) = \mathcal{LR}^{+}(\omega_{ij,t-365:t,}^{w}) + \mathcal{LR}^{-}(\omega_{ij,t-365:t,}^{w}) + \mathcal{LR}^{-}(\omega_{ij,t-365:t,}^{c}) - \mathcal{LR}^{+}(\omega_{ij,t-365:t,}^{c})
\end{equation}

The simple aggregated \emph{lead return} proxy is an effort to capture the different impacts from all the decomposed \emph{lead return} variables. However, the optimization of utilizing those variables is beyond this study.

\begin{table}[!htb]
  \small
  \setlength{\tabcolsep}{0pt}
  \begin{threeparttable}
    \caption{Portfolios by Sorting Preceding \emph{lead return} variables (EW)}\label{tab:portfolio_comparison_npp_daily}
    \begin{tabular*}{
      \linewidth}{@{\extracolsep{\fill}}>{\itshape}
      l
      *{5}{S[table-format=1.2, table-number-alignment = center]}
      *{2}{S[table-format=-1.2, table-number-alignment = center]}
      l
      *{2}{S[table-format=-1.2, table-number-alignment = center]}}
    \toprule
    \toprule
    \multicolumn{1}{c}{}  
    &\multicolumn{5}{c}{}
    &\multicolumn{2}{c}{FF3}  
    &\multicolumn{1}{c}{}  
    &\multicolumn{2}{c}{FF5}
    \\
    \cmidrule{7-8}
    \cmidrule{10-11}
    \multicolumn{1}{c}{Rank}  
    &\multicolumn{1}{c}{Mean} 
    &\multicolumn{1}{c}{SR} 
    &\multicolumn{1}{c}{\%MV} 
    % &\multicolumn{1}{c}{Size} 
    &\multicolumn{1}{c}{B/M} 
    &\multicolumn{1}{c}{Liquidty}
  
    &\multicolumn{1}{c}{$\alpha$}  
    &\multicolumn{1}{c}{$\alpha(t)$}  
    &\multicolumn{1}{c}{}  
    &\multicolumn{1}{c}{$\alpha$}  
    &\multicolumn{1}{c}{$\alpha(t)$}
    \\
    \midrule

      {}  & \multicolumn{1}{c}{(1)}  & \multicolumn{1}{c}{(2)}  & \multicolumn{1}{c}{(3)}   
      & \multicolumn{1}{c}{(4)}   & \multicolumn{1}{c}{(5)}  & \multicolumn{1}{c}{(6)}     
      & \multicolumn{1}{c}{(7)}   & {} & \multicolumn{1}{c}{(8)}     & \multicolumn{1}{c}{(9)} \\[0.15cm] % & {}
      
    \multicolumn{11}{c}{\textsc{Panel A: Sorting $\mathcal{LR}(\omega_{ij,t-365:t})$}}\\

    1                                                                       & 1.46  & 0.12  & 16.69 & 44.13 & 1.02 & -1.17 & -0.39 & {} & -1.16 & -0.38 \\
    2                                                                       & 1.70  & 0.16  & 21.14 & 40.35 & 0.87 & -0.95 & -0.37 & {} & -0.99 & -0.38 \\
    3                                                                       & 3.93  & 0.47  & 22.66 & 39.62 & 0.85 & 1.45  & 0.57  & {} & 1.38  & 0.55 \\
    4                                                                       & 3.69  & 0.43  & 21.88 & 39.54 & 0.86 & 1.24  & 0.47  & {} & 1.12  & 0.43 \\
    5                                                                       & 3.40  & 0.38  & 17.64 & 43.05 & 0.99 & 1.20  & 0.42  & {} & 1.07  & 0.39 \\
    5-1                                                                     & 1.94  & 0.30  & {}    & {}    & {}   & 2.37  & 1.00  & {} & 2.23  & 0.95 \\[0.15cm]
  
    \multicolumn{11}{c}{\textsc{Panel B: Sorting $\mathcal{LR}_{agg}(\omega_{ij,t-365:t})$}}\\
    
    % \multicolumn{3}{l}{A(2): $\mathcal{LR}_{agg}(\omega_{ij,t-365:t})$}\\
    % \cline{1-3}
  
    1                                                                       & 2.22  & 0.23  & 15.06 & 42.24 & 0.90 & -0.42 & -0.16 & {} & -0.48 & -0.19 \\
    2                                                                       & 2.05  & 0.21  & 20.37 & 40.65 & 0.89 & -0.69 & -0.26 & {} & -0.74 & -0.28 \\
    3                                                                       & 2.44  & 0.27  & 23.45 & 39.86 & 0.88 & -0.04 & -0.02 & {} & -0.09 & -0.04 \\
    4                                                                       & 3.30  & 0.37  & 22.24 & 40.42 & 0.90 & 0.98  & 0.37  & {} & 0.90  & 0.34 \\
    5                                                                       & 3.95  & 0.43  & 18.88 & 43.56 & 1.02 & 1.72  & 0.60  & {} & 1.62  & 0.57 \\
    5-1                                                                     & 1.73  & 0.35  & {}    & {}    & {}   & 2.14  & 1.27  & {} & 2.10  & 1.25 \\[0.15cm]
    % {}                                                                    & {}    & {}    & {}    & {}    & {}   & {}    & {}    & {} & {}    & {} \\
    Mkt                                                                     & 2.80  & 0.31  & {}    & {}    & {}   & 0.32  & 0.12  & {} & 0.25  & 0.10 \\
  
      \bottomrule
      \bottomrule
      % \multicolumn{11}{r}{\footnotesize\itshape continued}

    \end{tabular*}
    \begin{tablenotes}[flushleft]
      \setlength\labelsep{0pt} 
    \linespread{1}\small
    \item Both Panel A and Panel B report the daily return, $r_{t+1}$, performance metrics of the equal-weighted portfolios, and all portfolios are daily rebalanced.
    portfolios in Panel A are sorted by \emph{lead returns}, $\mathcal{LR}(\omega_{ij,t-365:t})$, whereas portfolios in Panel B are sorted by $\mathcal{LR}_{agg}(\omega_{ij,t-365:t})$ described in Equation \ref{eq:n_agg}.
    News networks are constructed with 1-year rolling window news linkages.
    Column (1) reports the simple average return denoted in bps.
    Columns (2)-(5) are average annualized Sharpe-ratio, average market value proportion in percentage, the average book-to-market ratio in percentage, the average volume-to-share ratio in percentage. 
    The last 4 columns report the alpha and robust Newey-West $t$-statistics in parentheses against Fama-French 3-factor and Fama-French 5-factor models, respectively.
  \end{tablenotes}
    \end{threeparttable}
\end{table}

We show that \emph{lead return} variables carry only marginal predictability of 1-day ahead stock return. the portfolios are rebalanced by the end of each trading day and holds for the next trading day. As reported in Panel A of Table \ref{tab:portfolio_comparison_npp_daily} sorting $\mathcal{LR}(\omega_{ij,t-365:t})$, although quintile 5 outperforms the equal-weighted market portfolio by a margin of $3.4$ bps (Sharpe-ratio=0.38) against $2.8$ bps (Sharpe-ratio=0.31), and the difference between quintile 5 and quintile 1 is positive and generates positive alphas with respect to FF3 and FF5 by $2.37$ bps and $2.23$ bps, respectively, nevertheless insignificant statistically. It is difficult to conclude the cross-sectional predictability in $\mathcal{LR}(\omega_{ij,t-365:t})$. 

\begin{table}[!htb]
  \small
  \setlength{\tabcolsep}{0pt}
  \begin{threeparttable}
    \caption{Residual Portfolios by Sorting Preceding \emph{lead return} (EW)}\label{tab:portfolio_comparison_excessivert_daily}
    \begin{tabular*}{
      \linewidth}{@{\extracolsep{\fill}}>{\itshape}
      l
      *{5}{S[table-format=1.2, table-number-alignment = center]}
      *{2}{S[table-format=-1.2, table-number-alignment = center]}
      l
      *{2}{S[table-format=-1.2, table-number-alignment = center]}}
    \toprule
    \toprule
    \multicolumn{1}{c}{}  
    &\multicolumn{5}{c}{}
    &\multicolumn{2}{c}{FF3}  
    &\multicolumn{1}{c}{}  
    &\multicolumn{2}{c}{FF5}
    \\
    \cmidrule{7-8}
    \cmidrule{10-11}
    \multicolumn{1}{c}{Rank}  
    &\multicolumn{1}{c}{Mean} 
    &\multicolumn{1}{c}{SR} 
    &\multicolumn{1}{c}{\%MV} 
    % &\multicolumn{1}{c}{Size} 
    &\multicolumn{1}{c}{B/M} 
    &\multicolumn{1}{c}{Liquidty}
  
    &\multicolumn{1}{c}{$\alpha$}  
    &\multicolumn{1}{c}{$\alpha(t)$}  
    &\multicolumn{1}{c}{}  
    &\multicolumn{1}{c}{$\alpha$}  
    &\multicolumn{1}{c}{$\alpha(t)$}
    \\
    \midrule
      {}  & \multicolumn{1}{c}{(1)}  & \multicolumn{1}{c}{(2)}  & \multicolumn{1}{c}{(3)}   
      & \multicolumn{1}{c}{(4)}   & \multicolumn{1}{c}{(5)}  & \multicolumn{1}{c}{(6)}     
      & \multicolumn{1}{c}{(7)}   & {} & \multicolumn{1}{c}{(8)}     & \multicolumn{1}{c}{(9)} \\[0.15cm] % & {}    
    
    \multicolumn{11}{c}{\textsc{Panel A: Residual Returns from FF3, $r_{t+1}^{resid}(FF3)$}}\\

    1   & 1.65  & 0.15  & 16.69 & 44.13 & 1.02 & 2.12  & 0.71  & {} & 2.06  & 0.69 \\
    2   & -0.53 & -0.14 & 21.14 & 40.35 & 0.87 & -0.01 & -0.00  & {} & -0.09 & -0.03 \\
    3   & -0.58 & -0.15 & 22.66 & 39.62 & 0.85 & 0.04  & 0.02  & {} & -0.03 & -0.01 \\
    4   & -2.80  & -0.47 & 21.88 & 39.54 & 0.86 & -2.23 & -0.85 & {} & -2.34 & -0.90 \\
    5   & -4.91 & -0.74 & 17.64 & 43.05 & 0.99 & -4.24 & -1.52 & {} & -4.37 & -1.58 \\
    5-1 & -6.56 & -1.49 & {}    & {}    & {}   & -6.35 & -2.98 & {} & -6.43 & -3.01 \\[0.15cm]
  
    \multicolumn{11}{c}{\textsc{Panel B: Residual Returns from FF3, $r_{t+1}^{resid}(FF5)$}}\\
    
    1   & 2.37  & 0.25  & 16.69 & 44.13 & 1.02 & 2.84  & 0.96  & {} & 2.79  & 0.94 \\
    2   & -0.32 & -0.11 & 21.14 & 40.35 & 0.87 & 0.18  & 0.07  & {} & 0.11  & 0.04 \\
    3   & -0.53 & -0.14 & 22.66 & 39.62 & 0.85 & 0.05  & 0.02  & {} & -0.02 & -0.01 \\
    4   & -2.95 & -0.49 & 21.88 & 39.54 & 0.86 & -2.40  & -0.92 & {} & -2.5  & -0.96 \\
    5   & -5.24 & -0.79 & 17.64 & 43.05 & 0.99 & -4.60  & -1.67 & {} & -4.71 & -1.72 \\
    5-1 & -7.61 & -1.75 & {}    & {}    & {}   & -7.44 & -3.54 & {} & -7.51 & -3.57 \\[0.15cm]
  
      \bottomrule
      \bottomrule
      % \multicolumn{11}{r}{\footnotesize\itshape continued}

    \end{tabular*}
    \begin{tablenotes}[flushleft]
      \setlength\labelsep{0pt} 
    \linespread{1}\small
    \item Both Panel A and Panel B report the performance metrics of equal-weighted portfolios sorted by \emph{lead returns}, $\mathcal{LR}(\omega_{ij,t-365:t})$, and all portfolios are daily rebalanced.
    Panel A reports the residual of daily return with respect to the Fama-French 3-factor model, i.e $r_{t+1}^{resid}(FF3)$, whereas Panel B reports the residual of daily return with respect to the Fama-French 5-factor model, i.e $r_{t+1}^{resid}(FF5)$.
    News networks are constructed with 1-year rolling window news linkages.
    Column (1) reports the simple average return denoted in bps.
    Columns (2)-(5) are average annualized Sharpe-ratio, average market value proportion in percentage, the average book-to-market ratio in percentage, the average volume-to-share ratio in percentage. 
    The last 4 columns report the alpha and robust Newey-West $t$-statistics in parentheses against Fama-French 3-factor and Fama-French 5-factor models, respectively.
  \end{tablenotes}
    \end{threeparttable}
\end{table}

Panel B reports the performance of assets sorted by $\mathcal{LR}_{agg}(\omega_{ij,t-365:t})$, which shows a more clear pattern compared with the previous proxy $\mathcal{LR}(\omega_{ij,t-365:t})$, implying that it is possible to improve the predictability by calibrating the combination of those \emph{lead return} variables. The quintile 5 portfolio generates $3.95$ bps with a Sharpe-ratio of $0.43$ compared with an average return of $1.73$ bps and Sharpe-ratio of $0.35$ from the equal-weighted market portfolio. Similar to Panel A, the difference between quintile 5 and quintile 1 offers positive alphas, and the Sharpe-ratio is only slightly better than the equal-weighted market portfolio. Note that we do not consider the transaction cost in the above portfolio test, which is essential for the daily rebalancing portfolio test.

Furthermore, we show that a reversal effect can be found in the stock residual returns $r_{t+1}^{resid}$ of the portfolios sorted by the 1-day lagged \emph{lead return}, $\mathcal{LR}(\omega_{ij,t-365:t})$. Motivated by the econometric results in Table \ref{tab:pred1_panel_res}, we compute the residual returns of the stocks in each of the portfolios, named \enquote{residual portfolios} of sorting $\mathcal{LR}(\omega_{ij,t-365:t})$ reported in Table \ref{tab:portfolio_comparison_excessivert_daily}. The portfolios are the same as the ones reported in Panel A of Table \ref{tab:portfolio_comparison_npp_daily}, only the performance is evaluated based on stock residual returns instead of simple stock returns.

Panel A in Table \ref{tab:portfolio_comparison_excessivert_daily} reports the FF3 residual return $r_{t+1}^{resid}(FF3)$ performance of equal-weighted portfolio, where one can see clear monotonic decreasing pattern in terms of average residual return, Sharpe-ratio, which is a clear evidence for the negative impact from \emph{lead return} on the 1-day ahead stock return. Also, the difference between returns in quintile 5 and quintile 1 has significant negative alpha $-6.35$ with $t$-statistics=$-2.98$ relative to FF3 and $-6.43$ with $t$-statistics=$-3.01$ relative to FF5. Consistent and more significant results is shown in Panel B of Table \ref{tab:portfolio_comparison_excessivert_daily} reporting the FF5 residual return $r_{t+1}^{resid}(FF5)$.

  %!TEX root = ../main.tex

\section[News \emph{Network Degree} and Cross-Sectional Stock Returns]{News \emph{Network Degree} and \\ Cross-Sectional Stock Returns}\label{sec:networkdegree_predictability}

In this section, we first show that the network attention proxy, \emph{network degree} $\mathcal{A}(\omega_{ij,T})$, defined in Section \ref{sec:network_degree} can be a good predictor for the cross-sectional stock returns with a monthly rebalancing trading rule. Then we demonstrate that the predictability of \emph{network degree} varies depending on the characteristics of lead firms, including firm size and stock liquidity. Finally, we conduct a series of robust checks to dissolve the suspicion on the predictability of \emph{network degree}.

\subsection{Hypothesis of News Network Attention}

It is an interesting problem that how would a company perform if it had more attention from the news media. On the one hand, greater attention could be a positive signal for the popularity of a company. Consider one company going "hype" as a result of news reports covering it either in the headline (lead) or content (follower). This media attention could be caused by a new product release that many of its customers find desirable, or delivering certain innovation that could expand its the market share. Such that an investor would expect positive returns on that company's stock. On the other hand, bad news of a company also draws much attention from the media. For instance, a controversial market strategy made by a company causing accusations or even boycotts would definitely get this company covered by many news reports. In this case, more attention could be a signal of worsening performance in its stock.

The null hypothesis is that the attention proxy does not carry predictability on its stock performance. Under the null hypothesis, portfolios sorted by the attention proxy would not generate significantly different returns. 

We conjecture that the media news are inclined to spend more attention on positive news instead of negative ones. In other words, a firm with higher attention from media would outperform those firms with lower attention from media. By employing \emph{network degree} to capture the attention effect as described in Section \ref{sec:network_degree}, we perform portfolio test by sorting \emph{network degree} variables.

We examine the cross-sectional predictability of monthly stock return with the method in Section \ref{sec:method_sorting}. The network attention proxy, \emph{network degree} $\mathcal{A}(\omega_{ij,t-30:t})$ is computed on 1-month aggregated news network allowing the strategy to be more feasible. All S\&P500 constituents are sorted into quintiles based on \emph{network degree} at the end of each month, and the quintile portfolios are rebalanced at the end of each month and held for one month. Within each of the quintile portfolios, stocks are equal-weighted, and the weights are rebalanced monthly. In Table \ref{tab:portfolio_comp_attention_ew_dropzeroatt}, we report the equal-weighted portfolios.

The main point of the portfolio test is to see whether a long-short trading strategy based on sorting \emph{network degree} can deliver abnormal returns. Under the efficient market hypothesis, one shall not see an abnormal return with this simple trading strategy. On the other hand, a positive abnormal return would suggest the existence anomaly returns with news media attention 
proxy, \emph{network degree}, derived from networks of news.

\subsection{Portfolios of Sorting Network Degree}\label{sec:portfolio_sort_nd}

Following the trading strategy described above, Table \ref{tab:portfolio_comp_attention_ew_dropzeroatt} reports the equal-weighted portfolios of sorting the preceding monthly \emph{network degree}, $\mathcal{A}(\omega_{ij,t-30:t})$. The quintile portfolios are ranked in ascending order, where quintile 1 consists of assets with the lowest \emph{network degree} value from the preceding month, and quintile 5 consists of assets with the highest network degree. As one can see the 1-month aggregated news network governed by the power-law distribution in Figure \ref{fig:degree_distribution} of Section \ref{sec:network_degree}, many firm have network degrees equal to 0, which are not informative in our prediction. Therefore, we focus on the portfolio performance of stocks with network degrees larger than 0, i.e dropping out the firms with 0 network degrees when constructing portfolios. The consistent result of keeping all the firms is reported in Table \ref{tab:portfolio_comp_attention_ew_keepzeroatt} in Appendix.

\begin{figure}[!htb]
  \centering
  \begin{minipage}{1\linewidth}
    \includegraphics[width=\textwidth]{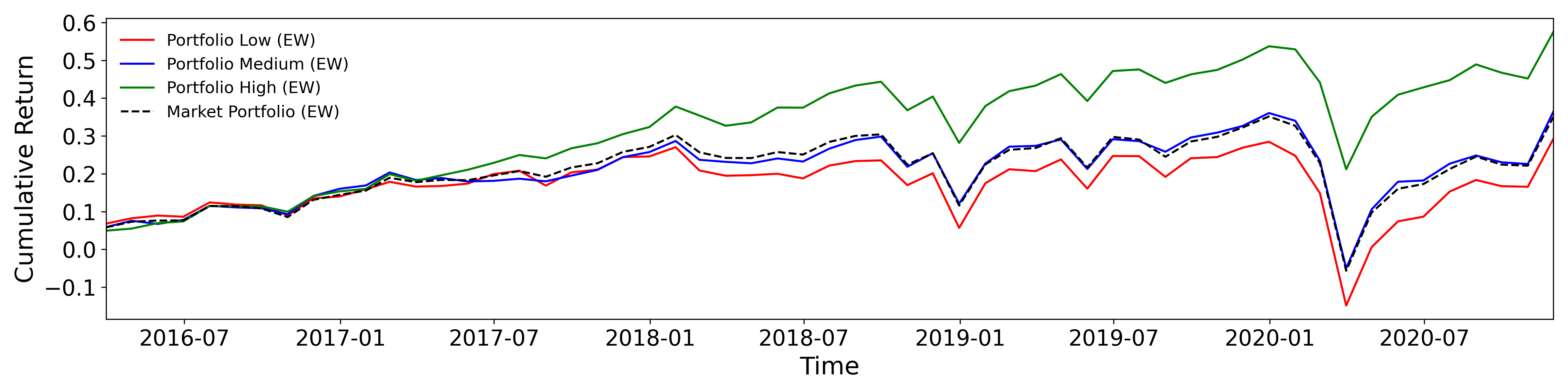}
    % \rule{\linewidth}{10em}
  \end{minipage}
  \caption{
    \textbf{Portfolios Cumulative Returns of Sorting Preceding Monthly Attention Proxy, \emph{Network Degree} $\mathcal{A}(\omega_{ij,t-30:t})$}
  \newline 
  \small
S\&P500 constituents are sorted into quintile equal-weighted portfolios based on $\mathcal{A}(\omega_{ij,t-30:t})$ from the preceding month. All portfolios are rebalanced at the end of each month and held for one month, and firms with zero network degrees are removed. The three solid lines in \textcolor{red}{red}, \textcolor{blue}{blue}, \textcolor{darkgreen}{green} refer to cumulative returns of portfolios with lowest, medium, and highest \emph{network degree} value. The black dash line is the benchmark cumulative return of an equal-weighted S\&P500 market portfolio.
}\label{fig:cumret_monthly_soreted_networkdegree}
\end{figure}

We first demonstrate the cumulative returns of portfolios constructed by sorting the preceding monthly $\mathcal{A}(\omega_{ij,t-30:t})$ in Figure \ref{fig:cumret_monthly_soreted_networkdegree}. The quintile 5 portfolio consists of stocks with highest network degree denoted in \textcolor{darkgreen}{green line} outperforms all the other portfolios significantly. The equal-weighted market portfolio in black dash line performs close to the quintile 3 portfolio in \textcolor{blue}{blue line}. Finally, the quintile 1 portfolio in \textcolor{red}{red line} has the lowest performance over the sample period.

\begin{table}[!htb]
  \small
  \setlength{\tabcolsep}{0pt}
  \begin{threeparttable}
    \caption{Portfolios by Sorting Preceding \emph{Network Degree} $\mathcal{A}$ \\\normalsize(Equal-Weighted, Drop $\mathcal{A}_{i,t}=0$)}\label{tab:portfolio_comp_attention_ew_dropzeroatt}
    \begin{tabular*}{\linewidth}{@{\extracolsep{\fill}}>{\itshape}
      l
      *{2}{S[table-format=1.2, table-number-alignment = center]}
      *{2}{S[table-format=2.2, table-number-alignment = center]}
      *{1}{S[table-format=1.2, table-number-alignment = center]}
      *{2}{S[table-format=-1.2, table-number-alignment = center]}
      l
      *{2}{S[table-format=-1.2, table-number-alignment = center]}
      } 
        \toprule
        \toprule
        \multicolumn{1}{c}{}  
        &\multicolumn{5}{c}{}
        &\multicolumn{2}{c}{FF3}  
        &\multicolumn{1}{c}{}  
        &\multicolumn{2}{c}{FF5}
        \\
        \cmidrule{7-8}
        \cmidrule{10-11}
        \multicolumn{1}{c}{Rank}  
        &\multicolumn{1}{c}{Mean} 
        &\multicolumn{1}{c}{SR} 
        &\multicolumn{1}{c}{\%MV} 
        % &\multicolumn{1}{c}{Size} 
        &\multicolumn{1}{c}{B/M} 
        &\multicolumn{1}{c}{Liquidty}
        &\multicolumn{1}{c}{$\alpha$}  
        &\multicolumn{1}{c}{$R^2$}  
        &\multicolumn{1}{c}{}  
        &\multicolumn{1}{c}{$\alpha$}  
        &\multicolumn{1}{c}{$R^2$}
        \\
        % \cline{1-2}
        \midrule
      % \multicolumn{2}{l}{Equal-Weighted} \\ 
      % {}  & {}   & {}   & {}    & {}    & {}   & {}      & {}    & {} & {}      & {} \\ %          & {}
      {}  & \multicolumn{1}{c}{(1)}  & \multicolumn{1}{c}{(2)}  & \multicolumn{1}{c}{(3)}   
      & \multicolumn{1}{c}{(4)}   & \multicolumn{1}{c}{(5)}  & \multicolumn{1}{c}{(6)}     
      & \multicolumn{1}{c}{(7)}   & {} & \multicolumn{1}{c}{(8)}     & \multicolumn{1}{c}{(9)} \\[0.15cm] % & {}
      1   & 0.51 & 0.23 & 8.38  & 42.86 & 0.89 & -0.75   & 91.17 & {} & -0.72   & 91.50\\
      {}  & {}   & {}   & {}    & {}    & {}   & (-2.55) & {}    & {} & (-2.50) & {} \\[0.15cm]
      2   & 0.41 & 0.17 & 10.21 & 42.29 & 0.89 & -0.87   & 92.89 & {} & -0.83   & 93.79\\
      {}  & {}   & {}   & {}    & {}    & {}   & (-3.75) & {}    & {} & (-3.86) & {} \\[0.15cm]
      3   & 0.64 & 0.31 & 13.15 & 42.35 & 0.91 & -0.64   & 90.96 & {} & -0.61   & 91.52\\
      {}  & {}   & {}   & {}    & {}    & {}   & (-2.75) & {}    & {} & (-2.70) & {} \\[0.15cm]
      4   & 0.51 & 0.22 & 19.28 & 38.90 & 0.93 & -0.81   & 87.79 & {} & -0.76   & 89.09\\
      {}  & {}   & {}   & {}    & {}    & {}   & (-2.81) & {}    & {} & (-2.69) & {} \\[0.15cm]
      5   & 1.01 & 0.58 & 48.98 & 36.82 & 1.00 & -0.27   & 93.03 & {} & -0.25   & 93.80\\
      {}  & {}   & {}   & {}    & {}    & {}   & (-1.48) & {}    & {} & (-1.39) & {} \\[0.15cm]
      5-1 & 0.50  & 0.79 & {}    & {}    & {}   & 0.48    & 37.51 & {} & 0.47    & 38.89\\
      {}  & {}   & {}   & {}    & {}    & {}   & (2.50)  & {}    & {} & (2.51) & {} \\[0.15cm]
      Mkt & 0.64 & 0.31 & {}    & {}    & {}   & -0.69   & 92.56 & {} & -0.65   & 93.34 \\ %       & {}
      {}  & {}   & {}   & {}    & {}    & {}   & (-3.03) & {}    & {} & (-2.97) & {} \\[0.15cm] %  & {}
      \bottomrule
        \bottomrule
    \end{tabular*}
  \begin{tablenotes}[flushleft]
   \setlength\labelsep{0pt} 
  \linespread{1}\small
  \item 
The table reports the performance of the equal-weighted portfolios sorted by the network attention proxy, \emph{network degree} $\mathcal{A}(\omega_{ij,t-30:t})$, computed with news linkages from the preceding month. Firms with zero network degrees are removed before forming the portfolios. Columns (1)-(5) are average simple daily return in percentage; the average annualized sharpe-ratio, average market value proportion in percentage, the average book-to-market ratio in percentage, and the average volume-to-share ratio in percentage. The last 4 columns report the alpha, $R^2$, and robust Newey-West $t$-statistics in parentheses with respect to the Fama-French 3-factor and the Fama-French 5-factor models, respectively.
\end{tablenotes}
  \end{threeparttable}
\end{table}

Column (1) in Table \ref{tab:portfolio_comp_attention_ew_dropzeroatt} reports the average simple return in percentage for each quintile portfolio across the 5-year sample period. Most importantly, quintile 5 generates $101$ bps per month with a Sharpe-ratio $0.58$, outperforming the other quintile portfolios and the equal-weighted market portfolio ($64$ bps and with Sharpe-ratio $0.31$). One can also see that the simple return is almost monotonic increasing, except for quintile 4, which generates a $51$ bps per month, just slightly higher than quintile 1 with $48$ bps.

As reported in the row with index "\emph{5-1}" of Table \ref{tab:portfolio_comp_attention_ew_dropzeroatt}, the zero-cost long-short portfolio generates a positive return of $50$ bps per month and a high Sharpe-ratio of $0.79$. Moreover, the long-short portfolio shows a significant positive alpha of $48$ bps ($t$-statistics=$2.07$) relative to FF3 and $47$ bps ($t$-statistics=$1.97$) with respect to FF5. A similar result of value-weighting portfolios is reported in Table \ref{tab:portfolio_comp_attention_vw_dropzeroatt} in Appendix.

The positive return and alphas from the simple zero-cost long-short portfolio provide evidence for the predictability of the attention proxy, \emph{network degree}, on stock return, and it is feasible for investors to form portfolios. Additionally, quintile 5 can be viewed as an enhanced market portfolio that consists of around 100 stocks from S\&P 500 constituents, for its much higher average return and Sharpe-ratio than the market index.

\subsection{What Are The Important News Linkages?}

Are all the detected linkages equally important, or are certain types of linkages more crucial for predicting the cross-sectional stock returns? In this subsection, we reveal that \emph{network degree} variables of different types of linkages carry different levels of predictability by studying network attention variants detailed in Section \ref{sec:network_degree}. 

% Those attention variants include \emph{lead(follower) attention}, \emph{within(cross)-sector attention}, \emph{big(small) size lead attention}, and \emph{high(low) liquid lead attention}.

% Specifically, we study the \emph{network degree} variables of the news network firms being lead and follower, the network degree of firms from the same and different sectors, the network degree accounting for the size effect and liquidity of the lead firm. Based on the news network decompositions described in Section \ref{sec:network_decomp}, \emph{network degree} can be derived further to account for some of the effects of our interests. 
% In the following portfolio tests, we still focus on the monthly aggregated news network. In other words, \emph{network degree} is noted as $\mathcal{A}(\omega_{ij,t-30:t})$, if not stated otherwise. Continuing with the previous portfolio test, we focus only on the firms with non-zero \emph{network degrees} for consistency.

\subsubsection{Lead and Follower Attention}

First of all, one might wonder whether the news attention of being lead firm is more important than being a follower firm for stock returns prediction. We analyze the predictability of the \emph{lead attention} and \emph{follower attention} defined in Section \ref{sec:network_degree}.

Panel A of Table \ref{tab:linkages_importance_test_ew_dropzeroatt} reports the portfolio test results on the two network attention variants, where Panel A(1) shows the portfolios of sorting $\mathcal{A}^{lead}(\omega_{ij,t-30:t})$, and Panel A(2) contains the $\mathcal{A}^{follower}(\omega_{ij,t-30:t})$ sorted results. The zero-cost long-shot trading strategy, indexed with "\emph{5-1}" in Panel A, of both sorting cases show similar significant abnormal returns with respect to FF3 and FF5. However, the performance of the quintile portfolio is not monotonically increasing in both of the sorting cases. In Panel A(1), we can see that quintile 5 outperforms any other quintile portfolios, resulting in the long-short strategy producing a significant alpha with respect to FF3 ($51$ bps, $t$-statistics=$2.92$) and FF5 ($51$ bps, $t$-statistics=$3.01$), as shown in Column (6)-(9). A similar result for the \emph{follower attention} sorting case in Panel A(2) reveals that quintile 5 stands out despite quintile 1-4 having similar performance. As such, the long-short strategy generates significant alpha with respect to FF3 ($39$ bps, $t$-statistics=$2.91$) and FF5 ($37$ bps, $t$-statistics=$2.82$). \emph{Lead attention} is relatively more important for the cross-sectional stock return prediction, but \emph{follower attention} also contributes to the predictability of using \emph{network degree}.

\subsubsection{Within-Sector and Cross-Sector Attention}

It is a stylized fact that firms from the same sector tend to have a closer relationship than those from different sectors. As shown in Section \ref{sec:linkage_valid}, comovement effect is more significant between the stocks from the same sector. We are motivated to understand whether the firms from the same sector are more informative in terms of stock return prediction. Base on the sector decomposed attention variants, we further test the predictability of \emph{within-sector attention} $\mathcal{A}(\omega_{ij,t-30:t}^{w})$, and \emph{cross-sector attention} $\mathcal{A}(\omega_{ij,t-30:t}^{c})$.

As reported in Panel B of Table \ref{tab:linkages_importance_test_ew_dropzeroatt}, we can see that \emph{within-sector attention} does provide stronger predictability compared with \emph{cross-sector attention} in terms of significant positive abnormal return from the long-short strategy. Specifically, the long-short portfolio by sorting $\mathcal{A}(\omega_{ij,t-30:t}^{w})$ produces an alpha of $59$ bps ($t$-statistics=$3.14$) with respect to FF3 and $57$ bps ($t$-statistics=$3.26$) relative to FF5. And the long-short portfolio sorted by $\mathcal{A}(\omega_{ij,t-30:t}^{c})$ has positive alphas with respect to FF3 and FF5 ($27$ bps, $26$ bps) but insignificant ($t$-statistics=$1.35$, $t$-statistics=$1.31$). Moreover, a clear monotonic increase pattern on the quintile portfolios performance is found with sorting $\mathcal{A}(\omega_{ij,t-30:t}^{w})$, see Columns (1)-(2) in Panel B(1). While assets sorted by $\mathcal{A}(\omega_{ij,t-30:t}^{c})$ do not have a monotonic increase performance pattern.

From the statistical point of view, one can conclude that \emph{within-sector attention} provides better and more stable cross-sectional predictability. However, it does not mean that one can always have higher performance with \emph{within-sector attention}. As one can see that quintile 5 sorted by $\mathcal{A}(\omega_{ij,t-30:t}^{w})$ in Panel B(1) with $89$ bps (Sharpe-ratio=$0.49$) underperforms quintile 5 sorted by $\mathcal{A}(\omega_{ij,t-30:t}^{c})$ in Panel B(2), which has $97$ bps (Sharpe-ratio=$0.52$).

\subsubsection{Big Size and Small Size Lead Firm Attention}

It is commonly believed that media are more likely to cover the news of larger firms. Here we would like to see how important it is to appear in the news content focusing on large firms compared to the news revolving around the small firms. The \emph{big lead attention} $\left[\omega_{ij,t-30:t}^{BigLead}\right]$ and \emph{small lead attention} $\left[\omega_{ij,t-30:t}^{SmallLead}\right]$ defined in Section \ref{sec:network_degree} are employed for the comparison.

Panel C in Table \ref{tab:linkages_importance_test_ew_dropzeroatt} reports the performance of the portfolios sorted by the two attention proxy variants, where we can see \emph{big lead attention} shows significant predictability by providing positive abnormal returns from the long-short strategy. In contrast, \emph{small lead attention} gives mixed results and negative insignificant abnormal returns from the long-short strategy. As reported in Column (1)-(2) of Panel C(1), quintile portfolio performance, mean return and Sharpe-ratio, increase monotonic along with the increase of $\mathcal{A}(\omega_{ij,t-30:t}^{BigLead})$, and Column (6)-(9) report that the long-short strategy indexed with "\emph{5-1}" provides an alpha of $68$ bps ($t$-statistics=$2.65$) with respect to FF3 and also an alpha of $64$ bps relative to FF5 ($t$-statistics=$2.62$). From the long-short strategy result of sorting $\mathcal{A}(\omega_{ij,t-30:t}^{SmallLead})$ reported in Column (6)-(9) of Panel C(2), it reveals that a firm with higher $\mathcal{A}(\omega_{ij,t-30:t}^{SmallLead})$ is likely to underperform, which is opposite to $\mathcal{A}(\omega_{ij,t-30:t}^{BigLead})$, but nevertheless insignificant. Note that the smaller firms have fewer news coverages, and we drop the stocks that have attention variables equal to zero. Hence, quintile portfolios sorted by $\mathcal{A}(\omega_{ij,t-30:t}^{SmallLead})$ have fewer stocks, averaging 38 stocks for each quintile portfolio in each month.

The above analysis shows that the news linkages formed with the big market capital lead firms can be more informative in terms of providing the cross-sectional stock return predictability. Specifically, the more linkages a firm connecting to big firms as a follower, the more likely it will have better performance in the next month. On the other side, the news linkages formed with smaller market capital lead firms give opposite implications, however insignificant. 

\begin{ThreePartTable}
  \renewcommand\TPTminimum{\textwidth}
  %% Arrange for "longtable" to take up full width of text block
  \setlength\LTleft{0pt}
  \setlength\LTright{0pt}
  \setlength\tabcolsep{0pt}
  
  \begin{TableNotes}[flushleft]
     \setlength\labelsep{0pt} 
    \linespread{1}\small
    \item 
This table reports the portfolio performance of sorting network attention proxies, \emph{network degree} $\mathcal{A}$, computed with different types of linkages. All portfolios are equal-weighted, and rebalanced monthly. We remove the firms that have the sorting variable equal to 0. In each panel, Columns (1)-(5) are the average simple daily return in percentage; average annualized sharpe-ratio, average market value proportion in percentage, average book-to-market ratio in percentage, average volume-to-share ratio in percentage. The last 4 columns report the alpha and robust Newey-West $t$-statistics with respect to the Fama-French 3-factor and the Fama-French 5-factor models, respectively.
\item Panel A reports the differences in portfolio performance between the \emph{lead attention} $\mathcal{A}^{lead}(\omega_{ij,t-30:t})$ (Panel A1) and the \emph{follower attention} $\mathcal{A}^{follower}(\omega_{ij,t-30:t})$ (Panel A2).
Panel B reports the differences in portfolio performance between the \emph{within-sector attention} $\mathcal{A}(\omega_{ij,t-30:t}^{w})$ (Panel B1) and the \emph{cross-sector attention} $\mathcal{A}(\omega_{ij,t-30:t}^{c})$ (Panel B2).
Panel C reports the differences in portfolio performance between the \emph{big lead attention} $\mathcal{A}(\omega_{ij,t-30:t}^{BigLead})$ (Panel C1) and the \emph{small lead attention} $\mathcal{A}(\omega_{ij,t-30:t}^{SmallLead})$ (Panel C2).
Panel D reports the differences in portfolio performance between the \emph{high lead attention} $\mathcal{A}(\omega_{ij,t-30:t}^{HighLead})$ (Panel D1) and the \emph{low lead attention} $\mathcal{A}(\omega_{ij,t-30:t}^{LowLead})$ (Panel D2).
All the attention variants are defined in Section \ref{sec:network_degree}.
  \end{TableNotes}
  
  \begin{xltabular}{\linewidth}{@{\extracolsep{\fill}}>{\itshape}
        l
        *{2}{S[table-format=1.2, table-number-alignment = center]}
        *{2}{S[table-format=2.2, table-number-alignment = center]}
        *{1}{S[table-format=1.2, table-number-alignment = center]}
        *{2}{S[table-format=-1.2, table-number-alignment = center]}
        l
        *{2}{S[table-format=-1.2, table-number-alignment = center]}
        }
      % \caption{Example table}\label{tab:example} \\
      \caption{Portfolio Tests on Linkage Types \\\normalsize(Equal-Weighted, Drop $\mathcal{A}_{i,t}=0$)}\label{tab:linkages_importance_test_ew_dropzeroatt}\\
      \toprule
      \toprule
          \multicolumn{1}{c}{}  
          &\multicolumn{5}{c}{}
          &\multicolumn{2}{c}{FF3}  
          &\multicolumn{1}{c}{}  
          &\multicolumn{2}{c}{FF5}
          \\
          \cmidrule{7-8}
          \cmidrule{10-11}
          \multicolumn{1}{c}{Rank}  
          &\multicolumn{1}{c}{Mean} 
          &\multicolumn{1}{c}{SR} 
          &\multicolumn{1}{c}{\%MV} 
          % &\multicolumn{1}{c}{Size} 
          &\multicolumn{1}{c}{B/M} 
          &\multicolumn{1}{c}{Liquidty}
          &\multicolumn{1}{c}{$\alpha$}  
          &\multicolumn{1}{c}{$\alpha(t)$}  
          &\multicolumn{1}{c}{}  
          &\multicolumn{1}{c}{$\alpha$}  
          &\multicolumn{1}{c}{$\alpha(t)$}
          \\
          \midrule
        {}                        & \multicolumn{1}{c}{(1)} & \multicolumn{1}{c}{(2)} & \multicolumn{1}{c}{(3)}
        & \multicolumn{1}{c}{(4)} & \multicolumn{1}{c}{(5)} & \multicolumn{1}{c}{(6)}
        & \multicolumn{1}{c}{(7)} & {}                      & \multicolumn{1}{c}{(8)} & \multicolumn{1}{c}{(9)} \\
  
          \addtocounter{table}{-1} 
      \endfirsthead
      %%%%
      \caption{continued} \\        
      
      \toprule
          \multicolumn{1}{c}{}  
          &\multicolumn{5}{c}{}
          &\multicolumn{2}{c}{FF3}  
          &\multicolumn{1}{c}{}  
          &\multicolumn{2}{c}{FF5}
          \\
          \cmidrule{7-8}
          \cmidrule{10-11}
          \multicolumn{1}{c}{Rank}  
          &\multicolumn{1}{c}{Mean} 
          &\multicolumn{1}{c}{SR} 
          &\multicolumn{1}{c}{\%MV} 
          % &\multicolumn{1}{c}{Size} 
          &\multicolumn{1}{c}{B/M} 
          &\multicolumn{1}{c}{Liquidty}
          &\multicolumn{1}{c}{$\alpha$}  
          &\multicolumn{1}{c}{$\alpha(t)$}  
          &\multicolumn{1}{c}{}  
          &\multicolumn{1}{c}{$\alpha$}  
          &\multicolumn{1}{c}{$\alpha(t)$}
          \\
          \midrule
        {}  & \multicolumn{1}{c}{(1)}  & \multicolumn{1}{c}{(2)}  & \multicolumn{1}{c}{(3)}   
        & \multicolumn{1}{c}{(4)}   & \multicolumn{1}{c}{(5)}  & \multicolumn{1}{c}{(6)}     
        & \multicolumn{1}{c}{(7)}   & {} & \multicolumn{1}{c}{(8)}     & \multicolumn{1}{c}{(9)} \\
  
      \endhead
      %%%%
      \midrule[\heavyrulewidth]
      \multicolumn{11}{r}{\footnotesize\itshape continued}
      \endfoot
      %%%%
      \bottomrule
      \bottomrule
      \insertTableNotes  % tell LaTeX where to insert the table-related notes
      \endlastfoot
      %%%%
        \multicolumn{11}{c}{\textsc{Panel A: Lead Attention v.s Follower Attention}}\\[0.15cm]
        \cline{3-9}
        \multicolumn{5}{l}{\textsc{Panel A(1): Sorted by $\mathcal{A}^{lead}$}}\\
        1   & 0.32 & 0.12 & 10.52 & 41.60 & 0.90 & -0.92 & -3.63 & {} & -0.90 & -3.57 \\
        2   & 0.78 & 0.37 & 12.57 & 41.11 & 0.88 & -0.51 & -1.62 & {} & -0.47 & -1.60 \\
        3   & 0.59 & 0.27 & 13.82 & 42.98 & 0.89 & -0.63 & -2.38 & {} & -0.57 & -2.29 \\
        4   & 0.51 & 0.24 & 19.57 & 39.59 & 0.94 & -0.81 & -3.27 & {} & -0.78 & -3.16 \\
        5   & 0.91 & 0.50  & 43.52 & 36.91 & 1.08 & -0.41 & -1.96 & {} & -0.39 & -1.79 \\
        5-1 & 0.60 & 1.07 & {}    & {}    & {}   & 0.51  & 2.92  & {} & 0.51  & 3.01 \\[0.15cm]
        \multicolumn{5}{l}{\textsc{Panel A(2): Sorted by $\mathcal{A}^{follower}$}}\\
        1   & 0.63 & 0.29 & 8.26  & 42.84 & 0.93 & -0.74 & -3.06 & {} & -0.70 & -3.03 \\
        2   & 0.51 & 0.21 & 9.96  & 41.24 & 0.91 & -0.8  & -2.92 & {} & -0.76 & -2.87 \\
        3   & 0.6  & 0.26 & 13.33 & 41.79 & 0.95 & -0.66 & -2.38 & {} & -0.62 & -2.28 \\
        4   & 0.53 & 0.26 & 20.90 & 38.88 & 0.96 & -0.73 & -3.69 & {} & -0.70 & -3.47 \\
        5   & 0.97 & 0.54 & 47.56 & 36.84 & 0.90 & -0.35 & -1.73 & {} & -0.33 & -1.70 \\
        5-1 & 0.34 & 0.61 & {}    & {}    & {}   & 0.39  & 2.91  & {} & 0.37  & 2.82 \\[0.15cm]
      % \midrule
  % \pagebreak\\
      \multicolumn{11}{c}{\textsc{Panel B: Within-Sector Attention v.s Cross-Sector Attention}}\\[0.15cm]
      \cline{3-9}
      \multicolumn{5}{l}{\textsc{Panel B(1): Sorted by $\mathcal{A}(\omega_{ij,t-30:t}^{w})$}}\\
      1   & 0.32 & 0.13 & 9.24  & 39.85 & 0.91 & -0.98 & -3.75 & {} & -0.95 & -3.74 \\
      2   & 0.54 & 0.24 & 11.14 & 41.59 & 0.90 & -0.71 & -2.34 & {} & -0.67 & -2.27 \\
      3   & 0.55 & 0.25 & 14.63 & 41.68 & 0.90 & -0.75 & -3.00 & {} & -0.72 & -2.97 \\
      4   & 0.61 & 0.29 & 22.00 & 40.63 & 0.91 & -0.68 & -3.12 & {} & -0.64 & -2.93 \\
      5   & 0.89 & 0.49 & 42.99 & 39.96 & 0.99 & -0.39 & -1.93 & {} & -0.38 & -1.92 \\
      5-1 & 0.56 & 1.02 & {}    & {}    & {}   & 0.59  & 3.14  & {} & 0.57  & 3.26 \\[0.15cm]
%   \pagebreak\\
      \multicolumn{5}{l}{\textsc{Panel B(2): Sorted by $\mathcal{A}(\omega_{ij,t-30:t}^{c})$}}\\
      1   & 0.50 & 0.22 & 9.73  & 44.56 & 0.89 & -0.66 & -2.48 & {} & -0.63 & -2.48 \\
      2   & 0.66 & 0.29 & 11.54 & 41.72 & 0.91 & -0.65 & -2.53 & {} & -0.63 & -2.52 \\
      3   & 0.63 & 0.31 & 13.91 & 39.39 & 0.93 & -0.68 & -3.06 & {} & -0.65 & -2.96 \\
      4   & 0.35 & 0.14 & 19.65 & 34.79 & 0.96 & -0.93 & -2.82 & {} & -0.87 & -2.77 \\
      5   & 0.97 & 0.52 & 45.17 & 36.45 & 1.04 & -0.38 & -1.89 & {} & -0.36 & -1.70 \\
      5-1 & 0.47 & 0.73 & {}    & {}    & {}   & 0.27  & 1.35  & {} & 0.26  & 1.31 \\[0.15cm]
  
  % \pagebreak\\
      \multicolumn{11}{c}{\textsc{Panel C: Lead Firm Size Attention}}\\[0.15cm]
      \cline{3-9}
      \multicolumn{5}{l}{\textsc{Panel C(1): Sorted by $\mathcal{A}(\omega_{ij,t-30:t}^{BigLead})$}}\\
      1   & 0.41 & 0.15 & 6.45  & 47.06 & 1.04 & -0.89 & -2.98 & {} & -0.85 & -2.91 \\
      2   & 0.54 & 0.21 & 8.06  & 43.72 & 1.03 & -0.85 & -2.48 & {} & -0.80 & -2.43 \\
      3   & 0.74 & 0.41 & 12.83 & 38.69 & 0.96 & -0.52 & -2.62 & {} & -0.51 & -2.57 \\
      4   & 0.83 & 0.51 & 22.95 & 33.09 & 0.72 & -0.44 & -2.50 & {} & -0.43 & -2.44 \\
      5   & 1.05 & 0.67 & 49.71 & 34.58 & 0.75 & -0.21 & -1.95 & {} & -0.22 & -1.79 \\
      5-1 & 0.64 & 0.59 & {}    & {}    & {}   & 0.68  & 2.65  & {} & 0.64  & 2.62 \\[0.15cm]
      \multicolumn{5}{l}{\textsc{Panel C(2): Sorted by $\mathcal{A}(\omega_{ij,t-30:t}^{SmallLead})$}}\\
      1   & 0.54  & 0.25  & 19.91 & 37.66 & 0.87 & -0.77  & -2.84 & {} & -0.71 & -2.76 \\
      2   & 0.78  & 0.42  & 20.04 & 39.26 & 0.89 & -0.48  & -2.29 & {} & -0.47 & -2.31 \\
      3   & 0.29  & 0.11  & 20.46 & 43.65 & 0.95 & -0.97  & -2.81 & {} & -0.94 & -2.72 \\
      4   & 0.54  & 0.20  & 21.21 & 47.73 & 1.08 & -0.80  & -1.82 & {} & -0.71 & -1.70 \\
      5   & -0.03 & -0.05 & 18.38 & 55.66 & 1.59 & -1.40  & -2.40 & {} & -1.30 & -2.31 \\
      5-1 & -0.57 & -0.52 & {}    & {}    & {}   & -0.62  & -1.42 & {} & -0.59 & -1.35 \\[0.15cm]

      \multicolumn{11}{c}{\textsc{Panel D: Lead Firm Liquidity Attention}}\\[0.15cm]
      \cline{3-9}
      \multicolumn{5}{l}{\textsc{Panel D(1): Sorted by $\mathcal{A}(\omega_{ij,t-30:t}^{HighLead})$}}\\
      1   & 0.54  & 0.28  & 15.87 & 38.22 & 0.80 & -0.66 & -2.73 & {} & -0.63 & -2.59 \\
      2   & 0.69  & 0.36  & 17.17 & 37.97 & 0.80 & -0.56 & -2.81 & {} & -0.56 & -2.79 \\
      3   & 0.57  & 0.24  & 18.62 & 42.15 & 0.95 & -0.92 & -3.57 & {} & -0.89 & -3.54 \\
      4   & 0.09  & -0.00  & 19.32 & 46.18 & 1.20 & -1.23 & -3.25 & {} & -1.16 & -3.20 \\
      5   & 0.33  & 0.09  & 29.01 & 47.47 & 1.71 & -1.17 & -2.38 & {} & -1.07 & -2.30 \\
      5-1 & -0.21 & -0.23 & {}    & {}    & {}   & -0.51 & -1.28 & {} & -0.44 & -1.16 \\[0.15cm]
        % \pagebreak\\
      \multicolumn{5}{l}{\textsc{Panel D(2): Sorted by $\mathcal{A}(\omega_{ij,t-30:t}^{LowLead})$}}\\
      1   & 0.32 & 0.12 & 8.07  & 43.42 & 1.04 & -0.96 & -4.43 & {} & -0.93 & -4.51 \\
      2   & 0.95 & 0.48 & 9.65  & 41.60 & 1.01 & -0.30 & -1.22 & {} & -0.28 & -1.18 \\
      3   & 0.51 & 0.24 & 13.41 & 38.52 & 0.84 & -0.80 & -2.87 & {} & -0.76 & -2.78 \\
      4   & 1.09 & 0.75 & 21.44 & 34.70 & 0.67 & -0.05 & -0.20 & {} & -0.03 & -0.14 \\
      5   & 1.05 & 0.71 & 47.44 & 34.37 & 0.59 & -0.11 & -0.73 & {} & -0.11 & -0.75 \\
      5-1 & 0.73 & 0.95 & {}    & {}    & {}   & 0.85  & 4.90  & {} & 0.81  & 4.99 \\[0.15cm]

  \end{xltabular}
  
  \end{ThreePartTable}

\subsubsection{More Liquid and Less Liquid Lead Firm Attention}

Stock liquidity is often investigated because firms with higher stock liquidity are deemed to incorporate news information faster into their stock prices. In comparison, less liquid stocks could carry material information ignored by many investors, then are informative for prediction. Similar to the analysis on the firm size of the lead firm, we further test how the linkages formed with high and low liquidity lead firms carry the cross-sectional predictability differently.
The \emph{high lead attention} $\left[\omega_{ij,t-30:t}^{HighLead}\right]$ and \emph{low lead attention} $\left[\omega_{ij,t-30:t}^{LowLead}\right]$ defined in Section \ref{sec:network_degree} are investigated.

Panel D of Table \ref{tab:linkages_importance_test_ew_dropzeroatt} reports the stocks sorting results, where one can see that the portfolio test result is mixed. On the one hand, as shown in Panel D(2), one can see the long-short strategy ("\emph{5-1}") by sorting $\mathcal{A}(\omega_{ij,t-30:t}^{LowLead})$ generates significant positive alphas with respect to FF3 ($85$ bps ($t$-statistics=$4.90$)) and FF5 ($81$ bps ($t$-statistics=$4.99$)). And the long-short strategy ("\emph{5-1}") by sorting $\mathcal{A}(\omega_{ij,t-30:t}^{HighLead})$ reported in Panel D(1) shows insignificant negative alphas with respect to FF3 and FF5, implying that news linkages of low stock liquidity lead firms carry higher predictability of monthly stock returns. On the other hand, neither of those two attention variants can sort portfolios with performance monotonic increase.

\subsection{Robust Check on Network Degree Predictability}

The portfolio test in Section \ref{sec:portfolio_sort_nd} showcases how the network attention proxy, \emph{network degree} can bring predictability on the cross-sectional stock returns. However, it is still a question if the predictability comes from some of the known characteristics that are also captured by the news \emph{network degree}. As reported in Columns (3)-(5) of Table \ref{tab:portfolio_comp_attention_ew_dropzeroatt}, where all the three characteristics, \%MV, B/M, and Liquidity, change monotonically, implying a strong correlation between those three characteristics and the portfolio ranking. For example, quintile 5 contains more large firms than quintile 1 and has higher liquidity. Hence, it is imperative to test the robustness of \emph{network degree} predictability to size, value, liquidity, momentum, and window size choice. We first employ the double-sorting method described in Section \ref{sec:double_sorting_method} to relieve the results from the first four possible effects mentioned above. After that, we check the robustness to the network formation window size. Note that the financial statement data, including the market share and book equity $B$, are two quarters lagged values when they are used to compute the sorting characteristics.

\begin{table}[!htb]
  \small
  \setlength{\tabcolsep}{0pt}
  \begin{threeparttable}
    \caption{Robustness Check on Size, Value, Liquidity, and Momentum}\label{tab:robust_check_maintable}
    \begin{tabular*}{
      \linewidth}{@{\extracolsep{\fill}}>{\itshape}
      l*{8}{S[table-format=2.3, table-number-alignment = center]}
      } 
  \toprule
  \toprule
  \multicolumn{1}{c}{Rank}  
  &\multicolumn{1}{c}{\emph{1}}
  &\multicolumn{1}{c}{\emph{2}} 
  &\multicolumn{1}{c}{\emph{3}} 
  &\multicolumn{1}{c}{\emph{4}} 
  &\multicolumn{1}{c}{\emph{5}} 
  &\multicolumn{3}{c}{\emph{5-1}}  
  \\
  \cmidrule{7-9}
  \multicolumn{1}{c}{}
  &\multicolumn{6}{c}{} 
  &\multicolumn{1}{c}{$\alpha$-FF3}
  &\multicolumn{1}{c}{$\alpha$-FF5}\\
  
  \midrule
  {}  & \multicolumn{1}{c}{(1)}  & \multicolumn{1}{c}{(2)}  & \multicolumn{1}{c}{(3)}   
  & \multicolumn{1}{c}{(4)}   & \multicolumn{1}{c}{(5)}  & \multicolumn{1}{c}{(6)}     
  & \multicolumn{1}{c}{(7)}   & \multicolumn{1}{c}{(8)}  \\

    \multicolumn{9}{c}{\textsc{Panel A: Controlling for Size}}\\[0.15cm]

\multicolumn{1}{c}{Mean} & 0.42 & 0.52 & 0.65 & 0.67 & 0.76 & 0.34 & 0.28 & 0.30\\[0.15cm]
\multicolumn{1}{c}{SR}   & 0.19 & 0.23 & 0.32 & 0.32 & 0.36 & 0.62 & (1.84) & (2.07)\\[0.15cm]
\midrule

\multicolumn{9}{c}{\textsc{Panel B: Controlling for Value}}\\[0.15cm]

\multicolumn{1}{c}{Mean} & 0.49 & 0.54  & 0.39  & 0.62  & 0.93 &  0.44  &  0.55 & 0.54  \\[0.15cm]
\multicolumn{1}{c}{SR}   & 0.21 & 0.25  & 0.16  & 0.28  & 0.52 &  0.73  &  (3.68) & (3.89)  \\[0.15cm]
\midrule

   \multicolumn{9}{c}{\textsc{Panel C: Controlling for Liquidity}}\\[0.15cm]

\multicolumn{1}{c}{Mean} & 0.51 & 0.44  & 0.67  & 0.52  & 0.84 &  0.33  &  0.36 & 0.35  \\[0.15cm]
\multicolumn{1}{c}{SR}   & 0.22 & 0.18  & 0.32  & 0.23  & 0.46 &  0.51  &  (1.88) & (1.86)  \\[0.15cm]
\midrule

   \multicolumn{9}{c}{\textsc{Panel D: Controlling for Momentum}}\\[0.15cm]

\multicolumn{1}{c}{Mean} & 0.45 & 0.54  & 0.54  & 0.56  & 0.94 &  0.49  &  0.48 & 0.47  \\[0.15cm]
\multicolumn{1}{c}{SR}   & 0.20 & 0.25  & 0.24  & 0.25  & 0.52 &  0.88  &  (2.96) & (3.00)  \\[0.15cm]

    \bottomrule
    \bottomrule
  
    \end{tabular*}
  \begin{tablenotes}[flushleft]
   \setlength\labelsep{0pt} 
  \linespread{1}\small
  \item 
The table reports the portfolio performance of sorting the network attention proxy, \emph{network degree} $\mathcal{A}(\omega_{ij,t-30:t})$, after controlling for the firm size (Panel A), firm value (Panel B), firm liquidity (Panel C), and momentum effect (Panel D). The double-sorting method is employed to control for those characteristics. Columns (1)-(5) in each panel are the average simple return and Sharpe-ratio of each quintile portfolio. Column (6) in each panel is the average simple return and Sharp-ratio of the long-short portfolio, i.e., the differences in returns between quintile 5 and quintile 1. Columns (7) and (8) in each panel are the alphas and robust Newey–West $t$-statistics with respect to the Fama-French 3-factor and 5-factor models, respectively.
\end{tablenotes}
  \end{threeparttable}
\end{table}

\subsubsection{Controlling for Firm Size}

News medias pay more attention on the firms with larger market capitalization, suggesting that \emph{network degree} is correlated with the firm market value at a certain level even though \emph{network degree} exclude most of the repetitive reports by counting the number of linkages.
In the sample period, within the S\&P500 constituents, the larger firms outperform the smaller ones, so it is reasonable to question that the predictability of news \emph{network degree} is related to the firm size. We relieve the size effect by employing the double-sorting method on firm market capital.

As reported in Panel A of Table \ref{tab:robust_check_maintable}, after controlling for firm size, average return of the long-short strategy is reduced from $50$ bps in Table \ref{tab:portfolio_comp_attention_ew_dropzeroatt} to $34$ bps, nevertheless \emph{network degree} still provides relatively strong predictability from two aspects.
First, the long-short strategy \emph{5-1} generates positive alphas with respect to FF3 ($28$ bps) and FF5 ($30$ bps), although alpha with respect to FF3 is only significant at 10\% level ($t$-statistics=$1.84$), but significant in FF5 at 5\% level ($t$-statistics=$2.07$) as reported in the last two columns.
The second aspect is that there is a monotonic increase pattern of portfolio performance in terms of average return and Sharpe-ratio as shown in the first five columns of Table \ref{tab:robust_check_maintable}.

\subsubsection{Controlling for Firm Value}

Much of the literature documents that small growth firms outperform those large firms. However, this phenomenon is not observed by sorting \emph{network degree} in our sample period as reported in Column (4) of Table \ref{tab:portfolio_comp_attention_ew_dropzeroatt}, where quintile 5 with the best performance has the lowest average book-to-market value.
Hence, a double-sorting is employed on book-to-value to test the robustness to firm value.

The firm value controlled portfolio result is reported in Panel B of Table \ref{tab:robust_check_maintable}, where average return of the long-short strategy is reduced to $44$ bps compared with the non-controlling result of $50$ bps. But the \emph{network degree} sorted long-short portfolio remains positive alphas of $55$ bps and $54$ bps with respect to FF3 and FF5 at 1\% level (with $t$-statistics=$3.68$ and $3.89$) as reported in the last two columns in Panel B of Table \ref{tab:portfolio_comp_attention_ew_dropzeroatt}.

\subsubsection{Controlling for Liquidity}

As shown in Column (5) of Table \ref{tab:portfolio_comp_attention_ew_dropzeroatt}, the liquidity proxy, volume/share, increases along with the portfolio rank. Such a pattern is consistent with the finding in \cite{Gervais_2001} that stocks tend to outperform when they experienced higher trading volume in the preceding month.

We control the liquidity effect with the double-sorting method applied to the turnover rate. As reported in Panel C of Table \ref{tab:robust_check_maintable}, which suggests that \emph{network degree} does capture quite amount of liquidity effect. An average return of the long-short strategy decreases to $33$ bps from $50$ bps in a non-controlled case. The long-short strategy still generates positive alphas with respect to FF3 ($36$ bps) and FF5 ($35$ bps), but only significant at 10\% level with $t$-statistics=$1.88$ and $1.86$.
Although abnormal returns from the long-short strategy are less significant compared with cases controlling other effects, one can still conclude that \emph{network degree} provides extra information in the cross-sectional return predictability.

\subsubsection{Controlling for Momentum}

It was well documented that the stocks that experienced higher returns in the past will tend to have higher returns, which is called momentum effect (\cite{JEGADEESH_1993}). One would suspect that \emph{network degree} correlates with the last monthly stock performance, in other words, the predictability of \emph{network degree} can be explained by momentum effect. Therefore, we conduct the robustness check by applying the double-sorting method on monthly return in the preceding month.

The long-short strategy produces an average return of $49$ bps, which is close to the average return of $50$ bps in the non-controlled case. 
The FF-3 and FF-5 alphas remain positive and significant at 1\% level with $t$-statistics=$2.96$ and $3.00$. Also, portfolio performance increases monotonically from quintile 1 to 5 by sorting \emph{network degree}. It's evidenced that the momentum effect can not account for the cross-sectional stock return predictability from \emph{network degree}.

\subsubsection{Robustness for Network Formation Window Size}

Note that the choice of network formation window size is not an outcome of pretests. We choose a 1-month window size to form the news network simply because the portfolio is rebalanced monthly, which is a convention in much of the literature that tests the cross-sectional predictability. The result is quite robust when we extend the 1-month window size 1-week and 2-week more. The long-short strategy with 1-more week window size still generates significant alphas of $47$ bps and $45$ bps relative to FF3 and FF5 with $t$-statistics=$2.39$ and $2.37$, compared with alphas of $48$ ($t$-statistics=$2.50$) bps and $47$ $t$-statistics=$2.51$ bps in 1-month window size case reported in Table \ref{tab:portfolio_comp_attention_ew_dropzeroatt}. As expected, with longer window size, the cross-sectional stock return predictability of \emph{network degree} decay. When the network formation window size is extended with 2 more weeks, the long-short strategy produces $36$ bps ($t$-statistics=$1.93$) with respect to FF3 and $35$ bps ($t$-statistics=$1.86$) with respect to FF5. Overall, the result is robust to the selection of network formation window size.

  %!TEX root = ../main.tex

\section{Conclusion}

This paper studies the news-based interdependency network between firms joint with cross-sectional returns and their consequences. By postulating the structure of firm-specific news, we define the lead-follower linkage embedded in the news. After that, we propose an algorithm to tackle the type-I error in identifying firm tickers from the news text so that we can construct the directed networks of S\&P500 firms from news articles.

By investigating the stock returns of lead firms, \emph{lead return}, and follower firms, we find that firms linked by news articles, regardless of the business relationship between lead and follower firms, tend to have strong comovement effects on daily returns, after controlling for factors including idiosyncratic alphas, market premium, size factor, profitability factor, etc. Moreover, the econometric model reveals that \emph{lead return} has a reversal effect on 1-day ahead stock return. However, in the cross-sectional portfolio test, the reversal effect is not strong enough to dominate the effects caused by other factors. Therefore, \emph{lead return} does not carry significant cross-sectional predictability.

We find that the monthly aggregated \emph{network degree}, capturing the attention of the news network, is a good predictor for the cross-sectional stock returns. A monthly rebalancing portfolio test reveals that \emph{network degree} provides significant positive alphas relative to the Fama-French 3-factor and 5-factor models. Among the different types of linkages, we show that \emph{network degree} of (1) both lead and follower networks, (2) within-sector network, (4) large size lead firm network, (5) and lower liquidity lead firm network, generate more significant positive alphas. Finally, we show the robustness of the cross-sectional predictability of \emph{network degree} against some of the known factors and window size of network formation. 

Our results provide two insights for portfolio management. First, investors can optimize their portfolios by accounting for the comovement effects between their assets. Second, the portfolio test on \emph{network degree} can be easily transformed into trading strategies, either a smart-beta index based on the high \emph{network degree} portfolios or an alpha investment strategy based on the long-short portfolios.
  
  % \input{sections/Comovement}
  % Bibliography.
  \clearpage

  \bibliographystyle{econometrica}
  \bibliography{nn_ref}

  \clearpage
    
%!TEX root = ../main.tex

\appendix

\section{}\label{app:algos}
%!TEX root = ../main.tex

\begin{algorithm}
    \SetKwFunction{sone}{S1}
    \SetKwFunction{stwo}{S2}
    \SetKwFunction{sthree}{S3}
    \SetKwFunction{clean}{cleanText}
    \SetKwFunction{pairsext}{pairsExtractor}
    \SetKwFunction{len}{length}
    \SetKwFunction{append}{append}
    \SetKwFunction{extend}{extend}
    \SetKwInOut{KwIn}{Input}
    \SetKwInOut{KwOut}{Output}
    \SetKwProg{Fn}{Function}{:}{End Function}
    \SetKwProg{parallelproc}{Parallel Processing}{}{End Parallel}

    \KwIn{A list of articles $[article_i]$, $i=0, 1, 2, \cdots, N$-1\;}
    \KwOut{A list of lead-follower pairs $resList = [pair_j]$, $i=0, 1, 2, \cdots, M$-1\;}

    \Fn{\pairsext{$article_i$}}{
        \KwIn{A list of article entry $article_i=[titleStr, contentStr, timeStr]$}
        Initialization $pairsList=[\ ]$\;
        \tcc{One might need to clean the text e.g special characters.} 
        \tcc{Refer to the code online for more details.}
        \clean{titleStr};
        \clean{contentStr}\;

        $lead$ = \sone{$titleStr$}; \tcp{S1 Strategy}
        \If{\len{$lead$} == 0}{
            $lead$ = \stwo{$titleStr$}; \tcp{S2 Strategy}
            \If{\len{$lead$ == 0}}{
                $lead$ = \sthree{$titleStr$}; \tcp{S3 Strategy}
            }
        }

        \If{\len{$lead$}$>1$}{
            $lead=[\ ]$; \tcp{We only identify single-lead articles}
        }

        $followersList$ = \sone{$contentStr$} + \stwo{$contentStr$} \tcp{S1, S2 Strategy}
        % \If{\len{$followersList$} == 0}{
        % $followersList$ = \stwo{$contentStr$} \tcp{S2 Strategy}
        % }
        
        \If{\len{$lead$}!=0 \& \len{$followersList$}!=0}{
            \For{$i=0$ \KwTo \len{$followersList$}}{
                $pairsList$.\append{$(lead, followersList[i], timeStr)$}
            }
        }

        \KwRet{$pairsList$}
    }

    \parallelproc{$article_i$, $i=0$ \KwTo $N$-1}{
        $resList.\extend(\pairsext(article_i))$\;
        }
    \KwRet{$newList$}

    \caption{Algorithm to Identify Firms }\label{algo:tickers_identi}
\end{algorithm}

\begin{algorithm}
    \SetKwFunction{len}{length}
    \SetKwFunction{intersect}{intersect}
    % \SetKwInput{Input}{Input}
    % \SetKwInput{Output}{Output}
    \SetKwInOut{KwIn}{Input}
    \SetKwInOut{KwOut}{Output}

    \KwIn{A list of full names of all the firms $[name_{i}]$, $i=0, 1, 2, \cdots, n$-1\;
            A list of suffix and redundant strings $[s_{m}]$, $m=0, 1, 2, \cdots, M$-1\;
            Threshold values $\eta_1$ and $\eta_2$}
    \KwOut{A list of lists of the segments for each firm: $segmentsList = [[segList_{i}]]$, $i=0, 1, 2, \cdots, n$-1}

    Initialization $segmentsList = [\ ]$\;

    \tcp{Construct the n-gram name segments for each firm}
    \For{$i=0$ \KwTo $n$-1}{
        Remove special characters from $name_i$\;
        \tcp{split $name_i$ into $K_i$-1 segments by space}
        $segList_i=[seg_k]$, $k=0,1,2\cdots, K_i$-1 \;
        \For{$k=1$ \KwTo $K_i$-1}{
            \tcp{replace each $seg_k$ with n-gram string}
            $seg_k \leftarrow segList_i[0, k].toString$\;
        }
        $segmentsList.append(segList_i)$
    }

    \tcp{Remove the common segments for different firms}
    \For{$i=0$ \KwTo $n-2$}{
        \For{$j=i$ \KwTo $n$-1}{
            $interList_{i,j} = \intersect(segList_i, segList_j)$\;
            \For{$p=0$ \KwTo $\len(interList_{i,j})$-1}{
                $segList_i \leftarrow segList_i.remove(interList[p])$\;
                $segList_j \leftarrow segList_j.remove(interList[p])$\;
            }
        }
    }

    \tcp{Further reduce the complexity}
    \For{$i=0$ \KwTo $n$-1}{
        \If{$\len(segList_i)>\eta_1$}{
            $segmentsList.remove(segList_i)$ \tcp*[f]{Remove the segments list of firm $i$}
        }
        \If{$\len({segList_i})>\eta_2$}{
            $segList_i.remove(seg_0)$ \tcp*[f]{Remove the first segment of firm $i$}
        }
    }

    \KwRet{$segmentsList$}
    \caption{Mapping Firm Names to n-gram Name Segments}\label{algo:fullname_seg_map}
\end{algorithm}
% \clearpage
\section{}\label{app:tables}

  \begin{table}[H]
    \small
    \setlength{\tabcolsep}{0pt}
    \begin{threeparttable}
      \caption{Infeasible Portfolios of Sorting Contemporaneous \emph{lead return} Sector Decomposition $\mathcal{LR}(\omega_{ij,t-365:t}^{w/c})$}\label{tab:portfolio_comparison_simul_sector}
      \begin{tabular*}{\linewidth}{@{\extracolsep{\fill}}>{\itshape}
        l*{5}{S[table-format=1.1, table-number-alignment = center]}*{2}{S[table-format=1.1, table-number-alignment = center]}l*{2}{S[table-format=1.1, table-number-alignment = center]}
        }       
    \toprule
    \toprule
    \multicolumn{1}{c}{}  
    &\multicolumn{5}{c}{}
    &\multicolumn{2}{c}{FF3}  
    &\multicolumn{1}{c}{}  
    &\multicolumn{2}{c}{FF5}
    \\
    \cmidrule{7-8}
    \cmidrule{10-11}
    \multicolumn{1}{c}{Rank}  
    &\multicolumn{1}{c}{Mean} 
    &\multicolumn{1}{c}{SR} 
    &\multicolumn{1}{c}{\%MV} 
    % &\multicolumn{1}{c}{Size} 
    &\multicolumn{1}{c}{B/M} 
    &\multicolumn{1}{c}{Liquidty}
  
    &\multicolumn{1}{c}{$\alpha$}  
    &\multicolumn{1}{c}{$R^2$}  
    &\multicolumn{1}{c}{}  
    &\multicolumn{1}{c}{$\alpha$}  
    &\multicolumn{1}{c}{$R^2$}
    \\
      \cline{1-2}
    \multicolumn{2}{l}{$\mathcal{LR}(\omega_{ij,t-365:t}^{w})$} \\
    1   & -90.35 & -9.72 & 18.60  & 44.21 & 1.03 & -92.98   & 15.26 & {} & -92.96   & 15.68 \\         %      & 23.83
    {}  & {}     & {}    & {}     & {}    & {}   & (-26.81) & {}    & {} & (-27.00) & {} \\[0.15cm]         %      & {}   
    2   & -27.87 & -3.76 & 20.54  & 40.55 & 0.86 & -30.4    & 17.69 & {} & -30.5    & 18.12 \\         %      & 23.92
    {}  & {}     & {}    & {}     & {}    & {}   & (-11.71) & {}    & {} & (-11.81) & {} \\[0.15cm]         %      & {}   
    3   & 3.73   & 0.46  & 20.65  & 39.19 & 0.84 & 1.24     & 18.47 & {} & 1.15     & 18.90 \\         %      & 23.94
    {}  & {}     & {}    & {}     & {}    & {}   & (0.50)   & {}    & {} & (0.47)   & {} \\[0.15cm]         %      & {}   
    4   & 33.86  & 4.44  & 20.69  & 39.70 & 0.86 & 31.36    & 18.12 & {} & 31.29    & 18.67 \\         %      & 23.94
    {}  & {}     & {}    & {}     & {}    & {}   & (12.10)  & {}    & {} & (12.18)  & {} \\[0.15cm]         %      & {}   
    5   & 90.38  & 10.30 & 19.51  & 43.10 & 0.99 & 88.14    & 16.56 & {} & 88.03    & 17.23 \\         %      & 23.86
    {}  & {}     & {}    & {}     & {}    & {}   & (26.37)  & {}    & {} & (26.62)  & {} \\[0.15cm]         %      & {}   
    % 5-1 & 180.74 & 25.47 & {}     & {}    & {}   & 181.13   & 0.28  & {} & 180.99   & 0.90 \\         %      & {}   
    % {}  & {}     & {}    & {}     & {}    & {}   & (46.80)  & {}    & {} & (47.09)  & {} \\         %      & {}   
  
    \cline{1-2}
    \multicolumn{2}{l}{$\mathcal{LR}(\omega_{ij,t-365:t}^{c})$} \\
    1   & -0.18 & -0.08 & 14.47  & 42.69 & 0.92 & -2.65   & 18.12 & {} & -2.75   & 18.67 \\                  % & 23.70
    {}  & {}    & {}    & {}     & {}    & {}   & (-1.00) & {}    & {} & (-1.04) & {} \\[0.15cm]                  % & {}   
    2   & 0.98  & 0.07  & 21.68  & 41.01 & 0.93 & -1.64   & 20.29 & {} & -1.72   & 20.79 \\                  % & 23.97
    {}  & {}    & {}    & {}     & {}    & {}   & (-0.63) & {}    & {} & (-0.67) & {} \\[0.15cm]                  % & {}   
    3   & 3.10  & 0.35  & 25.26  & 40.71 & 0.92 & 0.64    & 18.06 & {} & 0.59    & 18.50 \\                  % & 24.07
    {}  & {}    & {}    & {}     & {}    & {}   & (0.25)  & {}    & {} & (0.23)  & {} \\[0.15cm]                  % & {}   
    4   & 3.89  & 0.44  & 22.85  & 40.45 & 0.92 & 1.43    & 17.63 & {} & 1.38    & 18.09 \\                  % & 24.01
    {}  & {}    & {}    & {}     & {}    & {}   & (0.54)  & {}    & {} & (0.53)  & {} \\[0.15cm]                  % & {}   
    5   & 5.70  & 0.69  & 15.74  & 41.81 & 0.92 & 3.33    & 18.41 & {} & 3.27    & 18.94 \\                  % & 23.75
    {}  & {}    & {}    & {}     & {}    & {}   & (1.28)  & {}    & {} & (1.27)  & {} \\[0.15cm]                  % & {}   
    % 5-1 & 5.88  & 2.54  & {}     & {}    & {}   & 5.98    & 0.21  & {} & 6.02    & 0.50 \\                  % & {}   
    % {}  & {}    & {}    & {}     & {}    & {}   & (5.79)  & {}    & {} & (5.86)  & {} \\                  % & {}   

      \bottomrule
      \bottomrule
              
      \end{tabular*}
    \begin{tablenotes}[flushleft]
     \setlength\labelsep{0pt} 
    \linespread{1}\small
    \item     
    The upper panel reports the performance of the equal-weighted portfolios sorted by within-sector \emph{lead return} $\mathcal{LR}(\omega_{ij,t-365:t}^{w})$, and the bottom panel contains the performance of the equal-weighted portfolios sorted by cross-sector \emph{lead return} $\mathcal{LR}(\omega_{ij,t-365:t}^{c})$.
    Both news networks are formed with a 1-year rolling window. All the portfolios are infeasible.
    Column (1)-(5) are average simple daily return in bps; average annualized sharpe-ratio, average market value proportion in percentage, average book-to-market ratio in percentage, average volume-to-share ratio in percentage. 
    The last 4 columns report the alpha, $R^2$, and robust Newy-West $t$-statistics in parentheses against Fama-French 3-factor and Fama-French 5-factor, respectively.  
\end{tablenotes}
    \end{threeparttable}
  \end{table}

\begin{table}[H]
  \small
  \setlength{\tabcolsep}{0pt}
  \begin{threeparttable}
    \caption{Infeasible Portfolios of Sorting Contemporaneous \emph{lead return} Signs of Return Decomposition $\mathcal{LR}^{+/-}(\omega_{ij,t-365:t})$}\label{tab:portfolio_comparison_simul_sign}
    \begin{tabular*}{\linewidth}{@{\extracolsep{\fill}}>{\itshape}
      l*{5}{S[table-format=1.1, table-number-alignment = center]}*{2}{S[table-format=1.1, table-number-alignment = center]}l*{2}{S[table-format=1.1, table-number-alignment = center]}
      } 
  \toprule
  \toprule
  \multicolumn{1}{c}{}  
  &\multicolumn{5}{c}{}
  &\multicolumn{2}{c}{FF3}  
  &\multicolumn{1}{c}{}  
  &\multicolumn{2}{c}{FF5}
  \\
  \cmidrule{7-8}
  \cmidrule{10-11}
  \multicolumn{1}{c}{Rank}  
  &\multicolumn{1}{c}{Mean} 
  &\multicolumn{1}{c}{SR} 
  &\multicolumn{1}{c}{\%MV} 
  % &\multicolumn{1}{c}{Size} 
  &\multicolumn{1}{c}{B/M} 
  &\multicolumn{1}{c}{Liquidty}
  
  &\multicolumn{1}{c}{$\alpha$}  
  &\multicolumn{1}{c}{$R^2$}  
  &\multicolumn{1}{c}{}  
  &\multicolumn{1}{c}{$\alpha$}  
  &\multicolumn{1}{c}{$R^2$}
  \\
  \cline{1-2}
  
  \multicolumn{2}{l}{$\mathcal{LR}^{+}(\omega_{ij,t-365:t})$} \\
  1   & -54.11 & -7.63 & 12.85  & 43.72 & 0.91 & -56.18   & 16.13 & {} & -56.22   & 16.56 \\            %  & 23.62
  {}  & {}     & {}    & {}     & {}    & {}   & (-21.61) & {}    & {} & (-21.67) & {} \\            %  & {}   
  2   & -31.36 & -4.22 & 20.19  & 40.12 & 0.88 & -33.93   & 18.94 & {} & -33.95   & 19.38 \\            %  & 23.95
  {}  & {}     & {}    & {}     & {}    & {}   & (-13.25) & {}    & {} & (-13.35) & {} \\            %  & {}   
  3   & -5.13  & -0.72 & 24.18  & 39.64 & 0.87 & -7.80    & 18.43 & {} & -7.88    & 18.87 \\            %  & 24.06
  {}  & {}     & {}    & {}     & {}    & {}   & (-2.99)  & {}    & {} & (-3.04)  & {} \\            %  & {}   
  4   & 23.37  & 2.85  & 23.63  & 40.22 & 0.89 & 20.78    & 17.40 & {} & 20.69    & 17.88 \\            %  & 24.03
  {}  & {}     & {}    & {}     & {}    & {}   & (7.51)   & {}    & {} & (7.53)   & {} \\            %  & {}   
  5   & 77.31  & 8.40  & 19.15  & 42.99 & 1.03 & 74.81    & 17.32 & {} & 74.71    & 17.97 \\            %  & 23.85
  {}  & {}     & {}    & {}     & {}    & {}   & (21.99)  & {}    & {} & (22.23)  & {} \\            %  & {}   
  % 5-1 & 131.42 & 22.55 & {}     & {}    & {}   & 130.99   & 2.88  & {} & 130.93   & 3.45 \\            %  & {}   
  % {}  & {}     & {}    & {}     & {}    & {}   & (42.21)  & {}    & {} & (42.36)  & {} \\            %  & {}   
  
  \cline{1-2}
  \multicolumn{2}{l}{$\mathcal{LR}^{-}(\omega_{ij,t-365:t})$} \\
  1   & 55.67   & 7.87   & 12.72  & 42.79 & 0.90 & 53.73    & 16.82 & {} & 53.68    & 17.41 \\                   %   & 23.61
  {}  & {}      & {}     & {}     & {}    & {}   & (20.38)  & {}    & {} & (20.58)  & {} \\                   %   & {}   
  2   & 39.01   & 5.10   & 20.63  & 39.60 & 0.88 & 36.59    & 18.57 & {} & 36.50     & 19.20 \\                   %   & 23.96
  {}  & {}      & {}     & {}     & {}    & {}   & (13.80)  & {}    & {} & (13.91)  & {} \\                   %   & {}   
  3   & 13.04   & 1.64   & 24.52  & 39.50 & 0.87 & 10.57    & 18.05 & {} & 10.47    & 18.51 \\                   %   & 24.07
  {}  & {}      & {}     & {}     & {}    & {}   & (4.04)   & {}    & {} & (4.04)   & {} \\                   %   & {}   
  4   & -15.80   & -2.00  & 23.63  & 40.95 & 0.89 & -18.57   & 18.68 & {} & -18.68   & 19.16 \\                   %   & 24.03
  {}  & {}      & {}     & {}     & {}    & {}   & (-6.73)  & {}    & {} & (-6.82)  & {} \\                   %   & {}   
  5   & -74.98  & -8.14  & 18.50  & 43.88 & 1.04 & -77.75   & 15.91 & {} & -77.76   & 16.33 \\                   %   & 23.83
  {}  & {}      & {}     & {}     & {}    & {}   & (-23.37) & {}    & {} & (-23.52) & {} \\                   %   & {}   
  % 5-1 & -130.66 & -22.39 & {}     & {}    & {}   & -131.48  & 2.14  & {} & -131.44  & 2.56 \\                   %   & {}   
  % {}  & {}      & {}     & {}     & {}    & {}   & (-42.74) & {}    & {} & (-42.91) & {} \\                   %   & {}   

    \bottomrule
    \bottomrule
      \end{tabular*}
  \begin{tablenotes}[flushleft]
    \setlength\labelsep{0pt} 
  \linespread{1}\small
  \item 
  The upper panel reports the performance of the equal-weighted portfolios sorted by positive \emph{lead return} $\mathcal{LR}^{+}(\omega_{ij,t-365:t})$, and the bottom panel contains the performance of the equal-weighted portfolios sorted by negative \emph{lead return} $\mathcal{LR}^{-}(\omega_{ij,t-365:t})$.
  Both news networks are formed with a 1-year rolling window. All the portfolios are infeasible.
  Column (1)-(5) are average simple daily return in bps; average annualized sharpe-ratio, average market value proportion in percentage, average book-to-market ratio in percentage, average volume-to-share ratio in percentage. 
  The last 4 columns report the alpha, $R^2$, and robust Newy-West $t$-statistics in parentheses against Fama-French 3-factor and Fama-French 5-factor, respectively.  
\end{tablenotes}
  \end{threeparttable}
\end{table}

\begin{table}[H]
  \small
  \setlength{\tabcolsep}{0pt}
  \begin{threeparttable}
    \caption{Estimation of the Impacts from Sector and Sign of Returns Decomposed Network Price Proxies}\label{tab:pred1_panel_res_sign_sector}
    \begin{tabular*}{\linewidth}{@{\extracolsep{\fill}}>{\itshape}cl *{4}{S[table-format=-2.4,table-number-alignment = center]}} 
      \toprule\toprule
      
      \multicolumn{1}{c}{Depend. Var.} & {} & \multicolumn{1}{c}{$\mathcal{LR}^{+}(\mathcal{W}^{w}_{T})$} & \multicolumn{1}{c}{$\mathcal{LR}^{+}(\mathcal{W}^{c}_{T})$}  
      & \multicolumn{1}{c}{$\mathcal{LR}^{-}(\mathcal{W}^{w}_{T})$} & \multicolumn{1}{c}{$\mathcal{LR}^{-}(\mathcal{W}^{c}_{T})$} \\[0.15cm]
      \cline{1-2}
      \multirow{2}{*}{$r_{t+1}$} & {} & 0.031   & -0.005   & 0.007   & 0.015\\
      {}                         & {} & (4.701) & (-0.968) & (1.155) & (2.339)\\
      
      \bottomrule\bottomrule
        \end{tabular*}
  \begin{tablenotes}[flushleft]
   \setlength\labelsep{0pt} 
  \linespread{1}\small
  \item A 1-year time window for news network $T=[t-365:t]$.
  Estimation the impacts of \emph{lead return} decomped by sector and sign of return on 1-day ahead stock returns.
  Coefficients of \emph{lead return} variables are estimated by fixed effect panel model, and $t$-statistics reported in parentheses with the standard errors corrected by clustering on individual and time.
  Control variables include market value $\log(MV)$, book-market ratio $B/M$, and turnover rate $volume/share$ and 1-day lagged return $r_t$.
  All the financial statement related variables are 2 quarters lagged.
  \end{tablenotes}
  \end{threeparttable}
\end{table}

\begin{table}[H]
  \small
  \setlength{\tabcolsep}{0pt}
  \begin{threeparttable}
    \caption{Portfolios Sorted by the Preceding \emph{Network Degree} $\mathcal{A}$ \\\normalsize(Equal-Weighted, Keep $\mathcal{A}_{i,t}=0$)}\label{tab:portfolio_comp_attention_ew_keepzeroatt}
    \begin{tabular*}{\linewidth}{@{\extracolsep{\fill}}>{\itshape}
      l
      *{2}{S[table-format=1.2, table-number-alignment = center]}
      *{2}{S[table-format=2.2, table-number-alignment = center]}
      *{1}{S[table-format=1.2, table-number-alignment = center]}
      *{2}{S[table-format=-1.2, table-number-alignment = center]}
      l
      *{2}{S[table-format=-1.2, table-number-alignment = center]}
      } 
        \toprule
        \toprule
        \multicolumn{1}{c}{}  
        &\multicolumn{5}{c}{}
        &\multicolumn{2}{c}{FF3}  
        &\multicolumn{1}{c}{}  
        &\multicolumn{2}{c}{FF5}
        \\
        \cmidrule{7-8}
        \cmidrule{10-11}
        \multicolumn{1}{c}{Rank}  
        &\multicolumn{1}{c}{Mean} 
        &\multicolumn{1}{c}{SR} 
        &\multicolumn{1}{c}{\%MV} 
        % &\multicolumn{1}{c}{Size} 
        &\multicolumn{1}{c}{B/M} 
        &\multicolumn{1}{c}{Liquidty}
        &\multicolumn{1}{c}{$\alpha$}  
        &\multicolumn{1}{c}{$R^2$}  
        &\multicolumn{1}{c}{}  
        &\multicolumn{1}{c}{$\alpha$}  
        &\multicolumn{1}{c}{$R^2$}
        \\
        % \cline{1-2}
        \midrule
      % \multicolumn{2}{l}{Equal-Weighted} \\ 
      % {}  & {}   & {}   & {}    & {}    & {}   & {}      & {}    & {} & {}      & {} \\ %          & {}
      {}  & \multicolumn{1}{c}{(1)}  & \multicolumn{1}{c}{(2)}  & \multicolumn{1}{c}{(3)}   
      & \multicolumn{1}{c}{(4)}   & \multicolumn{1}{c}{(5)}  & \multicolumn{1}{c}{(6)}     
      & \multicolumn{1}{c}{(7)}   & {} & \multicolumn{1}{c}{(8)}     & \multicolumn{1}{c}{(9)} \\[0.15cm] % & {}
      1   & 0.48 & 0.21 & 8.66  & 44.55 & 0.89 & -0.77   & 91.52 & {} & -0.73   & 92.33 \\ %       & 23.37
      {}  & {}   & {}   & {}    & {}    & {}   & (-2.81) & {}    & {} & (-2.77) & {} \\[0.15cm] %  & {}
      2   & 0.58 & 0.27 & 10.12 & 42.57 & 0.88 & -0.78   & 91.71 & {} & -0.75   & 92.33 \\ %       & 23.54
      {}  & {}   & {}   & {}    & {}    & {}   & (-3.02) & {}    & {} & (-2.97) & {} \\[0.15cm] %  & {}
      3   & 0.64 & 0.31 & 12.85 & 42.40 & 0.90 & -0.74   & 92.25 & {} & -0.71   & 92.71 \\ %       & 23.71
      {}  & {}   & {}   & {}    & {}    & {}   & (-3.26) & {}    & {} & (-3.29) & {} \\[0.15cm] %  & {}
      4   & 0.51 & 0.22 & 18.97 & 39.84 & 0.91 & -0.80   & 88.40 & {} & -0.74   & 89.74 \\ %       & 24.02
      {}  & {}   & {}   & {}    & {}    & {}   & (-2.91) & {}    & {} & (-2.80) & {} \\[0.15cm] %  & {}
      5   & 0.98 & 0.57 & 49.40 & 36.56 & 1.00 & -0.37   & 92.12 & {} & -0.35   & 92.99 \\ %       & 24.82
      {}  & {}   & {}   & {}    & {}    & {}   & (-1.93) & {}    & {} & (-1.78) & {} \\[0.15cm] %  & {}
      5-1 & 0.51 & 0.82 & {}    & {}    & {}   & 0.40    & 43.07 & {} & 0.38    & 43.68 \\ %       & {}
      {}  & {}   & {}   & {}    & {}    & {}   & (2.07)  & {}    & {} & (1.97)  & {} \\[0.15cm] %  & {}
      Mkt & 0.64 & 0.31 & {}    & {}    & {}   & -0.69   & 92.56 & {} & -0.65   & 93.34 \\ %       & {}
      {}  & {}   & {}   & {}    & {}    & {}   & (-3.03) & {}    & {} & (-2.97) & {} \\[0.15cm] %  & {}
      \bottomrule
        \bottomrule
    \end{tabular*}
  \begin{tablenotes}[flushleft]
   \setlength\labelsep{0pt} 
  \linespread{1}\small
  \item 
  The table reports the performance of the equal-weighted portfolios sorted by \emph{network degree}, $\mathcal{A}(\omega_{ij,t-30:t})$, computed with news linkages from the preceding month.
  Firms with zero network degree are kept in the portfolios.
  Columns (1)-(5) are average simple daily return in percentage; the average annualized sharpe-ratio, average market value proportion in percentage, average book-to-market ratio in percentage, average volume-to-share ratio in percentage. 
  The last 4 columns report the alpha, $R^2$, and robust Newy-West $t$-statistics in parentheses with respect to the Fama-French 3-factor and the Fama-French 5-factor models, respectively.
\end{tablenotes}
  \end{threeparttable}
\end{table}

\begin{table}[H]
  \small
  \setlength{\tabcolsep}{0pt}
  \begin{threeparttable}
    \caption{Portfolios Sorted by the Preceding \emph{Network Degree} $\mathcal{A}$ \\\normalsize(Value-Weighted, Drop $\mathcal{A}_{i,t}=0$)}\label{tab:portfolio_comp_attention_vw_dropzeroatt}
    \begin{tabular*}{\linewidth}{@{\extracolsep{\fill}}>{\itshape}
      l
      *{2}{S[table-format=1.2, table-number-alignment = center]}
      *{2}{S[table-format=2.2, table-number-alignment = center]}
      *{1}{S[table-format=1.2, table-number-alignment = center]}
      *{2}{S[table-format=-1.2, table-number-alignment = center]}
      l
      *{2}{S[table-format=-1.2, table-number-alignment = center]}} 
        \toprule
        \toprule
        \multicolumn{1}{c}{}  
        &\multicolumn{5}{c}{}
        &\multicolumn{2}{c}{FF3}  
        &\multicolumn{1}{c}{}  
        &\multicolumn{2}{c}{FF5}
        \\
        \cmidrule{7-8}
        \cmidrule{10-11}
        \multicolumn{1}{c}{Rank}  
        &\multicolumn{1}{c}{Mean} 
        &\multicolumn{1}{c}{SR} 
        &\multicolumn{1}{c}{\%MV} 
        % &\multicolumn{1}{c}{Size} 
        &\multicolumn{1}{c}{B/M} 
        &\multicolumn{1}{c}{Liquidty}
        &\multicolumn{1}{c}{$\alpha$}  
        &\multicolumn{1}{c}{$R^2$}  
        &\multicolumn{1}{c}{}  
        &\multicolumn{1}{c}{$\alpha$}  
        &\multicolumn{1}{c}{$R^2$}
        \\
      \midrule
      {}  & \multicolumn{1}{c}{(1)}  & \multicolumn{1}{c}{(2)}  & \multicolumn{1}{c}{(3)}   
      & \multicolumn{1}{c}{(4)}   & \multicolumn{1}{c}{(5)}  & \multicolumn{1}{c}{(6)}     
      & \multicolumn{1}{c}{(7)}   & {} & \multicolumn{1}{c}{(8)}     & \multicolumn{1}{c}{(9)} \\[0.15cm] % & {}

      1                                     & 0.47 & 0.22 & 6.76  & 36.63 & 0.73 & -0.84   & 90.38 & {} & -0.82   & 90.8\\
      {}                                    & {}   & {}   & {}    & {}    & {}   & (-3.42) & {}    & {} & (-3.41) & {} \\[0.15cm]
      2                                     & 0.57 & 0.31 & 8.87  & 34.76 & 0.70 & -0.72   & 92.99 & {} & -0.7    & 93.27\\
      {}                                    & {}   & {}   & {}    & {}    & {}   & (-4.09) & {}    & {} & (-4.11) & {} \\[0.15cm]
      3                                     & 0.65 & 0.38 & 13.05 & 34.16 & 0.67 & -0.49   & 88.71 & {} & -0.48   & 88.79\\
      {}                                    & {}   & {}   & {}    & {}    & {}   & (-2.63) & {}    & {} & (-2.65) & {} \\[0.15cm]
      4                                     & 0.56 & 0.31 & 19.02 & 30.28 & 0.63 & -0.72   & 88.13 & {} & -0.7    & 88.47\\
      {}                                    & {}   & {}   & {}    & {}    & {}   & (-3.2 ) & {}    & {} & (-3.06) & {} \\[0.15cm]
      5                                     & 1.11 & 0.72 & 52.31 & 32.47 & 0.65 & -0.26   & 95.39 & {} & -0.25   & 95.64\\
      {}                                    & {}   & {}   & {}    & {}    & {}   & (-2.25) & {}    & {} & (-2.19) & {} \\[0.15cm]
      5-1                                   & 0.64 & 0.95 & {}    & {}    & {}   & 0.58    & 26.43 & {} & 0.56    & 27.46\\
      {}                                    & {}   & {}   & {}    & {}    & {}   & (2.53 ) & {}    & {} & (2.54 ) & {} \\[0.15cm]
      Mkt                                   & 0.85 & 0.52 & {}    & {}    & {}   & -0.53   & 95.19 & {} & -0.52   & 95.47 \\ %      & {}
      {}                                    & {}   & {}   & {}    & {}    & {}   & (-4.25) & {}    & {} & (-4.08) & {} \\[0.15cm] % & {}
      
      \bottomrule
        \bottomrule
    \end{tabular*}
  \begin{tablenotes}[flushleft]
   \setlength\labelsep{0pt} 
  \linespread{1}\small
  \item 
  The table reports the performance of the value-weighted portfolios sorted by \emph{network degree}, $\mathcal{A}(\omega_{ij,t-30:t})$, computed with news linkages from the preceding month.
  Firms with zero network degree are removed before forming the portfolios.
  Columns (1)-(5) are average simple daily return in percentage; the average annualized sharpe-ratio, average market value proportion in percentage, average book-to-market ratio in percentage, average volume-to-share ratio in percentage. 
  The last 4 columns report the alpha, $R^2$, and robust Newy-West $t$-statistics in parentheses with respect to the Fama-French 3-factor and the Fama-French 5-factor models, respectively.
\end{tablenotes}
  \end{threeparttable}
\end{table}

\section{}\label{app:figures}
%!TEX root = ../main.tex

\begin{figure}[H]
  \centering
  \begin{minipage}{1\linewidth}
    \noindent\fbox{%
    \parbox{\textwidth}{
        \begin{itemize}
            \tiny{
                \item Title: "3 Catalysts for Apple Stock in 2021"
                \item Timestamp: "2020-12-30 12:00:42"
                \item Content: "The NASDAQ-100 Technology Sector index has more than doubled the return of the S\&P 500 index over both the last one- and five-year periods, but Apple (NASDAQ: AAPL) has remained among the cream of the crop, surging roughly 84\% so far in 2020 and about 405\% over the last five years. Those heady gains put Apple's stock price at a premium valuation of 34 times forward earnings estimates, which looks expensive compared with a forward price-to-earnings multiple of 24 for the S\&P 500.
                However, Apple is currently entering one of its strongest product cycles in years, which could keep the business humming along and justify the stock's premium. Here are three growth catalysts to watch in 2021.
                Apple enters the new year after delivering a strong earnings report for the fiscal fourth quarter. While iPhone revenue was slightly down in fiscal 2020, sales of non-iPhone products grew 30\% year over year despite supply constraints on iPad, Mac, and Apple Watch.
                That momentum should continue into the first half of calendar 2021 with recent product releases. The new AirPods Max headphones have been sold out since their initial release in December. Other recent launches should provide sales momentum in the short term, including the HomePod mini, Apple Watch Series 6 and Watch SE, two new iPad models, and new MacBooks featuring Apple's internally developed M1 processor. Early signs are pointing to strong sales of iPhone 12. Even though 5G network coverage is spotty, customers seem to be scooping up the new iPhone at a record clip, which is great news since the iPhone generates half of Apple's total revenue. Apple is planning to increase its production of the iPhone 12 by 30\% in the first half of 2021, according to a report from Nikkei Asia that cited a source from a key Apple supplier. This would put Apple on pace to have its biggest year for iPhone sales since the record-breaking shipments of the iPhone 6 in 2015. Apple made a bold move earlier this year by announcing a two-year transition to use internally developed processors for its line of Macs, dropping Intel chips in the process. The M1 chip offers several features that will significantly enhance the user experience on Mac and could lead to further market share gains in the PC market for Apple.  For years, Apple has been working to bridge the user experience across its operating system for Mac and iOS. The M1 chip will take this a step further by allowing Mac users to run iPhone and iPad apps. This could be huge for Mac sales over the long term. By taking control over the development of its own chips, Apple can better plan its product road map and tailor future versions of the M1 for specific user experiences, such as enhanced image processing, security, and other cutting-edge features and technologies. Apple will likely unveil more benefits of the M1 chip over time, but for 2021, the significant boost to battery life and ability to run iOS apps directly on Mac should be enough to encourage more sales of MacBooks. Despite double-digit percentage growth from subscription services, sales of hardware products still make up nearly 80\% of Apple's total revenue. But with services growing 16\% year over year in the last quarter, this \$53 billion annual business could reach \$100 billion in the next five years.  After reaching an installed base of 1.5 billion active devices earlier this year, Apple reported that its installed base hit another record high in the fiscal fourth quarter. New services -- including Apple TV+, Arcade, News+, and Apple Card -- are attracting more users, and Apple is still adding new services and content to drive further growth. Apple TV+ continues to add new streaming content, which will be crucial to persuade users coming off their free trials that it's worth paying the relatively low monthly fee of \$4.99. The recent launch of Apple Fitness+ is yet another service with a lot of potential. Interactive fitness was already a fast-growing market before 2020, but it got an extra kick during the pandemic with more people looking for alternative workout solutions at home. Nike and Peloton Interactive have reported high engagement levels with their respective training apps lately. Apple's massive installed base of users should win a decent share of this booming market. Apple is on the verge of a major upgrade cycle across all its products. Moreover, this upgrade cycle could intensify as COVID-19 vaccines become available, encouraging more people to visit Apple stores as the year progresses. With these growth catalysts on the horizon for this top tech stock, Apple is well-positioned to outperform in 2021. 10 stocks we like better than AppleWhen investing geniuses David and Tom Gardner have a stock tip, it can pay to listen. After all, the newsletter they have run for over a decade, Motley Fool Stock Advisor, has tripled the market.\* David and Tom just revealed what they believe are the ten best stocks for investors to buy right now... and Apple wasn't one of them! That's right -- they think these 10 stocks are even better buys. See the 10 stocks \*Stock Advisor returns as of November 20, 2020 John Ballard owns shares of Apple, Nike, and Peloton Interactive. The Motley Fool owns shares of and recommends Apple, Nike, and Peloton Interactive. The Motley Fool recommends Intel. The Motley Fool has a disclosure policy."
            }
        \end{itemize}
            }
}\\
  \end{minipage}
  \caption{
    \textbf{Sample News on Apple, Intel, Nike, Peloton}
  \newline 
  \small
The sample news that is used in the demonstration of the algorithm flowchart in Figure \ref{fig:algo_example}.
Apple is identified in the headline, while Intel and Nike are identified in the content. Peloton is not identified as it is not in the S\&P500 index.
}\label{fig:newsexample}
\end{figure}

\begin{figure}[H]
  \centering
  \begin{minipage}{1\linewidth}
    \noindent\fbox{%
    \parbox{\textwidth}{
        \begin{itemize}
            \tiny{\item Title: "Dow Jones Industrial Average Rises 9 Points Despite \textcolor{red}{Apple} Stock Slump"\\
                \item Content: "MoneyMorning.com Report - For Jan. 5, 2016, here's the top stock market news and stocks to watch based on today's market moves. Dow Jones\\
                ...\\
                The Dow received a boost from \textcolor{blue}{DuPont Co. (DD)}, which gained 1.9\%. The top stock news came from \textcolor{red}{Apple Inc. (Nasdaq: AAPL)}, which saw shares slump 2.5\% after the Nikkei Asian Review reported the company will slash its iPhone production by roughly 30\% this quarter\\
                ..."\\
                \item Identified Firm Tickers: \textcolor{blue}{MON,TWTR,FB,CVX,XOM,F,FSLR,DIS,GS,AMZN,DD,GM}}
        \end{itemize}
    }
}\\
  \end{minipage}
  \caption{
    \textbf{Sample News of Market Summary Report}
  \newline 
  \small
  This sample news demonstrates the possibility of market report news having an iconic firm in the headline.
  }\label{fig:manyfollowerssample}
\end{figure}

\end{document}